\documentclass[longauth]{aa} 

\usepackage[varg]{txfonts}
\usepackage{graphicx}
\usepackage{rotating}
\usepackage{pdflscape}
\usepackage{lscape}
\usepackage{longtable}
\usepackage{enumerate}
\usepackage[flushleft]{threeparttable}
\usepackage{ulem}

\usepackage{natbib,twoopt}
\bibpunct{(}{)}{;}{a}{}{,}             
\usepackage{caption} 
\usepackage{subcaption}
\usepackage{xcolor} 
\usepackage{bibentry} 
\nobibliography*
\begin{document}

\title{The Gaia-ESO Public Spectroscopic Survey: Motivation, implementation, GIRAFFE data processing, analysis, and final data products\thanks{Based on observations collected at the ESO telescopes under programme 188.B3002, 193.B-0936, and 197.B-1074, the Gaia-ESO Public Spectroscopic Survey.}}

\author{G. Gilmore \inst{\ref{ioa}}
\and S. Randich\inst{\ref{oarcetri}}
\and C. C. Worley\inst{\ref{ioa}}
\and A. Hourihane\inst{\ref{ioa}}
\and A. Gonneau\inst{\ref{ioa}}
\and G. G. Sacco\inst{\ref{oarcetri}}
\and J. R. Lewis$^\dagger$\inst{\ref{ioa}}
\and L. Magrini\inst{\ref{oarcetri}}
\and P. Fran{\c c}ois\inst{\ref{oparis}}
\and R. D. Jeffries\inst{\ref{ukeele}}
\and S. E. Koposov\inst{\ref{ifa-roe},\ref{ioa}}
\and A. Bragaglia\inst{\ref{obologna}}
\and E. J. Alfaro\inst{\ref{iaa}}
\and C. Allende Prieto\inst{\ref{iac},\ref{ulaguna}}
\and R. Blomme\inst{\ref{obelgium}}
\and A. J. Korn\inst{\ref{uuppsala-oa}}
\and A. C. Lanzafame\inst{\ref{ucatania}}
\and E. Pancino\inst{\ref{oarcetri},\ref{ssdc}}
\and A. Recio-Blanco\inst{\ref{uca}}
\and R. Smiljanic\inst{\ref{ncac}}
\and S. Van Eck\inst{\ref{ulbruxelles}}
\and T. Zwitter\inst{\ref{uljubljana}}
\and T. Bensby\inst{\ref{olund}}
\and E. Flaccomio\inst{\ref{opalermo}}
\and M. J. Irwin\inst{\ref{ioa}}
\and E. Franciosini\inst{\ref{oarcetri}}
\and L. Morbidelli\inst{\ref{oarcetri}}
\and F. Damiani\inst{\ref{opalermo}}
\and R. Bonito\inst{\ref{opalermo}}
\and E. D. Friel\inst{\ref{uindiana}}
\and J. S. Vink\inst{\ref{oamagh}}
\and L. Prisinzano\inst{\ref{opalermo}}
\and U. Abbas\inst{\ref{otorino}}
\and D. Hatzidimitriou\inst{\ref{uathens},\ref{oathens}}
\and E. V. Held\inst{\ref{opadova}}
\and C. Jordi\inst{\ref{ubarcelona}}
\and E. Paunzen\inst{\ref{umasaryk}}
\and A. Spagna\inst{\ref{otorino}}
\and R. J. Jackson\inst{\ref{ukeele}}
\and J. Ma{\'i}z Apell{\'a}niz\inst{\ref{cda-vc}}
\and M. Asplund\inst{\ref{aas}}
\and P. Bonifacio\inst{\ref{oparis-m}}
\and S. Feltzing\inst{\ref{olund}}
\and J. Binney\inst{\ref{uoxford}}
\and J. Drew\inst{\ref{ucl}}
\and A. M. N. Ferguson\inst{\ref{ifa}}
\and G. Micela\inst{\ref{opalermo}}
\and I. Negueruela\inst{\ref{ualicante}}
\and T. Prusti\inst{\ref{esa}}
\and H.-W. Rix\inst{\ref{mpia}}
\and A. Vallenari\inst{\ref{opadova}}
\and M. Bergemann\inst{\ref{mpia},\ref{nbia}}
\and A. R. Casey\inst{\ref{umonash}}
\and P. de Laverny\inst{\ref{uca}}
\and A. Frasca\inst{\ref{ocatania}}
\and V. Hill\inst{\ref{uca}}
\and K. Lind\inst{\ref{ustockholm}}
\and L. Sbordone\inst{\ref{eso-chile}}
\and S. G. Sousa\inst{\ref{uporto-iace}}
\and V. Adibekyan\inst{\ref{uporto-iace}}
\and E. Caffau\inst{\ref{oparis-m}}
\and S. Daflon\inst{\ref{on-mcti}}
\and D. K. Feuillet\inst{\ref{olund},\ref{mpia}}
\and M. Gebran\inst{\ref{notredame}}
\and J. I. Gonz{\'a}lez Hern{\'a}ndez\inst{\ref{iac}}
\and G. Guiglion\inst{\ref{aip}}
\and A. Herrero\inst{\ref{iac},\ref{ulaguna}}
\and A. Lobel\inst{\ref{obelgium}}
\and T. Merle\inst{\ref{ulbruxelles}}
\and {\u S}. Mikolaitis\inst{\ref{uvilnius}}
\and D. Montes\inst{\ref{ucmadrid}}
\and T. Morel\inst{\ref{uliege}}
\and G. Ruchti$\dagger$\inst{\ref{olund}}
\and C. Soubiran\inst{\ref{ubordeaux}}
\and H. M. Tabernero\inst{\ref{cda-ta}}
\and G. Tautvai{\v s}ien{\. e}\inst{\ref{uvilnius}}
\and G. Traven\inst{\ref{uljubljana}}
\and M. Valentini\inst{\ref{aip}}
\and M. Van der Swaelmen\inst{\ref{oarcetri}}
\and S. Villanova\inst{\ref{uconcepcion}}
\and C. Viscasillas V{\'a}zquez\inst{\ref{uvilnius}}
\and A. Bayo\inst{\ref{uval1},\ref{uval2}}
\and K. Biazzo\inst{\ref{oroma}}
\and G. Carraro\inst{\ref{upadova}}
\and B. Edvardsson\inst{\ref{uuppsala-ta}}
\and U. Heiter\inst{\ref{uuppsala-oa}}
\and P. Jofr{\'e}\inst{\ref{udp}}
\and G. Marconi\inst{\ref{eso-chile}}
\and C. Martayan\inst{\ref{eso-chile}}
\and T. Masseron\inst{\ref{iac},\ref{ulaguna}}
\and L. Monaco\inst{\ref{uandresbello}}
\and N. A. Walton\inst{\ref{ioa}}
\and S. Zaggia\inst{\ref{opadova}}
\and V. Aguirre B{\o}rsen-Koch\inst{\ref{uaarhus}}
\and J. Alves\inst{\ref{uvienna}}
\and L. Balaguer-N{\'u}nez\inst{\ref{ubarcelona}}
\and P. S. Barklem\inst{\ref{uuppsala-ta}}
\and D. Barrado\inst{\ref{cda-vc2}}
\and M. Bellazzini\inst{\ref{obologna}}
\and S. R. Berlanas\inst{\ref{ualicante}}
\and A. S. Binks\inst{\ref{mit},\ref{ukeele}}
\and A. Bressan\inst{\ref{sissa}}
\and R. Capuzzo--Dolcetta\inst{\ref{uroma}}
\and L. Casagrande\inst{\ref{rsaa}}
\and L. Casamiquela\inst{\ref{ubordeaux}}
\and R. S. Collins\inst{\ref{ifa-roe}}
\and V. D'Orazi\inst{\ref{opadova}}
\and M. L. L. Dantas\inst{\ref{ncac}}
\and V. P. Debattista\inst{\ref{uclancashire}}
\and E. Delgado-Mena\inst{\ref{uporto-iace}}
\and P. Di Marcantonio\inst{\ref{otrieste}}
\and A. Drazdauskas\inst{\ref{uvilnius}}
\and N. W. Evans\inst{\ref{ioa}}
\and B. Famaey\inst{\ref{ustrasbourg}}
\and M. Franchini\inst{\ref{otrieste}}
\and Y. Fr{\'e}mat\inst{\ref{obelgium}}
\and X. Fu\inst{\ref{upeking}}
\and D. Geisler\inst{\ref{uconcepcion},\ref{iimct},\ref{userena}}
\and O. Gerhard\inst{\ref{mpiep}}
\and E. A. Gonz{\'a}lez Solares\inst{\ref{ioa}}
\and E. K. Grebel\inst{\ref{zfa-ari}}
\and M. L. Guti{\'e}rrez Albarr{\'a}n\inst{\ref{ucmadrid}}
\and F. Jim{\'e}nez-Esteban\inst{\ref{cda-vc}}
\and H. J{\"o}nsson\inst{\ref{umalmo}}
\and T. Khachaturyants\inst{\ref{uclancashire}}
\and G. Kordopatis\inst{\ref{uca}}
\and J. Kos\inst{\ref{uljubljana}}
\and N. Lagarde\inst{\ref{ubordeaux},\ref{ubfc}}
\and H.-G. Ludwig\inst{\ref{zfa-l}}
\and L. Mahy\inst{\ref{obelgium}}
\and M. Mapelli\inst{\ref{opadova}}
\and E. Marfil\inst{\ref{cda-vc}}
\and S. L. Martell\inst{\ref{unsw}}
\and S. Messina\inst{\ref{ocatania}}
\and A. Miglio\inst{\ref{ubologna},\ref{obologna}}
\and I. Minchev\inst{\ref{aip}}
\and A. Moitinho\inst{\ref{ulisbon}}
\and J. Montalban\inst{\ref{ubologna}}
\and M. J. P. F. G. Monteiro\inst{\ref{uporto-iace},\ref{uporto-dfa}}
\and C. Morossi\inst{\ref{otrieste}}
\and N. Mowlavi\inst{\ref{ugeneva}}
\and A. Mucciarelli\inst{\ref{ubologna},\ref{obologna}}
\and D. N. A. Murphy\inst{\ref{ioa}}
\and N. Nardetto\inst{\ref{uca}}
\and S. Ortolani\inst{\ref{upadova}}
\and F. Paletou\inst{\ref{utoulouse}}
\and J. Palou{\u s}\inst{\ref{cas}}
\and J. C. Pickering\inst{\ref{icl}}
\and A. Quirrenbach\inst{\ref{zfa-l}}
\and P. Re Fiorentin\inst{\ref{otorino}}
\and J. I. Read\inst{\ref{usurrey}}
\and D. Romano\inst{\ref{obologna}}
\and N. Ryde\inst{\ref{olund}}
\and N. Sanna\inst{\ref{oarcetri}}
\and W. Santos\inst{\ref{on-mcti}}
\and G. M. Seabroke\inst{\ref{mssl}}
\and L. Spina\inst{\ref{opadua}}
\and M. Steinmetz\inst{\ref{aip}}
\and E. Stonkut{\'e}\inst{\ref{uvilnius-ao}}
\and E. Sutorius\inst{\ref{ifa-roe}}
\and F. Th{\'e}venin\inst{\ref{uca}}
\and M. Tosi\inst{\ref{obologna}}
\and M. Tsantaki\inst{\ref{oarcetri}}
\and N. Wright\inst{\ref{ukeele}}
\and R. F. G. Wyse\inst{\ref{ujohnhopkins}}
\and M. Zoccali\inst{\ref{ucatolica}}
\and J. Zorec\inst{\ref{usorbonne}}
\and D. B. Zucker\inst{\ref{umacquarie}}
}

\institute{Institute of Astronomy, University of Cambridge, Madingley Road, Cambridge CB3 0HA, United Kingdom \\ \email{gil@ast.cam.ac.uk} \label{ioa}
\and INAF - Osservatorio Astrofisico di Arcetri, Largo E. Fermi, 5, 50125, Firenze, Italy\label{oarcetri}
\and GEPI, Observatoire de Paris, PSL Research University, CNRS, Univ. Paris Diderot, Sorbonne Paris Cit{\'e}, 61 avenue de l'Observatoire, 75014, Paris, France\label{oparis}
\and Astrophysics Group, Keele University, Keele, Staffordshire ST5 5BG, United Kingdom\label{ukeele}
\and Institute for Astronomy, Royal Observatory, University of Edinburgh, Blackford Hill, Edinburgh EH9 3HJ, UK\label{ifa-roe}
\and INAF - Osservatorio di Astrofisica e Scienza dello Spazio, via P. Gobetti 93/3, 40129 Bologna, Italy\label{obologna}
\and Instituto de Astrof{\'i}sica de Andaluc{\'i}a, CSIC, Glorieta de la Astronom{\'i}a s/n, Granada 18008, Spain\label{iaa}
\and Instituto de Astrof{\'i}sica de Canarias, V{\'i}a L{\'a}ctea s/n, E-38205 La Laguna, Tenerife, Spain\label{iac}
\and Departamento de Astrof{\'i}sica, Universidad de La Laguna, E-38205 La Laguna, Tenerife, Spain\label{ulaguna}
\and ROB - Royal Observatory of Belgium, Ringlaan 3, B-1180 Brussels, Belgium\label{obelgium}
\and Observational Astrophysics, Division of Astronomy and Space Physics, Department of Physics and Astronomy, Uppsala University, Box 516, 75120 Uppsala, Sweden\label{uuppsala-oa}
\and Dipartimento di Fisica e Astronomia, Sezione Astrofisica, Universit{\'a} di Catania, via S. Sofia 78, 95123, Catania, Italy\label{ucatania}
\and Space Science Data Center - Agenzia Spaziale Italiana, via del Politecnico, s.n.c., I-00133, Roma, Italy\label{ssdc}
\and Universit{\'e} C{\^o}te d'Azur, Observatoire de la C{\^o}te d'Azur, CNRS, Laboratoire Lagrange, Bd de l'Observatoire, CS 34229, 06304 Nice cedex 4, France\label{uca}
\and Nicolaus Copernicus Astronomical Center, Polish Academy of Sciences, ul. Bartycka 18, 00-716, Warsaw, Poland\label{ncac}
\and Institut d'Astronomie et d'Astrophysique, Universit{\'e} Libre de Bruxelles, CP 226, Boulevard du Triomphe, B-1050 Bruxelles, Belgium\label{ulbruxelles}
\and Faculty of Mathematics and Physics, University of Ljubljana, Jadranska 19, 1000 Ljubljana, Slovenia\label{uljubljana}
\and Lund Observatory, Department of Astronomy and Theoretical Physics, Box 43, SE-22100 Lund, Sweden\label{olund}
\and INAF - Osservatorio Asronomico di Palermo, Piazza del Parlamento, 1 90134 Palermo, Italy\label{opalermo}
\and Astronomy Department, Indiana University, 727 East 3rd St, Bloomington, IN 47405, USA\label{uindiana}
\and Armagh Observatory and Planetarium, College Hill, Armagh BT61 9DG, United Kingdom\label{oamagh}
\and INAF - Osservatorio Astrofisico di Torino, via Osservatorio 20, I-10025 Pino Torinese, Italy\label{otorino}
\and Section of Astrophysics, Astronomy and Mechanics, Department of Physics, National and Kapodistrian University of Athens, GR15784, Athens, Greece\label{uathens}
\and IAASARS, National Observatory of Athens, GR15236, Penteli, Greece\label{oathens}
\and INAF - Osservatorio Astronomico di Padova, Vicolo dell’Osservatorio 5, I-35122, Padova, Italy\label{opadova}
\and Institut de Ci{\`e}ncies del Cosmos (ICCUB), Universitat de Barcelona (IEEC-UB), Mart{\'i} i Franqu{\`e}s 1, E-08028 Barcelona, Spain\label{ubarcelona}
\and Department of Theoretical Physics and Astrophysics, Faculty of Science, Masaryk University, Kotlarska 2, 611 37 Brno, Czech Republic\label{umasaryk}
\and Centro de Astrobiologia (CSIC-INTA), Departamento de Astrofisica, campus ESAC. Camino bajo del castillo s/n. 28 692 Villanueva de la Canada, Madrid, Spain.\label{cda-vc}
\and Australian Academy of Science, Box 783, Canberra ACT 2601, Australia\label{aas}
\and GEPI, Observatoire de Paris, Universit{\'e} PSL, CNRS, 5 Place Jules Janssen, 92190 Meudon, France\label{oparis-m}
\and Rudolf Peierls Centre for Theoretical Physics, Clarendon Laboratory, Parks Road, Oxford OX1 3PU, United Kingdom\label{uoxford}
\and Department of Physics and Astronomy, University College London, Gower Street, London WC1E 6BT, United Kingdom\label{ucl}
\and Institute for Astronomy, University of Edinburgh, Blackford Hill, Edinburgh EH9 3HJ UK\label{ifa}
\and Departamento de F{\'i}sica Aplicada, Facultad de Ciencias, Universidad de Alicante, 03690 San Vicente del Raspeig, Alicante, Spain\label{ualicante}
\and European Space Agency (ESA), European Space Research and Technology Centre (ESTEC), Keplerlaan 1, 2201 AZ Noordwijk, The Netherlands\label{esa}
\and Max-Planck-Institut fur Astronomie, K{\"o}nigstuhl 17, D-69117 Heidelberg, Germany\label{mpia}
\and Niels Bohr International Academy, Niels Bohr Institute, Blegdamsvej 17, DK-2100 Copenhagen {\O}, Denmark\label{nbia}
\and School of Physics and Astronomy, Monash University, Wellington Road, Clayton 3800, Victoria, Australia\label{umonash}
\and INAF - Osservatorio Astrofisico di Catania, Via S. Sofia 78, 95123 Catania, Italy\label{ocatania}
\and Department of Astronomy, Stockholm University, AlbaNova University Center, SE-106 91 Stockholm, Sweden\label{ustockholm}
\and ESO - European Organisation for Astronomical Research in the Southern Hemisphere, Alonso de C{\'o}rdova 3107, Vitacura, 19001 Casilla, Santiago de Chile, Chile\label{eso-chile}
\and Instituto de Astrof{\'i}sica e Ci{\^e}ncias do Espa{\c c}o, Universidade do Porto, CAUP, Rua das Estrelas, 4150-762 Porto, Portugal\label{uporto-iace}
\and Observat{\'o}rio Nacional - MCTI (ON), Rua Gal. Jos{\'e} Cristino 77, S{\~a}o Crist{\'o}v{\~a}o, 20921-400, Rio de Janeiro, Brazil\label{on-mcti}
\and Department of Chemistry and Physics, Saint Mary’s College, Notre Dame, IN 46556, USA\label{notredame}
\and Leibniz-Institut f{\"u}r Astrophysik Potsdam (AIP), An der Sternwarte 16, 14482 Potsdam, Germany\label{aip}
\and Institute of Theoretical Physics and Astronomy, Vilnius University, Sauletekio av. 3, LT-10257 Vilnius, Lithuania\label{uvilnius}
\and Departamento de F{\'i}sica de la Tierra y Astrof{\'i}sica and IPARCOS-UCM (Instituto de F{\'i}sica de Partículas y del Cosmos de la UCM), Facultad de Ciencias F{\'i}sicas, Universidad Complutense de Madrid, E-28040 Madrid, Spain\label{ucmadrid}
\and Space Sciences, Technologies, and Astrophysics Research (STAR) Institute, Universit{\'e} de Li{\`e}ge, Quartier Agora, B{\^a}t B5c, All{\'e}e du 6 ao{\^u}t, 19c, 4000 Li{\`e}ge, Belgium\label{uliege}
\and Laboratoire d'astrophysique de Bordeaux, Univ. Bordeaux, CNRS, B18N, all{\'e}e Geoffroy Saint-Hilaire, 33615 Pessac, France\label{ubordeaux}
\and Centro de Astrobiologia (CSIC-INTA), Carretera de Ajalvir km 4, E-28850 Torrejon de Ardoz, Madrid, Spain\label{cda-ta}
\and Departamento de Astronom{\'i}a, Casilla 160-C, Universidad de Concepci{\'o}n, Concepci{\'o}n, Chile\label{uconcepcion}
\and Instituto de F{\'i}sica y Astronom{\'i}a, Facultad de Ciencias, Universidad de Valpara{\'i}so, Chile\label{uval1}
\and N{\'u}cleo Milenio Formaci{\'o}n Planetaria - NPF, Universidad de Valpara{\'i}so, Chile\label{uval2}
\and INAF - Osservatorio Astronomico di Roma, Via Frascati 33, I-00040 Monte Porzio Catone (Roma), Italy\label{oroma}
\and Department of Physics and Astronomy, University of Padova, v. dell'Osservatorio 2, 35122, Padova, Italy\label{upadova}
\and Theoretical Astrophysics, Department of Physics and Astronomy, Uppsala University, Box 516, SE-751 20 Uppsala, Sweden\label{uuppsala-ta}
\and N{\'u}cleo de Astronom{\'i}a, Facultad de Ingenier{\'i}a y Ciencias, Universidad Diego Portales, Av. Ej{\'e}rcito 441, Santiago, Chile\label{udp}
\and Departamento de Ciencias Fisicas, Universidad Andres Bello, Fernandez Concha 700, Las Condes, Santiago, Chile\label{uandresbello}
\and Stellar Astrophysics Centre, Department of Physics and Astronomy, Aarhus University, Ny Munkegade 120, DK-8000 Aarhus C, Denmark\label{uaarhus}
\and University of Vienna, Dept. Astrophysics, T{\"u}rkenschanzstrasse 17, 1180 Vienna, Austria\label{uvienna}
\and Centro de Astrobiolog{\'i}a (INTA-CSIC), Camino Bajo del Castillo s/n, 28692, Villanueva de la Ca{\~n}ada, Madrid, Spain\label{cda-vc2}
\and Massachusetts Institute of Technology, Kavli Institute for Astrophysics and Space Research, 77 Massachusetts Ave., Cambridge, MA 02139, USA\label{mit}
\and SISSA, via Bonomea, 265 - 34136 Trieste, Italy\label{sissa}
\and Dep. of Physics, Sapienza, University of Roma, Roma, Italy\label{uroma}
\and Research School of Astronomy and Astrophysics, Australian National University, ACT 2611, Australia\label{rsaa}
\and Jeremiah Horrocks Institute, University of Central Lancashire, Preston PR1 2HE, United Kingdom\label{uclancashire}
\and INAF - Osservatorio Astronomico di Trieste, Via G.B Tiepolo, 11 I-34143 Trieste, Italy\label{otrieste}
\and Universit{\'e} de Strasbourg, CNRS, Observatoire Astronomique de Strasbourg, UMR 7550, F-67000 Strasbourg, France\label{ustrasbourg}
\and The Kavli Institute for Astronomy and Astrophysics at Peking University, 100871, Beijing, PR China\label{upeking}
\and Instituto de Investigaci{\'o}n Multidisciplinario en Ciencia y Tecnolog{\'i}a, Universidad de La Serena, Avenida Ra{\'u}l Bitr{\'a}n s/n, La Serena, Chile\label{iimct}
\and Departamento de Astronom{\'i}a, Facultad de Ciencias, Universidad de La Serena, Av. Juan Cisternas 1200, La Serena, Chile\label{userena}
\and Max-Planck-Institute for Ex. Physics, Giessenbachstr.1, 85748 Garching, Germany\label{mpiep}
\and Astronomisches Rechen-Institut, Zentrum f{\"u}r Astronomie der Universit{\"a}t Heidelberg, M{\"o}nchhofstr.\ 12--14, 69120 Heidelberg, Germany\label{zfa-ari}
\and Materials Science and Applied Mathematics, Malm{\"o} University, SE-205 06 Malm{\"o}, Sweden\label{umalmo}
\and Institut UTINAM, CNRS UMR6213, Univ. Bourgogne Franche-Comt{\'e}, OSU THETA Franche-Comt{\'e}-Bourgogne, Observatoire de Besan{\c c}on, BP 1615, 25010 Besan{\c c}on Cedex, France\label{ubfc}
\and Zentrum fur Astronomie der Universit{\"a}t Heidelberg, Landessternwarte, K{\"o}nigstuhl 12, 69117 Heidelberg, Germany\label{zfa-l}
\and School of Physics, University of New South Wales, Sydney 2052, Australia\label{unsw}
\and Dipartimento di Fisica e Astronomia, Universit{\`a} degli Studi di Bologna, Via Gobetti 93/2, I-40129 Bologna, Italy\label{ubologna}
\and CENTRA, Faculdade de Ci\^encias, Universidade de Lisboa, Ed. C8, Campo Grande, 1749-016 Lisboa, Portugal\label{ulisbon}
\and Departamento de F{\'i}sica e Astronomia, Faculdade de Ci{\'e}ncias da Universidade do Porto, Portugal\label{uporto-dfa}
\and Department of Astronomy, University of Geneva, 51 chemin Pegasi, 1290 Versoix, Switzerland\label{ugeneva}
\and Universit{\'e} de Toulouse, Observatoire Midi-Pyr{\'e}n{\'e}es, CNRS, IRAP, 14 av. E. Belin, F-31400 Toulouse\label{utoulouse}
\and Astronomical Institute, CAS, Bo{\u c}n{\'i} II 1401, 141 00 Prague 4, Czech Republic\label{cas}
\and Physics Department, Imperial College London, Prince Consort Road, London SW7 2BZ, United Kingdom\label{icl}
\and University of Surrey, Physics Department, Guildford, GU2 7XH, UK\label{usurrey}
\and Mullard Space Science Laboratory, University College London, Holmbury St Mary, Dorking, Surrey, RH5 6NT, United Kingdom\label{mssl}
\and INAF - Astronomical Observatory of Padua, Vicolo dell’Osservatorio, 5, 35122 Padua PD, Italy\label{opadua}
\and Astronomical Observatory, Institute of Theoretical Physics and Astronomy, Vilnius University, Sauletekio av. 3, 10257 Vilnius, Lithuania\label{uvilnius-ao}
\and Department of Physics and Astronomy, Johns Hopkins University, Baltimore, MD 21218, USA\label{ujohnhopkins}
\and Institute of Astrophysics, Pontificia Universidad Cat{\'o}lica de Chile, Av. Vicu{\~n}a Mackenna 4860, Macul, Santiago, Chile\label{ucatolica}
\and Sorbonne Universit{\'e}, CNRS, UPMC, UMR7095 Institut d'Astrophysique de Paris, 98bis Bd. Arago, F-75014 Paris, France\label{usorbonne}
\and Department of Physics and Astronomy, Macquarie University, Sydney, NSW 2109, Australia\label{umacquarie}
}


 \date{Received  01/2022; Accepted}
 \abstract
{
The Gaia-ESO Public Spectroscopic Survey is an ambitious project
  designed to obtain astrophysical parameters and elemental abundances
  for 100,000 stars, including large representative samples of the  stellar populations 
  in the Galaxy, and a well-defined sample of $~60$ (plus 20 archive) open clusters. We provide internally consistent results calibrated on benchmark stars and star clusters, extending across a very wide range of abundances and ages. This  provides a legacy data set of intrinsic value,  and equally a large wide-ranging dataset that is of value for homogenisation of other and future stellar surveys and Gaia's astrophysical parameters. } 
{This article provides an overview 
  of the survey methodology, the scientific aims, and the implementation, including a description of the data processing for the GIRAFFE spectra. A companion paper introduces the survey results.
} 
{ Gaia-ESO aspires to quantify both random and systematic contributions
to measurement uncertainties. Thus  all available spectroscopic analysis techniques are
  utilised, each spectrum being analysed by up to several different analysis pipelines, 
with considerable effort being made to homogenise and calibrate the resulting parameters. 
We describe here the sequence of activities up to delivery of
  processed data products to the ESO Science Archive Facility for
  open use. }
{ The Gaia-ESO Survey obtained 202,000 spectra of 115,000 stars using 340 allocated VLT nights between  December 2011 and January 2018 from GIRAFFE and UVES. }
{  The full consistently reduced final data set of spectra was released through the ESO Science Archive Facility in late 2020, with the full astrophysical parameters sets following in 2022. A companion article reviews the survey implementation, scientific highlights, the open cluster survey, and data products.
}
\authorrunning{Gilmore, Randich et al.}
\titlerunning{The Gaia-ESO Survey}
\maketitle
\section {Introduction  \label{Intro}}

This is one of two papers providing the overview of the Gaia-ESO Public Spectroscopic Survey of 
stellar populations. This survey has utilised the ESO VLT and multi-object FLAMES facility, and both GIRAFFE and UVES spectrographs,
 to derive astrophysical parameters and elemental abundances for some 100,000 stars.
The companion survey overview paper, \citet{Randich21}, describes the primary scientific results, illustrating the outcomes of the many activities described in this paper. That paper also provides additional summary information on the open cluster aspects of Gaia-ESO. It is our intent in these two papers to document the origins, motivation, original case, structure as implemented, operation, and a summary of the outputs from the Survey. We also take care to identify those key individuals who led the work packages during the survey operation. In any large consortium it is a challenge to acknowledge due credit. In Gaia-ESO we have attempted to do this by identifying these individuals in this paper, and importantly by authorship policy - each key work package is described in its own paper, with lead authors the key individuals, as listed in this section below. 

This paper is organised as follows. Following this introduction (Section~\ref{Intro}), which includes the list of reference articles describing the Gaia-ESO Survey, in Section~$\ref{scicase}$ we
provide a summary overview of the science case for the 
survey, and introduce the envisioned legacy. 

Section~$\ref{strategy}$
introduces our ambition to define both measurement precision and accuracy, which
explains our motivation for involving the many different spectroscopic analysis methodologies available in the community, 
with significant effort in subsequent homogenisation to provide a single recommended parameter set per star. In Section~$\ref{structure}$ we
describe the survey working group (WG) structure and organisation.  The match of scientific ambition to practical target selection for each aspect of the survey is described in Section~\ref{obs-strategy}.
Section~$\ref{targets}$ provides the top-level overview of
 the target selection and sky coverage. It then describes the practical implementation effort, provided by WGs 0-6. As part of that effort we introduce our
approach to maximise the Survey legacy, which leads to substantial calibration effort.   Sub-section~$\ref{calibration}$ describes
the efforts to calibrate the survey, ensuring consistency which makes the survey results valuable for Gaia and allows cross-calibration with other present and future ground-based large
spectroscopic surveys and the asteroseismology space missions.   Section~$\ref{JRLpipeline}$
describes the GIRAFFE data reduction pipeline developed and enhanced
as part of the Gaia-ESO Survey, which generated the reduced
spectra which are publicly released as a survey product. Within this,
sub-section~$\ref{sky}$ provides a more detailed discussion of sky
subtraction, considering the various GIRAFFE settings we used and
relevant astrophysical background sources.
Section~$\ref{wg8}$ describes the calculation of radial velocities, and
their accuracies. This process also determines
first-pass astrophysical parameters for each spectrum, useful as a
starting point in later more detailed analysis. 
Section~$\ref{spectrum_analyses}$ provides an overview of the
several approaches to spectrum analyses, and determination of the
published recommended parameters. More detailed articles
describing each method are referenced as appropriate. 
The astrophysical parameters and elemental abundances output from the various pipelines, nodes, and Working Groups must be calibrated onto a single internally consistent system, which is consistent with 
the calibration effort. This very challenging task is described in Section~\ref{WG15}. Overview of the Survey, and the structure and operation of the data flows through the working and then survey databases is described in Section~\ref{monitor}.
The data
released to the public through ESO Science Archive Facility (ESO SAF) data releases
are explained in Section~$\ref{ESODR}$. Section~$\ref{Conclusions}$
summarises the scientific and operational status of the Gaia-ESO Survey.

 The Gaia-ESO Survey is an ESO public spectroscopic survey, targeting 
$10^5$ stars, systematically covering all the major components of the
Milky Way, from halo to star forming regions, providing an
homogeneous overview of the distributions of kinematics and elemental
abundances. The Survey utilises both medium ($R \simeq 20,000$) and high ($R \simeq 50,000$) 
resolution spectroscopy, and reaches faint enough to explore a significant range of Galacto-centric distances. This alone will contribute to our knowledge of Galactic and
stellar evolution: when combined with Gaia astrometry, the Survey helps
quantify the formation history and evolution of young, mature and
ancient Galactic populations. With well-defined samples, based
primarily on ESO-VISTA photometry for the field stars, and  from a
variety of photometric surveys of open clusters, the Survey quantifies
the kinematic-multi-element abundance distribution functions of the
bulge, the thick and the thin discs and the halo stellar components,
as well as a significant sample of $\sim 60$ open star clusters,
covering all accessible cluster ages and stellar masses.
A brief overview is available at \citet{Mess1}, with an early progress
report at \citet{Mess2}. These outline the pre-history of the project,
and the partnership between ESO and the Gaia-ESO Survey team in
developing and implementing this ambitious Public Spectroscopic Survey.

The Survey has obtained VLT/FLAMES spectra to quantify individual
elemental abundances; yield precise radial velocities for a 4-D
kinematic phase-space; map kinematic gradients and abundance -
phase-space structure throughout the Galaxy; and follow the formation,
evolution and dissolution of open clusters as they populate the
disc. Several GIRAFFE settings, optimised for the astrophysical
parameters of each target group, and parallel UVES spectra have been
obtained for each surveyed open cluster. GIRAFFE spectra, with two
settings, have been obtained for statistically significant samples of
stars in all major stellar populations. These are supplemented by UVES
spectra of an unbiased sample of G-stars within $\geq 1$kpc of the
Sun, providing the abundance distribution function for the local thin
disc, thick disc and halo. The open cluster survey is described in a
companion article \citep{Randich21}. The Survey is designed to provide
a legacy dataset that adds enormous value to the Gaia mission and
ongoing ESO imaging surveys.

The Gaia-ESO Survey delivers the data to support a wide variety of
studies of stellar populations, the evolution of dynamical systems,
and stellar evolution. Gaia-ESO complements Gaia by using high-resolution spectra from UVES to
measure the metallicity and detailed abundances for several chemical
elements in $\sim5000$ field stars with $V\leq 15$ and in $\sim 2000$
open cluster members down to $V\sim 16.5$. Depending on target
signal-to-noise ratio (SNR), and astrophysical parameters, the Survey
typically probes the two fundamental nucleosynthetic channels, nuclear
statistical equilibrium (V, Cr, Mn, Fe, Co), and $\alpha$-chain (Si,
Ca, Ti). [Fe/H], $[\alpha$/Fe], and some other element abundance
ratios have been  obtained from the lower resolution GIRAFFE spectra. The radial velocity (RV)
precision for this sample is $\simeq 0.1$\,km\,s$^{-1}$ to $\leq
5$\,km\,s$^{-1}$, depending on target, with in each case the
measurement precision being that appropriate for a range of relevant
astrophysical analyses. 

The Gaia-ESO dataset supports analyses which aim to identify, on both chemical and kinematic
grounds, phase-space substructures that bear witness to specific
merger or starburst events. The dataset also allows mapping the
dissolution of clusters, and the Galactic migration of field
stars. The Survey not only supplies homogeneously determined element
abundances, but also complementary astrophysical information for large
samples of members of clusters with precise distances from
Gaia. This information can be used to challenge models of stellar
structure and evolution, as well as to test models of mass accretion
from circumstellar discs into the star \citep{Randich18}.

A substantial observational effort has been devoted to abundance
calibration, establishing targets in common with other spectroscopic
surveys, including observations of CoRoT and Kepler stars, and expanding the grid of ``benchmark stars'' which act as primary spectroscopic
standards, to ensure maximal legacy value. The Gaia-ESO Survey
additionally invested considerable effort in re-analyses of spectra already available in the ESO SAF, where scientifically complementary to the primary Survey.

The Survey consortium and operations  was  structured into a set of working 
groups, with the whole overseen by a Steering Committee. As a partial motivation
for the Gaia-ESO Survey was to build an ESO-wide community ready to
reap the vast scientific potential of Gaia, we explicitly included all
groups in the ESO community active, at the time the Survey was designed, in precision stellar
spectroscopy, with their ranges of expertise and methods. This provided
the opportunity to cross-calibrate the various available methods, and
identify possible systematic differences, particularly based on
analyses of the well-quantified benchmark calibrator stars. The outcomes 
of all these methods were then homogenised into the recommended astrophysical parameters for each star which became available publicly through the ESO SAF (Sect~\ref{ESODR}), and also
through the Gaia-ESO Survey archive hosted at the Wide Field Astronomy Unit, Edinburgh (WFAU) (Sect~\ref{monitor}).

The big themes in astronomy require complementary space and ground based observations. 
To maximise the scientific output it is necessary to coordinate the  European
efforts. ESO and ESA have recognised this coordination necessity in various topics which can and must be addressed both from the ground and in space. Joint working groups have been nominated for selected topics and the fourth such group was central to this project. This ESA-ESO working group (chaired by Catherine Turon) addressed the topic of "Galactic Populations, Chemistry and Dynamics". The report of the
working group was published in 2008 and remains up to date today \citep{Turon2008}. Many recommendations  from that study are of relevance to this project, but the key ones can be summarised in two words covering both space and ground: Gaia and spectroscopy. Gaia began its science operations in July 2014. The Gaia-ESO public survey aimed to support the European stellar spectroscopy community  to deliver full value from the Gaia potential for our Milky Way. The Gaia-ESO Survey had very ambitious goals. It included spectral types from O to M, all stellar populations, field and clusters, open and globular, all stellar ages. This ambition  made it a particularly challenging endeavour. The spectroscopic data
products are made available to the community in the same time frame as the
 intermediate Gaia catalogues. This allows the European and global - all survey products, like those of Gaia, are fully open-access -  scientific community to
address a multitude of galactic astronomy topics with the combined spectroscopic and Gaia data sets.

While this Survey was a substantial effort in its own right, and was the first dedicated stellar spectroscopic survey using 8-m class telescopes, it has been clear since early planning for Gaia that dedicated highly-multiplexed wide field spectroscopic facilities were needed for effective Gaia science exploitation. These are arriving, with many major surveys, with a range of spectral resolutions. These include (alphabetically) APOGEE \citep{APOGEE2014}, Gaia-RVS \citep{Gaia-mission}, GALAH \citep{GALAH2015}, LAMOST \citep{LAMOST2012},  and RAVE \citep{RAVE2006}, among many others. 
MOONS \citep{MOONS-mess}, WEAVE \citep{WEAVE2016} and 4MOST \citep{4MOST2019} are among those in construction with major European involvement. 

As a consortium-building exercise in preparation for these major long-term projects, a secondary ambition for the Gaia-ESO Public Spectroscopic Survey was to bring together the many high-quality stellar spectroscopy groups across Europe. Having all these groups applying their specific expertise, and learning to communicate and compare data and analyses, helped to build the successful consortia now carrying the subject forward. 

Additionally, the project aspired to help identify and reduce the impact of the factors which lead to systematic scale differences between survey pipeline outputs. For this, a very major effort was committed to developing a homogeneous atomic and molecular line list relevant to abundance analyses of FGK-type stars in the relevant Gaia-ESO passbands 480-680\,nm and 850-900\,nm \citep{Heiter2021}. To ensure the highest homogeneity possible in the
quantities derived, all the different Gaia-ESO spectrum
analysis methods adopted the same atomic and molecular
data, as well as the same set of
model atmospheres to the extent possible - see sect~\ref{spectrum_analyses}. 

In order to widen consistent cross-calibration possibilities between ground-based spectroscopic, asteroseismic and Gaia analyses, considerable overlaps of targets with CoRoT and Kepler stars were ensured, while significant samples of stars close to the equator (SDSS Stripe 82) were included. Substantial efforts were also invested in developing the Gaia Benchmarks calibrator stars, and ensuring consistency between that calibrator set and the overall Gaia-ESO elemental abundance and astrophysical parameter scales. Of course, the fundamental design of Gaia-ESO, with a substantial sample of open clusters of all available ages and abundances, extending to field stars with a wide range of ages and abundances, ensures the Gaia-ESO calibration is fundamentally tied to observationally-tested isochrones on stars with Gaia and spectroscopic data to ensure reliable cluster membership.

Publication of the full Survey overview in a single data release paper makes it difficult to ensure full credit is given to the leaders and members of the work packages who have invested considerable efforts to deliver this survey. Hence we decided not to follow the model of a single survey description paper at each of the data releases, but rather to describe the technical work in a series of specific articles.
At the time of writing there are over 100 Gaia-ESO Survey articles already published by the consortium. We list here (Table~\ref{Gaia-ESO-refs}) those which are reference and methods articles of direct relevance to understanding the final data products. The Gaia Benchmark Stars is a joint project to provide high-precision calibrator stars. Those references are also included in the Table.

\begin{table*}
\begin{center}
\caption{Gaia-ESO Survey Methods description papers}
\begin{tabular}{l|c}
\hline
Article title & published reference   \\
\hline
The Gaia-ESO Public Spectroscopic Survey & \citet{Mess1} \\
     The Gaia-ESO Large Public Spectroscopic Survey. & \citet{Mess2} \\
     The Gaia-ESO Survey: processing FLAMES-UVES spectra. & \citet{Sacco-UVES} \\
     Fe I oscillator strengths for the Gaia-ESO survey. & \citet{Ruffoni14} \\
     The Gaia-ESO Survey: The analysis of high-resolution \\
    \hskip0.5truecm UVES spectra of FGK-type stars. & \citet{Smiljanic14} \\
     The Gaia-ESO Survey: Extracting diffuse interstellar bands \\
     \hskip0.5truecm from cool star spectra.  &\citet{Puspitarini15} \\
     Gaia-ESO Survey: Analysis of pre-main sequence stellar spectra. & \citet{Lanzafame15} \\ 
     The Gaia-ESO Survey: Empirical determination of the \\
     \hskip0.5truecm precision of stellar radial velocities and projected rotation velocities. & \citet{Jackson15} \\
     The Gaia-ESO Survey Astrophysical Calibration. & \citet{Pancino17} \\
     The Gaia-ESO Survey: the selection function of the Milky Way field stars. &\citet{Stonkute16} \\
     A Grid of NLTE Corrections for Sulphur Lines in Atmospheres \\
     \hskip0.5truecm  of Cool Stars for the Gaia-ESO Survey. & \citet{Korotin17} \\
     The Gaia-ESO Survey: double-, triple-, and quadruple-line \\
     \hskip0.5truecm  spectroscopic binary candidates. & \citet{Merle17} \\
     Gaia-ESO Survey: INTRIGOSS{\textemdash}A New Library of \\
     \hskip0.5truecm  High-resolution Synthetic Spectra. & \citet{Franchini18} \\
     Atomic data for the Gaia-ESO Survey. A line list \\
      \hskip0.5truecm  for analysis of UVES and GIRAFFE observations of cool stars. & \citet{Heiter2021} \\
     The Gaia-ESO Survey: detection and characterization \\
     \hskip0.5truecm  of single-line spectroscopic binaries. & \citet{MerleSB1} \\
     The Gaia-ESO Survey: Spectroscopic-Asteroseismic analysis \\
      \hskip0.5truecm  of the K2@Gaia-ESO stars. & \citet{Worley2020} \\
     The Gaia-ESO Survey: Target Selection of Open Cluster Stars. & \citet{Bragaglia_prep} \\
     The Gaia-ESO Survey: The analysis of hot-star spectra.  & \citet{Blomme21} \\
     The Gaia-ESO Survey: The analysis of the medium-resolution \\
     \hskip0.5truecm  GIRAFFE spectra of FGK stars. & \citet{Worley19_wg10} \\
     The Gaia-ESO Survey: Homogenisation of the multi-node \\
      \hskip0.5truecm astrophysical parameters sets. & \citet{Hourihane21} \\
     The Gaia-ESO Survey: Spectroscopic-Asteroseismic analysis of the \\
     \hskip0.5truecm  CoRoT@Gaia-ESO Stars. & \citet{Masseron_prep} \\
     The Gaia-ESO Survey: Survey implementation, data products, \\
     \hskip0.5truecm  open cluster survey, and legacy. & \citet{Randich21} \\
     The Gaia-ESO Public Spectroscopic Survey: motivation, implementation, \\
     \hskip0.5truecm  GIRAFFE data processing, analysis, and final data products. & this paper \\
       \\
\hline
\hline
    \\
     Gaia FGK benchmark stars: Metallicity. & \citet{Jofre14} \\
     The Gaia FGK benchmark stars. High resolution spectral library. &\citet{BC14} \\
     Gaia FGK benchmark stars: Effective temperatures and surface gravities. & \citet{Heiter15} \\
     Gaia FGK benchmark stars: abundances of {\ensuremath{\alpha}} and iron-peak elements. & \citet{Jofre15} \\
     Gaia FGK benchmark stars: new candidates at low metallicities. & \citet{Hawkins16_bench} \\
     Gaia FGK benchmark stars: a bridge between spectroscopic surveys. & \citet{Jofre17_conf} \\
     Gaia FGK benchmark stars: \\
     opening the black box of stellar element abundance determination. & \citet{Jofre17} \\
     The Gaia FGK Benchmark Stars Version 2.1. & \citet{Jofre18}  \\
     Benchmark ages for the Gaia benchmark stars. &\citet{Sahlholdt19} \\
\hline
\label{Gaia-ESO-refs}
\end{tabular}
\end{center}
\end{table*}

All Gaia-ESO Survey processed, calibrated and reduced data are available for unrestricted public access through the ESO Science SAF Facility web interface (Table~\ref{ESO-DRs}). 

Each data release has extensive documentation\footnote{ available for example, DR4, at https://www.eso.org/rm/api/v1/public/releaseDescriptions/152}.
In addition to a summary of the release content - that for DR4 is below - the release documentation provides a summary overview of the observing and reduction operations, the types of targets, and explanation of the (many) keywords.

DR4.0, Dec 2020, delivers about 190,000 stacked, quality-controlled, 1-d spectra (R between 18000 and 54000) of 114,500 unique stellar targets. These stars were observed with GIRAFFE and UVES from 31.12.2011 to 26.01.2018, during the entire time execution of the survey. These targets were selected from all the major structural components of the Milky Way: bulge, thick and thin discs, halo, including open star clusters of all ages and masses. The 1-D spectra from Gaia-ESO DR4 augment or update the spectra published in the previous data releases.

The DR5 release of the final derived abundances and astrophysical parameters associated with the spectra published in DR4 was published on May 16 2022.  The
Advanced Products are described further below, and in the companion paper \citet{Randich21}.

\begin{table*}
\begin{center}
\caption{Gaia-ESO Survey ESO Science Archive Facility Data Releases}
\begin{tabular}{l|lccc}
\hline
Name & Date  & Unique targets & Spectra & Advanced Products  \\
\hline
DR1 & 10-2013 &  3834  & 5654 & no \\
DR2 & 02-2015 &  15093  & 27359 & yes \\
DR3 & 08-2016 &  25534  & 44214  & yes \\
DR4 & 12-2020 &  114500  & 190200 & yes \\
DR5 & 05-2022 &  114324  &   202233  & all products \\
DR5.1 & late 2022 & 500 & 500 & remaining stars \\
\hline
\label{ESO-DRs}
\end{tabular}
\end{center}
\end{table*}

\section{Gaia-ESO Survey - proposal top-level science case  \label{scicase}}

\begin{figure}
\centering
\includegraphics[width=8.0cm]{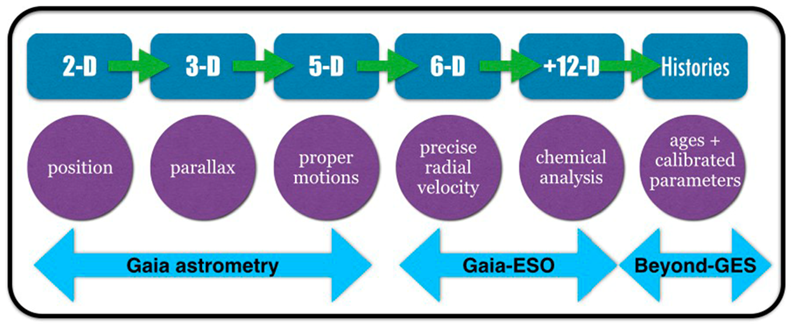}
\caption{The complementary contributions of astrometry, spectroscopy, field star and cluster isochrone and other age calibrations, and asteroseismology, to populating a high-dimensional description of Galactic stars.}
\label{dim}
\end{figure}

 This section of this article essentially reproduces the key section of the original Gaia-ESO Survey proposal to ESO. We do this to bring into the public record the early context and history which led to the survey. We also in this way present the survey's original goals, to allow comparison of ambition and outcome. We avoid post-hoc justifications. This context also established the survey optimisation. As the choice of instrumentation was fixed, the variables available for optimisation are the wavelength settings, the signal-noise per pixel per object, and most significantly the number of targets required to meet the basic science goals. The text also puts in context the status of stellar spectroscopic surveys at the time, which was very much less rich than the situation now.

 How disc galaxies form and evolve, and how their component stars and 
stellar populations form and evolve, are among the most fundamental
questions in contemporary astrophysics. The Gaia-ESO survey has been formulated
to contribute to those key questions, by enhancing our knowledge of
the formation and evolution of the Galaxy and the stars that populate
it. Gaia-ESO is a high statistical weight ($\simeq 10^5$ stars)
spectroscopic survey which utilises the opportunity of a large
telescope in the Southern Hemisphere to sample all the main components of the
Galaxy, from star-forming regions to ancient halo stars. This
survey has enormous value in its own right. However, its products are
even further enhanced by Gaia astrometry and Gaia
spectrophotometry and improved stellar parameters.

Understanding how galaxies actually form and evolve within our
$\Lambda$CDM universe continues to be an enormous challenge
\citep{Peebles11, Kormendy10}.
Simulations of the aggregation of cold dark matter, complemented by direct studies of galaxies at high redshifts, suggest that
galaxies grow through a sequence of merger/accretion
events. However, theoretical models of galaxy formation, which
necessarily involve modelling star formation and stellar evolution,
rely more heavily on phenomenological models than on physical theory.
Thus, these models require calibration with well-studied (nearby) test
cases. 

For example, star formation involves turbulence, magnetic
reconnection, collisionless shocks, and radiative transfer through a
turbulent medium. Similarly, the treatment of convection, mixing,
equations of state at high density, opacities, rotation and magnetic
fields can all significantly affect stellar luminosities, radii, and
lifetimes at different evolutionary phases. We are also far from being
able to simulate the coupled evolution of CDM and baryons from
ab-initio physics.  Observations are crucial to learning how galaxies
and stars were formed and evolved, and what their structure now
is \citep{Bland10}. Observations of objects at high redshifts and long look-back
times are important for this endeavour, as is detailed examination of
our Galaxy, because such ``near-field cosmology'' gives insights into
key processes that cannot be obtained by studying faint, poorly
resolved objects with uncertain futures. 

Just as the history of life
was deduced by examining rocks, we expect to deduce the history of our
Galaxy by examining stars.  Stars record the past in their ages,
compositions, and in their kinematics. For example, individual
accretion and cluster dissolution events can be inferred by detecting
stellar streams from accurate phase-space positions. Correlations
between the chemical compositions and kinematics of field stars 
enable us to deduce the history of star formation and even the past
dynamics of the disc. The kinematic structure of the bulge reveals
the relative importance in its formation of disc instability and an
early major merger.  

The study of open clusters is crucial to
understanding fundamental issues in stellar evolution, the star
formation process, and the assembly and evolution of the Milky Way
thin disc. Theories of cluster
formation range from the highly dynamic through to quasi-equilibrium
and slow contraction scenarios. These different routes lead to
different initial cluster structures and kinematics. Subsequent
evolution depends on many factors, including the initial conditions,
star formation efficiency and tidal interactions. Whilst hydrodynamic
and N-body simulations are developing, a fundamental requirement is an
extensive body of detailed observations. A complete comparison
requires precise position and velocity phase-space information
resolving the internal cluster kinematics, $\leq 0.5$km/s. Even more
sophisticated studies follow combination with Gaia astrometry.
The velocity fields within the youngest clusters reveal
their formation history, whilst the kinematics of the older clusters and the
age dependence of their mass functions test theories of cluster
destruction. Each star cluster provides a (near-)coeval snapshot of the stellar
mass function. This survey contributes to testing  stellar evolution
models from pre-main sequence phases right through to advanced
evolutionary stages. Much of the input physics in stellar models can
be tested by its effects on stellar luminosities, radii and the
lifetimes of different evolutionary phases. Homogeneous spectroscopy
will provide estimates of stellar parameters and reddening for large
samples of stars over a wide range of masses, in clusters with a wide
range of ages and mean chemical compositions. Such data are essential
in testing, calibrating, and refining both evolutionary tracks and stellar
parameters derived from spectra. 

When combined with Gaia astrometry, and
supplemented by asteroseismology, these data isolate and probe all the
theoretical uncertainties, whilst simultaneously identifying and
quantifying important perturbing factors such as binarity, rotation,
accretion and magnetic activity.  The interplay of these difficult-to-model physical phenomena can only be dissected by studying a wide range of clusters the properties of which make one or the other effect dominate.

An important focus of the Gaia-ESO Survey is to ensure consistently derived 
(to the extent possible) and consistently calibrated astrophysical parameters 
on the widest possible range of stellar populations. These range from young 
open star clusters, through clusters with a range of ages and metallicities, 
to field stars young and old, metal-rich to metal-poor. By ensuring consistency 
across this wide range, the age-calibrations and element abundance ratios 
can be made consistent, quantifying evolution.  The goal is to be able to 
combine Gaia-ESO age calibration using open cluster isochrones, asteroseismology 
calibrations, spectroscopic age indicators such as the [C/N] ratio in Red Giants, and Gaia data.

Of course, many ground-based surveys address these same general goals,
while the {\it par excellence} step forward is the ESA Gaia mission,
currently in its extended-mission phase. 

A convenient way of picturing the Gaia – ground complementarity is looking at the dimensionality of data which
can be obtained on an astrophysical object. Larger amounts of
information of higher quality are the goal, to allow increasing
understanding. Fig~\ref{dim} (adapted from \citep{Mess1}) gives a
cartoon view of this information set. There are four basic thresholds
which we must pass. The first is to know a source exists, its
position, and basic photometric data. Photometric surveys, such as
those undertaken at VISTA and VST, which are source photometry for 
this survey, deliver this information. The
second is to add the time domain – motions, including parallax,
providing distances and speeds. Here Gaia is revolutionary. The
third threshold is radial velocity, turning motions into orbits.
While Gaia will provide radial velocities, the magnitude limit is
several magnitudes brighter than that of the astrometry and the
precision at fainter magnitudes is much below that of Gaia's proper
motions. Gaia-ESO, together with other major spectroscopic surveys, is
crucial to supplement Gaia spectroscopy. The fourth threshold is
chemistry, and astrophysical parameters. These latter two both require
spectroscopy, which is the key information from Gaia-ESO.

We must quantify the scale of the challenge for a stellar spectroscopic survey. 
The key to decoding the history of
galaxy evolution involves chemical element mapping, which quantifies
timescales, mixing and accretion length scales, and star formation
histories; spatial distributions, which relate to structures and
gradients; and kinematics, which relates to both the felt but unseen
dark matter, and dynamical histories of clusters and merger events
\citep{FBH02}.  With Gaia, and calibrated stellar models, one can
also add ages. 

 Manifestly, very large samples are required to define
all these distribution functions and their spatial and temporal
gradients. Orbit space is (only) three-dimensional because generic
orbits in typical galaxy potentials admit three isolating integrals.
The number of objects required to determine the underlying probability
density of objects grows rapidly with the dimensionality of the
space. So in the present case, if we have ten bins along each axis in
integral space, corresponding to a resolution in velocity as coarse as
$\sim6$ km/s, we have 1000 bins in integral space. Then we wish to
distinguish at a minimum between young stars, stars of intermediate
age and old stars, and similarly, between stars with solar abundances,
stars with abundances similar to those of disc clusters and of halo
clusters. Thus each of the age, [Fe/H], and [$\alpha$/H] axes must be
divided into at least three bins, giving us $27\,000$ bins in the
minimal six-dimensional space. Even with perfectly adapted bin sizes,
an estimate of the density of stars in this space will have Poisson
noise of order unity unless we have in excess of $10^5$
stars. Similarly, defining the information content in the open cluster
system requires adequate sampling of the four dimensional (age,
metallicity, position in the Galaxy, mass/density) parameter
space. Even considering the inhomogeneous (mostly abundance)
measurements available in the literature, only a homogeneous
survey of $\simeq$ 70 clusters, containing $\simeq 5\times10^4$ stars,
will have sufficient statistical power. 

\begin{figure}
\begin{center}
\includegraphics[width=8.0cm]{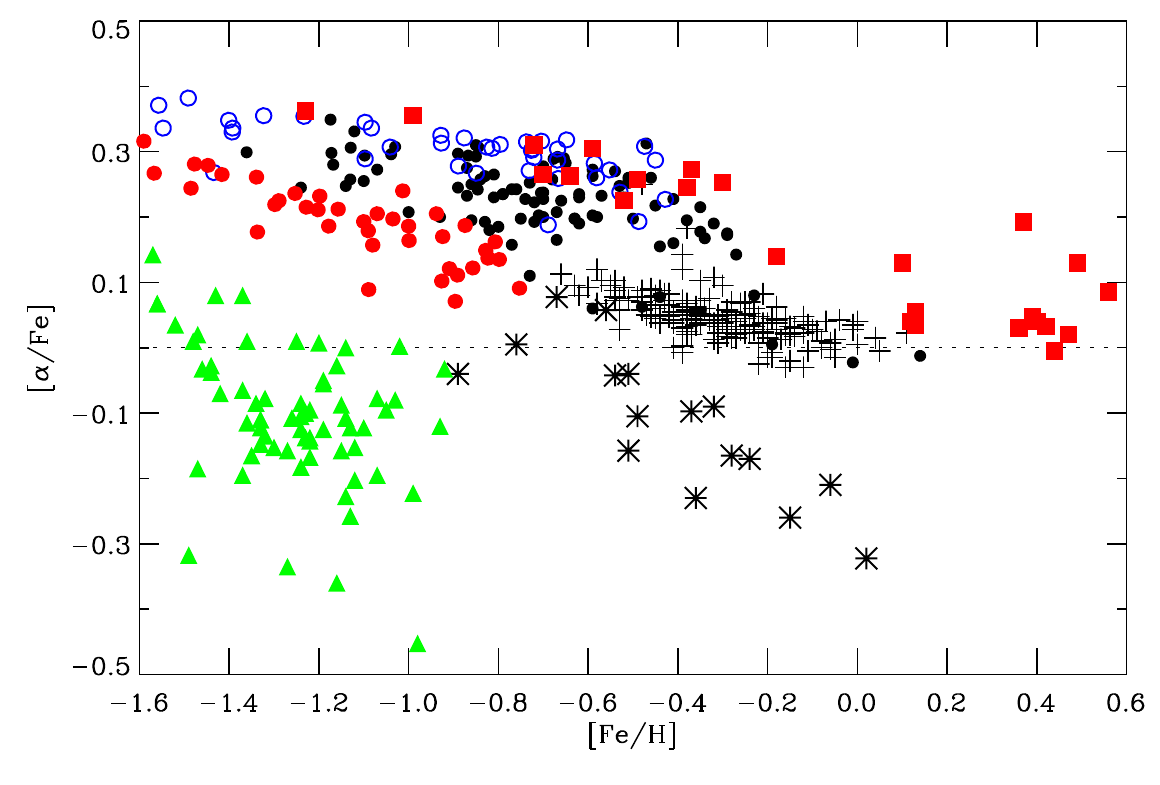}
\includegraphics[width=8.0cm]{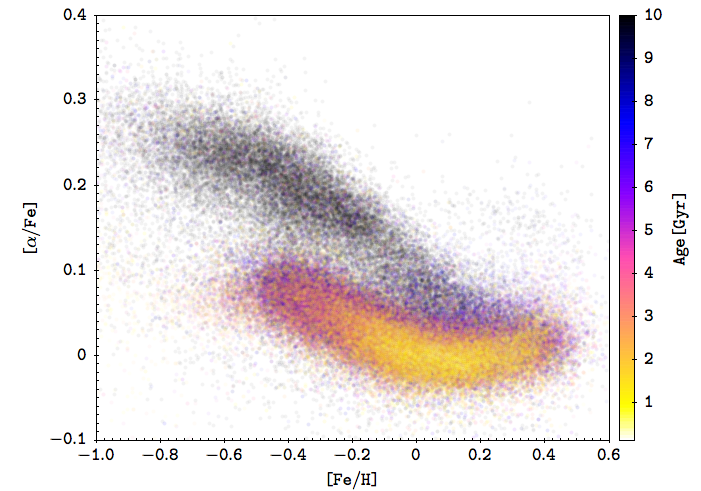}
\caption{ [$\alpha$/Fe] vs. [Fe/H] for various stellar populations.
Upper: thin-disc stars from \citet{Reddy03} are shown with plus symbols.
Filled circles refer to thick-disc stars from \citet{Reddy06} and \citet{Nissen_Schuster10}.  
Filled (red) squares are microlensed bulge stars from \citet{Bensby11}. 
Open (blue) circles are high-$\alpha$ and filled (red) circles
low-$\alpha$ halo stars from \citet{Nissen_Schuster10}.
Asterisks refer to stars in the Sagittarius dSph galaxy \citep{Sbordone07},
and filled (green) triangles show data for stars in the Sculptor dSph galaxy
from \citet{Kirby09}, for which the precision of [$\alpha$/Fe] is
better than 0.15\,dex.  This figure is taken from \citep{Nissen13}.
Lower: [$\alpha$/Fe] vs. [Fe/H] for various stellar populations, 
colour coded by age derived from the [C/N] chronometer calibrated 
from Gaia-ESO observed open clusters.
The distinction between the high-alpha thick disc and the low-alpha 
thin disc is manifest. Also apparent is the age gradient down the 
thick disc sequence, continued in age after a discontinuous jump 
in [Fe/H] to [Fe/H]=-0.6  This identifies the last major merger 
in the Milky Way. This figure is taken from \citet{Casali_CN}.}
\label{Niss_psss}
\end{center}
\end{figure}

An illustration of the information content in abundance-kinematic
surveys already available prior to the Gaia-ESO Survey is provided 
by the extensive review by  \citet{Nissen13}. 
Fig 16 of that review is shown as the  top panel of  Fig~\ref{Niss_psss}  here. 

As a direct example of the evolution of the progress in observational 
constraints on Galactic evolution from stellar spectroscopic surveys 
between 2011, when the proposal for the Gaia-ESO Survey was submitted, 
and 2022, when the survey final data release was made public, we contrast 
the 2011-vintage upper panel Fig~\ref{Niss_psss} with the lower panel Fig, 
topical results from a recent Gaia-ESO Survey analysis paper  \citep{Casali_CN}.

This figure 
illustrates the ability of chemical abundances and kinematic population 
assignments to sample the clear thin-thick disc distinction, the complexity of 
the halo populations, the considerable difference between Galactic
halo satellites and Galactic halo field stars, and the very metal-rich inner-Galaxy 
stars. Additionally, with newly developed age calibrations, the temporal 
evolution of the Milky Way begins to become quantified.

\section{Survey strategy: defining both precision and accuracy for stellar abundances\label{strategy}}

Gaia-ESO includes stars with almost the full observationally available range of astrophysical parameters, hot to cool, young to old, metal-rich to metal-poor, pre-main sequence to evolved giants. Since a primary goal is to derive high-quality astrophysical parameters and elemental abundances across this wide range, it is clear that a range of different analysis pipelines is required. This immediately raises the challenge of homogenising and calibrating the outputs on to a consistent (set of) scales. 

The challenges of stellar spectroscopy in the limiting regimes has been well summarised in two relevant recent review articles: 
"High-precision stellar abundances" \citep{Nissen_Gustafsson} and 
 "Accuracy and precision of industrial stellar abundances" \citep{Jofre19}.  We retain the full titles of these articles since they very clearly define the complementarity they both provide. 

The need for multiple approaches to maximise reliability is not a new concept. We note as one example the Segue Stellar Parameter Pipeline I - which involves 11 methods for $T_{\rm eff}$, 10 for $\log g$, 12 for $\rm Fe/H$, though with considerable overlap among them. [\cite{Lee2008-1}, \cite{Lee2008-2}, \cite{Lee2008-3}].

\begin{figure*}
\begin{center}
\includegraphics[width=15.0cm]{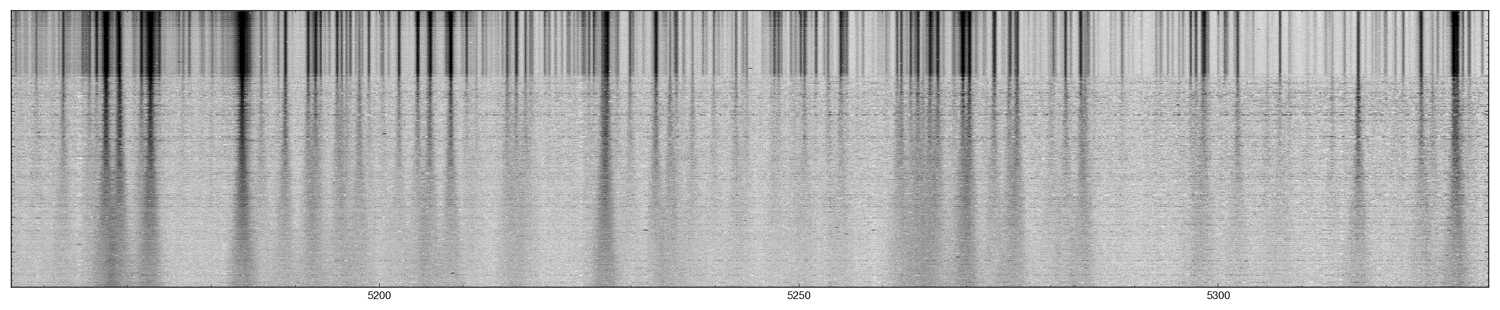}
\includegraphics[width=15.0cm]{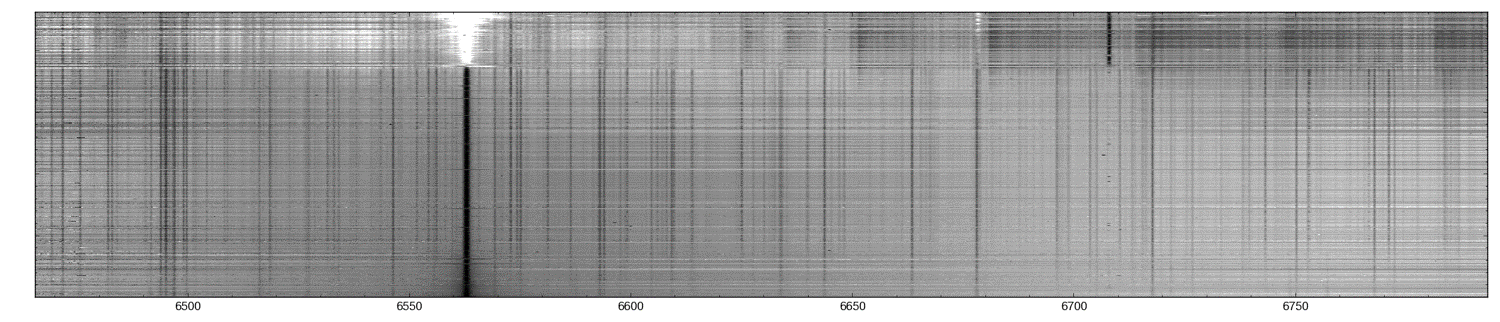}
\caption{Gaia-ESO Survey stars in a typical young open cluster. This presents a single Giraffe setup of about 100 stars with spectra scaled to illustrate the dynamic range, and sorted in temperature (hottest at the top). The top panel is Giraffe HR9B, 514-535nm, the lower HR15N, 644-682nm. This illustrates the wide range of astrophysical parameters, and hence appearances of the spectrum, which must be managed in a wide-ranging stellar survey.}
\label{Giraffe-stars}
\end{center}
\end{figure*}

Our need to cater for a very wide range in stellar astrophysical parameters is illustrated in Fig~\ref{Giraffe-stars}.  This presents spectra for a set of about 100 stars in a young open cluster. The spectra are sorted by temperature (hottest at the top) with naturally the coolest stars also illustrating lower signal-noise data. The changing emission (H$\alpha$ and Lithium in the lower panel) and absorption lines, and the range of astrophysical parameters which must be handled is apparent. The top panel is Giraffe HR9B, the lower HR15N (see Table~\ref{tab:giraffe_modes}).  

Calibration of spectroscopic analyses between authors and across methods has been a requirement ever since the subject began. Typically in very large surveys iterative increases in accuracy and precision are achieved with experience, and by detailed studies of individual data sets. For example,
 \citet{Hawkins16} reanalyse the APOGEE DR12 APOKASC subsample, to show the importance of linelist selection and the treatment of microturbulence to ensure robust metallicity scales and elemental abundance ratios. The lesson learned is that careful treatment of the important astrophysical parameters involved in abundance determinations, augmented by analysis of a set of "benchmark" stars, provides abundance results which do not need later empirical scaling to match independent literature studies. Another example of systematics which require careful treatment is given by  \citet{Venn12}, their Fig 5, showing the large and wavelength-dependant continuum scattering corrections required. 
 
 In spite of best efforts, some analysis limitations are discovered only at science verification analysis level. For example,
 \citet{Piatti19} notes the apparent but entirely spurious dispersion in a study of NGC188, indicating an apparent dispersion of 0.16dex in [Na/Fe]. He concludes "Therefore I warn users of large spectroscopic surveys to be extra careful when finding peculiar abundance results". This caveat emptor applies fully to Gaia-ESO Survey results.

Robust astrophysical analyses require that parameters derived from several different pipelines are as much as is feasible on a single consistent scale. This is a fundamental requirement which does not require new justification.
It is not always completely easy to do. We illustrate this by considering the 
successive (re-)analyses of the star Boo-1137, a metal-poor red giant member of the Bootes-I ultra-faint dwarf galaxy, based on a single VLT spectrum with the same settings and typical quality as those relevant to the Gaia-ESO Survey. This star was discovered by \cite{NWG2010}, and subsequently studied  by \cite{NYG2010}, and by \cite{GNM2013}. The analysis of \cite{GNM2013} was a full double-blind study, comparable in methodology and methods with the analyses of Gaia-ESO stellar spectra by several nodes acting independently. 

The starting point of this example is that \cite{NYG2010} derived an abundance ratio $\rm [Mg/Fe] = +0.47$, while \cite{GNM2013} derived $\rm [Mg/Fe] = +0.26$ from the same spectrum. Clearly the difference between the derived values is entirely a consequence of different analyses. How do such differences arise, and what lessons should be {\b learned} for multi-node survey calibration and homogenisation? 

The first point to note is the importance of data selection. The double-blind analysis of \cite{GNM2013} showed the desirability to restrict the wavelength range under consideration.
This generated changes in both Mg and Fe abundances, with opposite sign.
There is a further change of note: the adopted Solar abundances. The 2010 study adopted the Solar abundance from \cite{Asplund2006}, while the 2013 study adopted those of \cite{Asplund2009}. These individually small changes all affect the derived elemental abundance ratio cumulatively. 

This simple example illustrates the desirability of independent double-blind analyses to identify the parameter ranges in which specific analysis systems are optimised. It also illustrates the need to isolate and fix parameters which can be controlled, such as line lists, model atmospheres and adopted scale or benchmark references, that being the Sun in this case.

It also of course highlights the need to bring all the node results together into a consistent homogenised whole. For Gaia-ESO Working Group~15 carried out this last critical task. In the rest of this paper we describe the many steps and very substantial investment of effort which was required to implement the multi-analysis approach introduced above.

\begin{figure}
\begin{center}
\includegraphics[width=9.0cm]{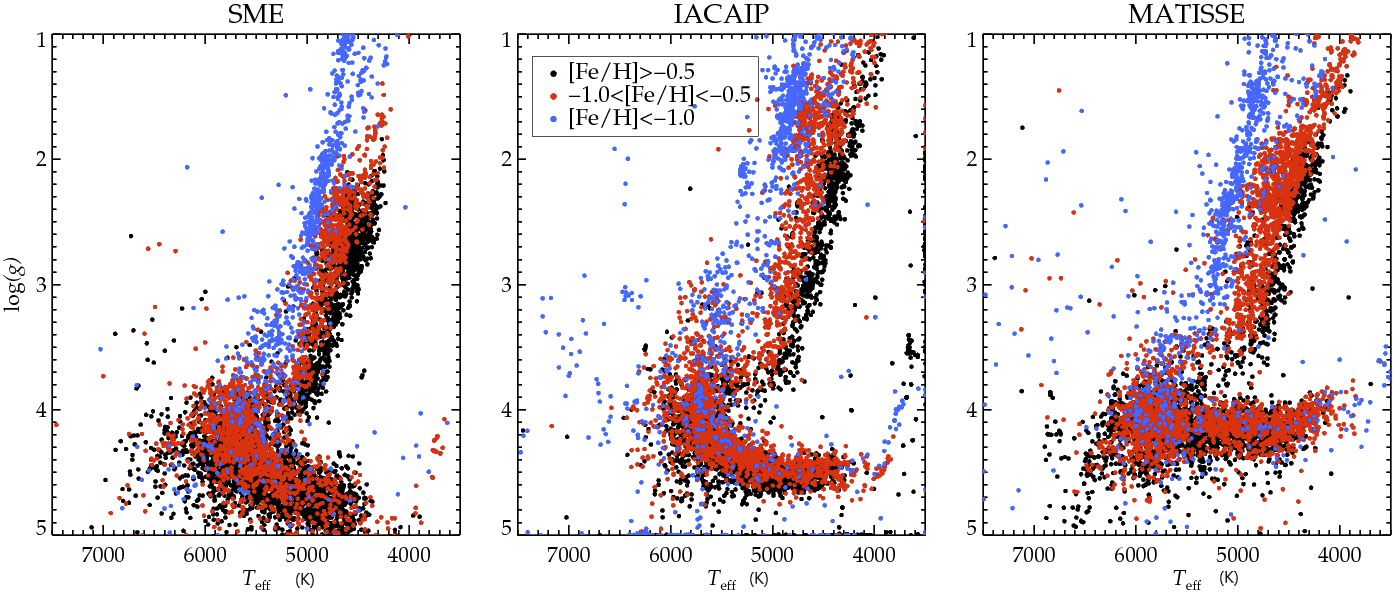}
\includegraphics[width=6.0cm]{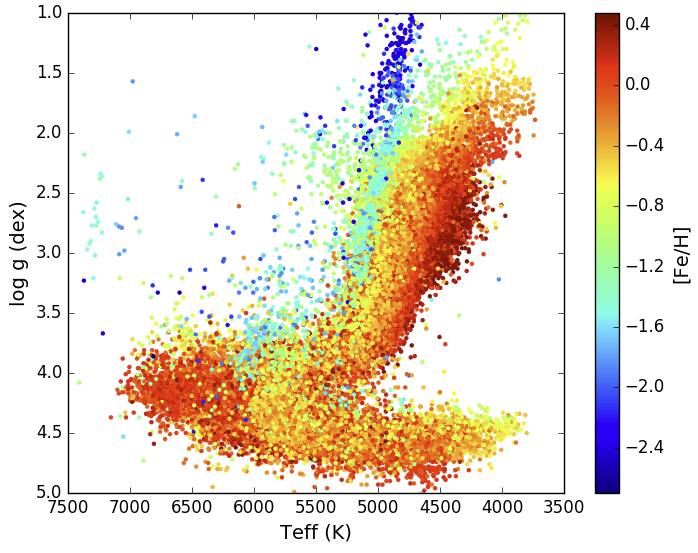}
\caption{Before homogenisation and after - initial method-specific systematic differences can be homogenised away if one has many methods available in a single study.  The three top panels present results from specific nodes (see Sect 9 for detail), while the lower panel shows the outcome of homogenisation, further colour-coded by [Fe/H] to make clear the dispersion is now real astrophysical dispersion. The systematic effects apparent in the top three panels have been substantially reduced.}
\label{hrd-homog}
\end{center}
\end{figure}

Homogenisation of analyses from several independent studies of the same spectra improves more than (just) reliability of the derived parameters. An additional advantage, and indeed a robust sanity check on the whole process, is its effect on the homogenised H-R diagram. This is described more in section~\ref{WG15}, but is illustrated here by an example in Fig~\ref{hrd-homog}. Reductions of method-specific systematics, essentially calibrating away limits on the relevant parameter range in which a method is robust, are apparent. A full description of how the Gaia-ESO approach was implemented, showing node-level and homogenised HRDs, is available in \cite{Randich21}, especially figures 10 and 11 of that paper.

\section{Structure of the Gaia-ESO Survey project team}\label{structure} 

As with all ESO public surveys, the Gaia-ESO Spectroscopic Survey Consortium role and responsibilities are defined through a Survey Management Plan, which is a Memorandum of Understanding between ESO and the Co-PIs, Gilmore and Randich. This defines what is essentially described in this paper, the structure and responsibilities
of the survey team, and the deliverable data products for release through the ESO SAF.

In order to implement those formal requirements, in a consortium with originally some 400 Co-Investigators from some 100 Institutes, 
with a commitment to 300 VLT nights of data, a formal structured set of internal Working Groups with clearly defined responsibilities
was established. An overview Steering Group supervised survey management, while annual reports to and bi-annual reviews by ESO panels
monitored technical and scientific progress. Internal communications relied heavily on a dedicated wiki system, newsletters, and annual whole-Consortium meetings (pre-Covid).

Among the minor but significant challenges in bringing together such a large number of established spectroscopic analysis groups 
was efficient data exchange.  Indeed it is a valuable legacy product of the survey that the community has learned to adopt a standard data format for both results and data - FITS - and has become familiar with handling very large and complex data files.

\subsection{How it came together}

Development of the large and ambitious Gaia-ESO Survey project involved many people and much time and effort. The range of expertise required to be able to select and observe suitable targets, reduce the data, analyse the spectra and deliver science verification for the ambitious very wide range of astrophysical targets was assembled and organised. The two branches of community interest were brought together, with one group with more expertise focussed on young stars and star clusters, the other on field stars. This ensured a single ambition, to build internal survey-wide consistency, with analyses anchored from clusters and their isochrones, then extending to the widest accessible range of ages and abundances. 

All interested European spectroscopic analysis groups were invited to join, ensuring that essentially all widely-used analysis packages were involved, and the requisite range of expertise and effort was available. Although the observational approach for the open clusters was already optimised, from community experience with FLAMES, for the field stars many simulations were implemented, to optimise the selection of available observing setups, and targeted signal-noise ratios. The considerable effort to develop the data reduction pipelines and data management system was identified. All this took place prior to and shortly after submission of the survey proposal to ESO in March 2011.

Following acceptance of the survey by ESO, the detailed implementation plans, from target selection, through observing, data reduction, data analysis and science verification, to the expected survey products leading to a substantial public archive, were specified in a Survey Management Plan between ESO and the Co-Principal Investigators.

In the early stages of planning to ensure maximal value from the range of approaches involved, it was clear that a small number of critical issues needed to be addressed. This included from the astrophysical view adoption of an optimal single astrophysical line-list \citep{Heiter2021}, use of the (developing) Gaia Benchmark Stars as prime calibrators, dedicated data processing pipelines (UVES at Arcetri, GIRAFFE at Cambridge) and data archives, and efficient and effective ways to move spectra and derived parameters to and from the analysis teams and homogenisation Working Groups. For this last activity adopting a FITS structure for all files proved very beneficial.
Regular specialist and team-wide meetings facilitated the smooth operation of this whole process. The rest of this section describes the implemented structure in more detail.

\subsection{Project organisation}

\begin{figure*}
\centering
\includegraphics[width=16truecm]{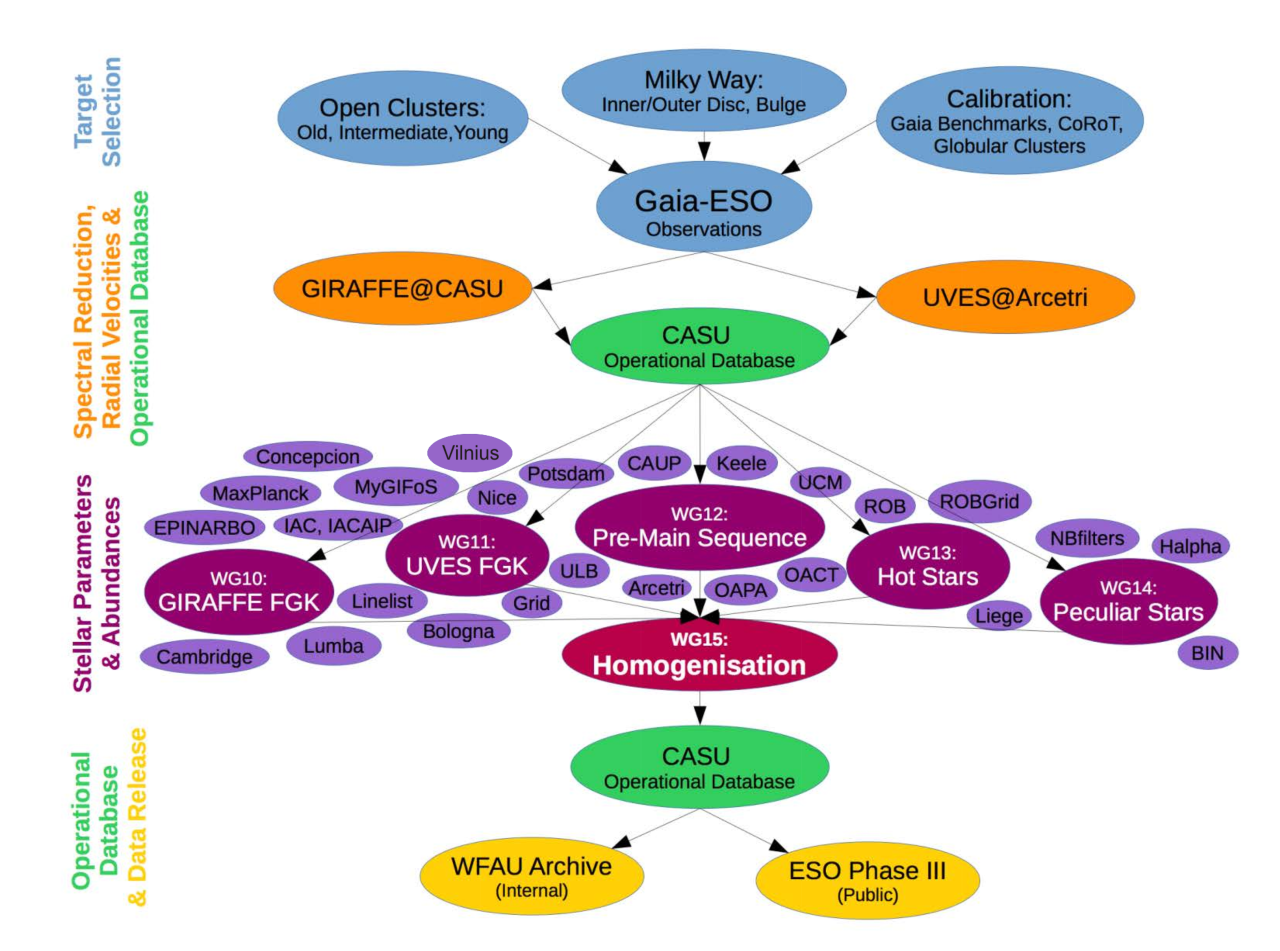}
\caption{A science data processing perspective of the Gaia-ESO Survey structure. The small purple ovals identify the 26 processing and analysis nodes which are described in later sections of this paper.
 \label{fig-data}}
\end{figure*}

An overview of the Gaia-ESO Survey data flow process is presented in Fig~\ref{fig-data}.  The tasks are distributed among 19 Working Groups (WGs), WG0
to WG18, each of which has a coordinator - see  Table~\ref{WG-summary}, and an active membership.

\begin{table*}
\tiny
\caption{Gaia-ESO Survey Working Group organisational structure}
\begin{tabular}{l|lcc}
\hline\hline
Working Group & Task & Coordinator(s) & Reference article \\
\hline\hline
WG 0 & Paranal Observing Team & Thomas Bensby & this paper \\
WG 1 & Cluster Membership Analysis & Emilio Alfaro & \cite{Bragaglia_prep} \\
WG 2 & Auxiliary Data for Cluster Target Selection & merged with WGs
1 \& 4 \\
WG 3 & Galactic Field and Plane Target Selection & Carine Babusiaux, Sergey Koposov &
this paper \\
WG 4 & Cluster Stars Target Selection & Angela Bragaglia &  \cite{Bragaglia_prep} \\
WG 5 & Calibrators \& Standards & Elena Pancino & \cite{Pancino17} \\
WG 6a & OB/fposs generation [Field] & Thomas Bensby & this paper \\
WG 6b &  OB/fposs generation [clusters] &  Ettore Flaccomio & \cite{Bragaglia_prep} \\
WG 7a & Raw Data Pipelines: GIRAFFE & Jim Lewis, Mike Irwin & this
paper \\
WG 7b & Raw Data Pipelines: UVES &  Germano Sacco & \citet{Sacco-UVES} \\
WG 8a & Radial Velocities: GIRAFFE & Sergey Koposov, Rob Jeffries &
this paper;  \cite{Jackson15} \\  
WG 8b & Radial Velocities: UVES & Germano Sacco & \citet{Sacco-UVES} \\
WG 9  & Discrete Classification & Sergey Koposov & this paper \\
WG 10 & GIRAFFE FGK-star Analyses & Clare Worley;  Carlos
Allende-Prieto &  \cite{Worley19_wg10} \\
WG10  & WG 10 - initial team & Alejandra Recio-Blanco & \cite{Recio-Blanco14} \\
WG 11 & UVES FGK-star Analyses & Rodolfo Smiljanic; Andreas Korn  &
\citet{Smiljanic14} \\
WG 12 & PMS-Star Spectrum Analyses & Alessandro Lanzafame & \citet{Lanzafame15} \\
WG 13 & OBA-Star Spectrum Analyses & Ronny Blomme & \cite{Blomme21} \\
WG 14 & Non-standard Objects, Dictionary & Sophie Van Eck, Tomaz Zwitter & \cite{SVE19} \\
WG 15  & Survey Parameter Homogenisation & Patrick Francois & \cite{Hourihane21} \\
WG 16 & Survey Progress Monitoring & Gerry Gilmore \& Sofia Randich &
this paper \& \cite{Randich21} \\
WG 17  & CASU Operational Datacentre & CASU/Mike Irwin & this paper \\
WG 18 & Survey Internal Archive & WFAU/Nigel Hambly & this paper \\
\hline \hline
\end{tabular}
\label{WG-summary}
\end{table*}

The tasks of the Working Groups are to implement the data flow,
from target selection and characterisation, through preparation of the ESO OB observing files, observing, pipeline data processing, detailed spectrum
analyses, astrophysical parameter quality/sanity checking and
homogenisation, to science quality control, through to preparation,
documentation and delivery of external data products to both ESO and a
dedicated public archive.

\begin{table}
\tiny
\caption{Gaia-ESO Survey Builders}
\begin{tabular}{l|l|l|l}
\hline
Name & Name & Name & Name  \\
\hline
& G Gilmore & S Randich &\\
M Asplund & J Binney & P Bonifacio & J Drew \\
S Feltzing & A Ferguson & R Jeffries & G Micela \\
I Negueruela & T Prusti & H-W Rix & A Vallenari \\
E Alfaro & C Allende Prieto & C Babusiaux & T Bensby\\
R Blomme & A Bragaglia & E Flaccomio & P Francois\\
N Hambly & M Irwin & S Koposov & A Korn \\
A Lanzafame & E Pancino & A Recio-Blanco & R Smiljanic \\
S Van Eck & N Walton & A Bayo & M Bergemann \\
K Biazzo & G Carraro & A Casey & M Costado \\
F Damiani & B Edvardsson & E Franciosini & A Frasca  \\
A Gonneau & U Heiter & V Hill & A Hourihane \\
R Jackson & P Jofr\'e & C Lardo & P de Laverny \\
J Lewis & K Lind & L Magrini & G Marconi \\
C Martayan & T Masseron & L Monaco & L Morbidelli \\
L Prisinzano & G Sacco & L Sbordone & S Sousa \\
C Worley & S Zaggia & T Zwitter\\
\hline
\end{tabular}
\label{Builders}
\end{table}
 
In addition to the Working Group leads, a significant number of individuals have provided exceptional efforts to deliver the Gaia-ESO Survey, they are listed in Table~\ref{Builders}. These have been credited by identification as "Builders", a recognition which provides co-authorship rights on survey papers. Among their duties has been to provide internal refereeing of survey papers prior to their journal submission.

\begin{table}
\caption{Gaia-ESO Survey Project Office Team}
\begin{tabular}{lc}
\hline
\hline
              &  Institute \\
\hline
Anais Gonneau & Cambridge \\
Anna Hourihane & Cambridge \\
Germano Sacco & Arcetri \\
Clare Worley & Cambridge \\
\hline
\end{tabular}
\label{PO}
\end{table}

The two Co-PIs, Gerry Gilmore \& Sofia Randich, led the survey jointly, with Gilmore being specifically responsible for the field star
and calibration efforts, and Randich for the open cluster work. Both were assisted and supported by a dedicated Project Office team listed in Table~\ref{PO}, whose work was critical to successful delivery of the final survey products. In addition, survey-wide and management issues were supported by and overseen by a Steering Group (Table~\ref{Steer}) of senior scientists representing the broad range of Institutes and subjects involved in the survey.

\begin{table}
\tiny
\caption{Gaia-ESO Survey steering group}
\begin{tabular}{l|lcc}
\hline
Name & Function & Affiliation & Country  \\
\hline
Gerry Gilmore & Co-PI &  Institute of Astronomy  & UK \\
Sofia Randich & Co-PI &  INAF Obs Arcetri & I \\
\hline
Martin Asplund & Steering Group & ANU/MPA  & Aus/D \\
James Binney & Steering Group & Oxford  & UK \\
Piercarlo Bonifacio & Steering Group & Paris  & Fr \\
Janet Drew & Steering Group & Hertfordshire/UCL  & UK \\
Sofia Feltzing & Steering Group & Lund  & Se \\
Annette Ferguson  & Steering Group & Edinburgh  & UK \\
Rob Jeffries & Steering Group & Keele  & UK \\
Giusi Micela & Steering Group & Palermo  & I \\
Ignacio Negueruela & Steering Group & Alicante  & Sp \\
Timo Prusti & Steering Group & ESA  & ESA \\
Hans-Walter Rix & Steering Group & MPIA  & D \\
Antonella Vallenari & Steering Group & Padova  & I \\
\hline
\end{tabular}
\label{Steer}
\end{table}

\section{Gaia-ESO Survey observational strategy}\label{obs-strategy} 

The Gaia-ESO Survey observing strategy has been designed to deliver
the top-level survey goals. The observations  include
Milky Way (MW) field observations, Open Cluster observations, and calibration observations of
different targets, such as radial velocity standard stars, benchmark stars, globular clusters,
CoRoT red giants and Kepler K2 red giants.

The Gaia-ESO Survey observations were performed with the multi-object
optical spectrograph FLAMES mounted on UT2 at the VLT \citep{Pasquini02}. 
FLAMES is a multi-object system, feeding an intermediate (GIRAFFE) and
a high resolution (UVES)
spectrograph with a field of view 25 arcmin in diameter.

The choice of setups and integration times were optimised through extensive simulations and tradeoffs for the field stars, while previous observational experience dictated the open cluster settings and requirements. 
The field star simulations were led by Vanessa Hill, and proved to be reliable and accurate by the survey observational results.  The essential decision was that, given that the amount of observing time per star was fixed by the requirement to observe $10^5$ stars,  and the amount of survey observing time ESO was making available, field star observations were optimised by two equal-length observations, one HR10, one HR21. This maximised the number of stars for which element abundance ratio data could be derived.  This observing approach was planned to deliver median signal-to-noise ratio per pixel (SNR) spectra of 25 in HR21 and 10 in HR10, given system performance and the apparent brightness distribution of the targets. For the field-star UVES parallel sample the comparable SNR target was 40. In the event these requirements were delivered. Table 1 and section 2 in \cite{Randich21} provide a detailed discussion of delivered system performance, which closely matched expectation.  
The field star observational approach was therefore not changed during the survey. 

Considerable experience with the VLT system was available for the open cluster aspects of the survey. There median SNR targets ranging from 30 to 75 were planned for the various GIRAFFE settings, and 75 to 150 for UVES. Corresponding exposure times were of course target specific, depending on age, distance and extinction. 
In some cases for open cluster targets more pointings were required to cover the whole cluster area and complete membership candidates, slightly reducing the 
initially planned number of cluster targets with the reward of higher quality results. A full discussion of the overall survey delivered performance, with additional focus on the open cluster science is available in \cite{Randich21}
Observations were restricted to
$+10^{\circ} \geq \rm DEC \geq -60^{\circ}$ whenever possible to
minimise airmass effects.

The observing strategy for the Gaia-ESO Survey was tailored to match the requirements of the
individual populations being observed. Table~\ref{tab:giraffe_modes} 
shows the GIRAFFE
setups employed in Gaia-ESO Survey observations, their wavelength coverage
and their resolutions. 
Table~\ref{UVES_modes} presents comparable data for UVES.
Table~\ref{tab:modesbyproject} describes the 
stellar populations 
observed and enumerates the GIRAFFE setups employed. Calibration and 
standard stars were observed in all the setups listed in the two tables.

Observation blocks were implemented in a manner which splits a single observation into 
two equal length exposures. This aids in the detection of transient features
such as cosmic rays. For most observing modes a much shorter exposure is 
inserted with the simultaneous calibration ("simcal") arc lamp switched 
on. This allows for the wavelength zeropoint drift to be monitored over the
course of the night. One observing mode (HR21) does not use simcal observations
as there are a number of very bright arc lines in that wavelength region
which saturate even in short exposures. For this one mode wavelength drift
can be monitored using the night sky lines.

\begin{table}
\caption{Properties of the GIRAFFE setup modes that are used in the
  Gaia-ESO Survey. Resolution improved in February 2015 following refocus efforts.}
\begin{tabular}{lcccl}
\hline
\hline
Setup  &  central & wavelength & resolution & resolution\\
          &     wavelength       &   range & < 02-2015 & >02-2015\\
       &     (nm)   &    (nm)    &   & \\
\hline
HR3    &    412.4   & 403-420    & 24800 & 31400\\
HR4    &    429.7  & 419-439  &  20350 & 24000\\
HR5A   &    447.1   & 434-458    & 18470 & 20250\\
HR5B   &    446.4   & 437-455   & 26000 & archive\\
HR6    &    465.6   & 453-475    & 20350 & 24300\\
HR9B   &    525.8   &  514-535   & 25900 & 31750 \\
HR10   &    548.8   & 533-561    & 19800 & 21500\\
HR14A  &    651.5   & 630-669    & 17740 & 18000\\
HR14B  &    650.5   & 638-663    & 28800 & archive \\
HR15N  &    665.0   & 644-682    & 17000 & 19200\\
HR21   &    875.7   & 848-898    & 16200 & 18000\\ \hline
\end{tabular}
\label{tab:giraffe_modes}
\end{table}

\begin{table}
\caption{Properties of the UVES setup modes that are used in the
  Gaia-ESO Survey}
\begin{center}
\begin{tabular}{lccc}
\hline
\hline
Setup  &  central & wavelength  & resolution \\
       &    wavelength (nm)   &   range (nm)    &    \\
\hline
U520    &    519   & 414-621    & 47000 \\
U580    &    580  & 476-684  &  47000\\
\hline
\end{tabular}
\label{UVES_modes}
\end{center}
\end{table}

\begin{table}
\caption{GIRAFFE modes used for each project within the Gaia-ESO Survey}
\begin{tabular}{lcc}
\hline
\hline
Survey Area                &  Typical   & GIRAFFE setups \\
                           &  magnitudes & used \\
\hline
Bulge                      & $I = 15$                & HR10,21               \\
Halo/Thick Disc            & $15 \leq r \leq 18$       & HR10,21              \\
Outer thick disc           & $r \leq 18$              & HR10,21              \\
Open clusters (early type) & $I \leq 19$              & HR3,4,5A,6,14A      \\
Open clusters (late type)  & $I \leq 19$              & HR15N             \\
Calibrator clusters        & various                  & all modes    \\
Benchmark Stars             & $I \leq 15$             &  all modes   \\
Radial Velocity standards  & various               & all modes  \\
\hline
\end{tabular}
\label{tab:modesbyproject}
\end{table}

\section{Target selection and observations [WG0-WG6]}\label{targets}

\begin{figure}
\begin{center}
\includegraphics[width=9.0cm]{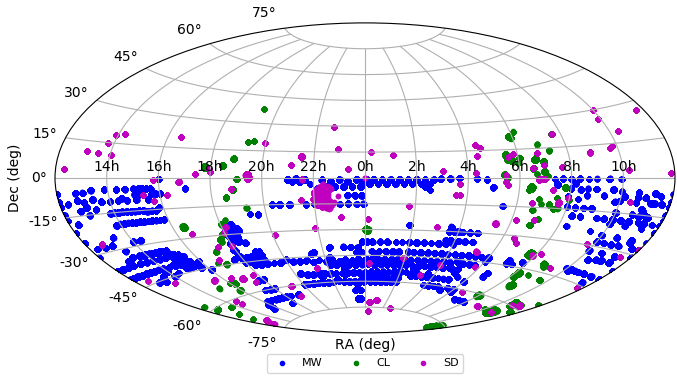}
\caption{The sky distribution of Gaia-ESO Survey observed fields for all observations broken down by field-type: Milky Way fields (MW), clusters (CL), standards and calibration fields (SD). The clump of SD fields around 22h, -10 deg includes the Kepler K2 red giants introduced as standards in July 2016.
}
\label{skyplot}
\end{center}
\end{figure}

 The survey includes the Galactic inner and
outer bulge, inner and outer thick and thin discs, the halo and known
halo streams. There is special focus on open clusters at all ages, and
on solar neighbourhood field stars, as these trace both stellar and
Galactic evolution, complement Gaia astrometry, and will benefit most
from the most precise Gaia data. The sky coverage achieved is shown in Fig~\ref{skyplot}.

\noindent {\bf Open clusters.}  
Cluster selection is optimised to fine-sample the age-[Fe/H]-radial
 distance-mass parameter space. Open clusters in all phases of evolution
 (except embedded), from $\sim 10^6$~Myr up to $\sim 8$~Gyr are included,
 sampling different environments and star formation conditions.  For all clusters we 
 use GIRAFFE to target faint cluster members (down to V$=19$), while
 UVES fibers are fed with brighter or key objects (down to
 V$=16.5$), to be used for accurate abundance determination or for
 which better precision in RV is required.  The several GIRAFFE set-ups 
 employed are listed in Table~\ref{tab:modesbyproject}.  Setups HR03/04/05A/06/14A contain a large
 number of spectral features used to derive RVs and
 astrophysical characteristics (e.g., temperature, gravity, wind) of early-type
 stars; HR15N is instead the most commonly used grating for
 late-type stars; this accesses a large enough number of lines to derive
 RVs, as well as to retrieve key information on the star
 characteristics (e.g., temperature, Li, accretion rates, chromospheric
 activity, rotation).  For UVES, CD3 is most suitable both
 for early-type (520~nm setting) stars and late-type members (580~nm
 setting). Cluster FPOSS configurations
are typically observed at least twice to identify binaries. 

\noindent {\bf Bulge
  survey.} Here the prime targets are K giants, including the red
clump ($I=15$ typically). These dominate the relevant CMD selection.
Two GIRAFFE settings are needed (HR21, HR10), implying up to 4H/fibre
setup, depending on the field and the extinction. This  measures
Mg, Ca, Ti for most stars, and Si, Cr, Mn, and Ni for many stars. The
bulge RGB is clearly visible in CMDs at $b \leq 45^{\circ}$, so this
survey extends that far. In low extinction regions, brighter gK
stars can be observed with UVES 580-nm parallels to sample both bulge
and inner Galaxy populations. In the event scheduling limitations (the focus on the VLTI SgrA* science, 
and overrides to implement spectroscopy of micro-lensed bulge dwarfs)
reduced the available effort on the inner Galaxy and Bulge. \\ 
\textbf{Halo/thick disc survey.}
Primary targets are $r$=17-18 F+G stars, with the bluer, fainter F
stars probing the halo, brighter, redder F/G stars probing the thick
disc. SDSS analyses show a clear thick disc/halo transition in the
range $17 \leq r \leq 18$. The spectra allow measurement of both iron-peak
elements and alpha elements, for stars down to [M/H]$\leq-1.0$. In
fields crossing known halo streams (eg Sgr), stream K giant candidates
are also observed. A subset of fibres was allocated to specially
selected candidate members of rare but astrophysically important
stellar populations, such as extremely metal poor stars.  The fields
are distributed in the whole sky, but predominately in the Galactic
cap (SGC, NGC) and bulge regions. This minimises scheduling clashes with the
cluster targets, and ensures southern and northern fields for
scheduling, and photometric overlap with SDSS, PS1, and
ESO/VST. \\ 
\textbf{Outer thick/thin disc, 2-4 kpc from the Sun.} These fields
 have distant F/G stars as prime targets, and 2 settings, as for
the halo in both requirements and measurables. This well-defined low
latitude sample probes 2-4~kpc, more than a radial scale length.  In
addition, 25\% of the fibres are allocated to candidate K giants
(r$\leq18$), which probe the far outer disc, warp, flare and
Monoceros stream.  \\ 
\textbf{Solar Neighbourhood.}  UVES parallels for
the field surveys are dedicated to an unbiased sample of order 5000
G-stars extending $\geq 1$ ~kpc from the Sun, to quantify the local
detailed elemental abundance distribution functions. The sample is
photometrically-selected to ensure all possible ages and metallicities
for unevolved stars and subgiants are sampled. UVES 580-nm setting is
adopted. These are parallel observations, requiring no dedicated time.

\subsection{Observational Working Group activities: WG0 to WG6} 

In these sections we provide a brief summary of the efforts required to obtain observations at the VLT. These efforts include target identification and selection,
creation of the ESO Observing system OB and FPOSS files, which control the observations
 and the fibre allocations respectively, and operation of the VLT and data quality real-time checking during observations. These processes were required for the three
main target classes: field stars, cluster stars, and calibrator and standard stars.

\subsection{WG0 - Paranal observing team}

The Working Group 0 Coordinator was Thomas Bensby. In addition to coordinating the observing team, and carrying out many runs, he checked all necessary observing files prior to each run, and ensured suitable records were available to optimise completion of sets of observations.

Observations for the Gaia-ESO Survey took place during 59 scheduled observing runs in the period 31.12.2011-26.01.2018. The observers were drawn (largely) from a small experienced team, Table~\ref{Paranal}, including ESO Support astronomers when they were available. This minimised needless travel and ensured fully experienced and efficient observers were available.

\begin{table}
\caption{The core Paranal Observing Team.}
\begin{tabular}{lc}
\hline
\hline
Observer              &  Institute \\
\hline
T. Bensby & Lund Observatory  \\
A. Bayo & Universidad de Valparaiso \\
G. Carraro & Universita' di Padova \\
G. Marconi & ESO \\
C. Martayan & ESO \\
L. Monaco & Universidad Andres Bello \\
L. Sbordone & ESO \\
S. Zaggia & Osservatorio Astronomico di Padova \\
\hline
\end{tabular}
\label{Paranal}
\end{table}

A report on the observing outcomes, including sky conditions, is included in \citet{Randich21}.

\subsection{WG3, WG6a - Field star target selection and observation preparation}

Field star target selection was initially defined by Carine Babusiaux and Gerry Gilmore, with  the regular target files creation being done by Sergey Koposov. The selection function for the Gaia-ESO Survey field star observations is described in additional detail in \citet{Stonkute16}.  We provide an overview here.

\begin{figure}[!h]
\begin{center}
\includegraphics[width=9.0cm]{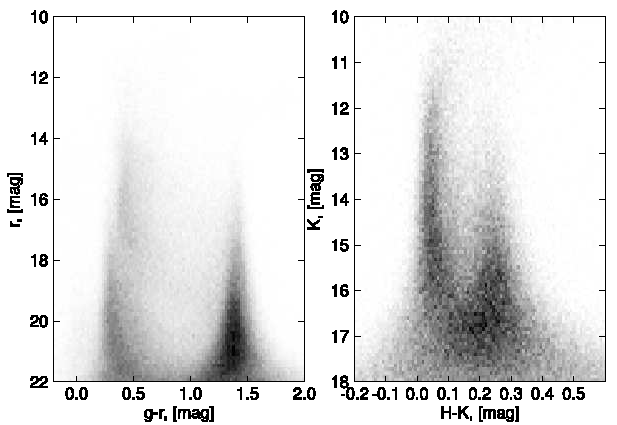}
\caption{LHS: SDSS high-latitude photometry illustrating the two blue-star populations, $0.3 \leq (g-r) \leq 0.6$. The bluer, becoming dominant at fainter magnitudes, is the halo. The less blue, becoming less dominant at $r \sim17$, is the thick disc. RHS: the same stars observed in VHS [K$_s$, H-K$_s$] photometry, which is the source of photometry applied by Gaia-ESO.}
\label{cmds}
\end{center}
\end{figure}

 Field stars were primarily selected for observation with GIRAFFE from $(J, H, K_s)$
VISTA imaging, using the photometry derived by the  Vista Hemisphere Survey (VHS; \citep{VHS2013}), ensuring excellent recent astrometry, and thus further increasing the value of the ESO VISTA surveys.  
For  the brighter UVES parallel sample,  2MASS \citep{Skrutskie06}  was used. The VHS near-IR photometry was chosen because there was, at the time, no publically available recent optical photometry covering a sufficiently wide sky area to ensure targets could be prepared with reliable astrometry and photometry. Reliable close-epoch astrometry was a critical requirement given that the observations were through relatively small fibres, so that the brighter acquisition and guide stars must be on the same astrometric system as the fainter target stars, with proper motions either known, or irrelevant through epoch matching. To ensure that the JHK selection matched the halo-thick disc division apparent from SDSS optical photometry, fields with both SDSS and VHS photometry were used to define the actual selection, as illustrated in Fig~\ref{cmds}.

\begin{figure}[!h]
\begin{center}
\includegraphics[width=0.98\hsize,angle=0]{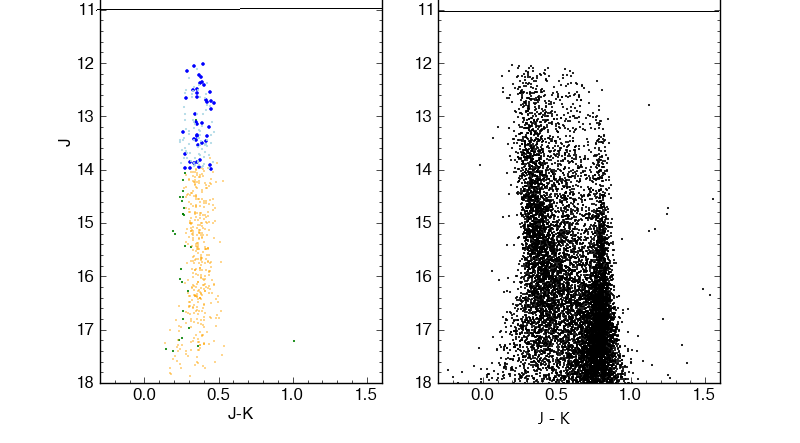}
\caption{\label{fig:field-targets} Target selection for the field-star survey.
The left colour-magnitude diagram shows the selected stars for both GIRAFFE and UVES, the right CMD shows the parent sample in that field from which the targets are selected.  The larger blue points brighter than J=14 are the primary UVES target. Fainter yellow points are the primary GIRAFFE targets}.

\label{cmds-2}
\end{center}
\end{figure}

The primary GIRAFFE targets are $r$=17-18 F+G stars, with the bluer, fainter F
stars probing the halo, brighter, redder F/G stars probing the thick
disc. SDSS analyses show a clear thick disc/halo transition in the
range $17 \leq r \leq 18$ -- Gaia-ESO uses the equivalent selection
from VISTA JHK photometry, as illustrated in Fig~\ref{cmds} and Fig~\ref{cmds-2}.  

 The main GIRAFFE target selection selecting these is made from stars with
$$0.00 \leq (J-K_s) \leq 0.45; \\ 14.0 \leq J \leq 17.5. $$
This selection is complemented with candidate red-clump giants from the same population, which are selected in
$$ 0.40 \leq (J-K_s) \leq 0.70; 12.5 \leq J \leq 15.0. $$
Stars are selected with an equal number per magnitude and per 0.1mag colour bin in these ranges, as illustrated in Fig~\ref{cmds-2}. Additional targets are selected in same CMD-selection to allow for fibre-positioner limits on target acquisition.

In parallel UVES fibres are dedicated to a special selection.
These field-star UVES parallel targets are chosen according to their near-infrared colors to be FG-dwarfs/turn-off
stars with magnitudes down to J = 14 mag. The goal was to observe a sample of  approximately 5000 FG-type
stars within 2 kpc of the Sun to derive the detailed kinematic-multi-element distribution
function of the solar neighborhood. This sample includes mainly thin and thick disc stars, with a sufficiently broad colour selection to include stars of
all ages and metallicities, including also a small fraction of local halo stars. For these targets the VHS photometry is saturated, so 2MASS is used. 
The target selection is based on 2MASS photometry (point sources with quality flags "AAA"). The
CMD box is defined by $12 \leq J \leq 14$ and $0.23 \leq (J-K) \leq (0.45+0.5\times E(B-V))$, the Schlegel et al. (1998) map being
used to determine the extinction E(B-V). The targets selected before April 2012 had a brightest
cut on J of 11 instead of 12 . When there were not enough targets the red edge was extended.
When there were too many potential targets an algorithm selected roughly an equal number of
stars per magnitude bin with the rest being marked as lower priority.
 Further illustration of the field star target selection, including more discussion of higher density and higher extinction fields is  available in \citet{Stonkute16}.

 \subsection{WG1, WG4, WG6b - open cluster star target selection and observation preparation}

The considerable work on open cluster star selection was coordinated by Angela Bragaglia, and described in \citeauthor[][in prep]{Bragaglia_prep}.

The open cluster survey has the very ambitious aim to cover the age-metallicity-distance-mass parameter space which is realistically available observationally.
Depending on the stellar spectral type, open cluster stars are observed with different GIRAFFE setups and two UVES settings. For
the hot/massive stars, the GIRAFFE setups HR3, HR4 (introduced only in 2016), HR5A, HR6, HR9B, and HR14N are employed, while HR15N is used for cool stars on the main sequence, pre-main sequence, and giant candidates. The
corresponding choices for UVES are the U520 and U580 setups. 
The final data release includes
spectra, radial velocities, rotational velocities, stellar parameters, metallicity, and detailed abundances for 62\footnote{One cluster is actually double, NGC~2451A and NGC~2451B; one candidate, Loden~165, turned out not to be a cluster after all.} open clusters observed by Gaia-ESO (plus approximately 20 clusters reanalysed from the ESO SAF). 

The open clusters show a large variety of cluster ages, evolutionary phases, spectral types and luminosities, and the choice of instrument/setup needs to take this into account. For cool stars, normally the fainter cluster members ([pre-]main sequence or turn-off stars) are observed using
GIRAFFE HR15N (sometimes HR09B for turn-off stars), while for the brighter stars (typically evolved giants or bright [pre-]main sequence
cluster candidates) UVES parallels are employed (with the U580 setup). Limiting magnitudes for cool stars (later than
A-type) are V=16.5 and V=19 mag for UVES and GIRAFFE respectively. Different magnitude ranges
are covered in clusters where hot stars are observed with the blue gratings. An overlap in
magnitude between the GIRAFFE and UVES samples is present normally and a number of stars
are observed with both instruments for inter-calibration purposes. For more details, see \citet{Bragaglia_prep}.

Within each cluster, the target selection procedure was implemented differently between
GIRAFFE and UVES, but uniformly across clusters. Namely, for GIRAFFE, with which we aim to
observe unbiased and inclusive samples, cluster candidates are essentially selected on the basis of
photometry. We used proper motions and radial velocities and other membership indicators (like e.g., X-ray emission)
only to define the photometric sequences and the spatial extent of the clusters. In other words, we employed existing information on membership only to
discard secure non-members, lying far from the cluster sequences. For UVES, with which we aim to target more secure cluster
members, we instead employed membership information from the literature (e.g. $v_{rad}$, Li abundance, H$\alpha$),
whenever available. More details on the target selection within clusters and preparation of the OBs can be found in \citet{Bragaglia_prep}.

\subsection{WG5 - Calibration and standard star target selection and observation preparation \label{calibration}}

The Gaia-ESO Survey primary calibration strategy, based on targeted observations of globular clusters, 
open clusters, and field stars, was coordinated by Elena Pancino, and is described 
more fully in \citet{Pancino17}.

Gaia-ESO dedicated considerable effort to define calibration stars -
clusters, special fields, CoRoT and Kepler-K2 fields, stars which are Gaia
calibrators, etc, to ensure Gaia-ESO is optimally calibrated, and that
other major surveys can be calibrated onto compatible parameter
scales. We recall that a primary goal of Gaia-ESO is to ensure all stellar populations, from young to old, hot to cool, metal-rich to metal-poor,
are on as consistent a calibrated scale as is feasible. This of course imposes severe challenges on the calibration strategy.
In addition, especially for field stars where only a single star could be observed  by the full fibre system, 
 there is obvious pressure to identify bright targets, so that twilight observations could be used as much as possible.

The survey team also analysed the
ESO SAF for abundance calibrations, and complementary data. The
large-scale (AMBRE-\citet{Worley16})  re-reduction of the ESO SAF is being done
consistently with the Gaia-ESO Survey, to ensure maximum value. 
Relevant archive data, specifically high signal-noise ratio observations with the Gaia-ESO setups of
targets consistent with the Gaia-ESO selection, were re-analysed as part of this survey, to
ensure maximum consistency across all datasets. Open cluster selection
is based on a critical analysis of available data in the archive, in
order not to re-observe cluster members for which spectra with the
required set-up and signal-noise ratio are already available. Calibration targets
were deliberately selected to optimise the archive value, by allowing
available spectra to be re-calibrated onto our abundance system.

The Standard cluster fields included in the survey, with spectra released as part of the public data release, are: calibration observations of stars in the globular
clusters M12, M15, M2, NGC104, NGC1261, NGC1851, NGC1904, NGC2808, NGC362, NGC4372,
NGC4590, NGC4833, NGC5927 and NGC6752 which meet our data quality selection threshold for inclusion;
and  calibrating open clusters observed in a range of setups to aid in
inter-setup calibration (Berkeley 32, M67, Melotte 71, NGC2243, NGC2420, NGC2477, NGC3532,
NGC6253, NGC6553).

In addition, much work was devoted to observing and improving the parameters of the Gaia Benchmark stars. The Gaia Benchmarks are a much broader project to support Gaia, but which also provides a set of primary calibrators adopted and extended by Gaia-ESO. The project is intended to provide a global set of well-understood standards of value to all spectroscopic surveys \citep{Jofre17_conf}.

To add further calibration weight, with specific ambition to ensure the asteroseismic and spectroscopic $ \log g$ scales are consistent, and also to include age calibration, special efforts in the calibration programme were focussed
on stars observed by CoRoT (\citep{Masseron_prep} and as part of the Kepler K2 mission \citep{Worley2020}.

\section{GIRAFFE data reduction pipeline [WG7]}\label{JRLpipeline} 

Gaia-ESO uses two spectrographs, both fed by the FLAMES fibre system. 
The higher resolution UVES data (table~\ref{UVES_modes}) are processed at Arcetri observatory by a team lead by Germano Sacco (WG7b). 
Their processing pipeline is based on the pipeline provided by ESO, and improved for this survey. 
This work is described in full in \citet{Sacco-UVES}.

\begin{figure}[!h]
\begin{center}
\includegraphics[width=\hsize]{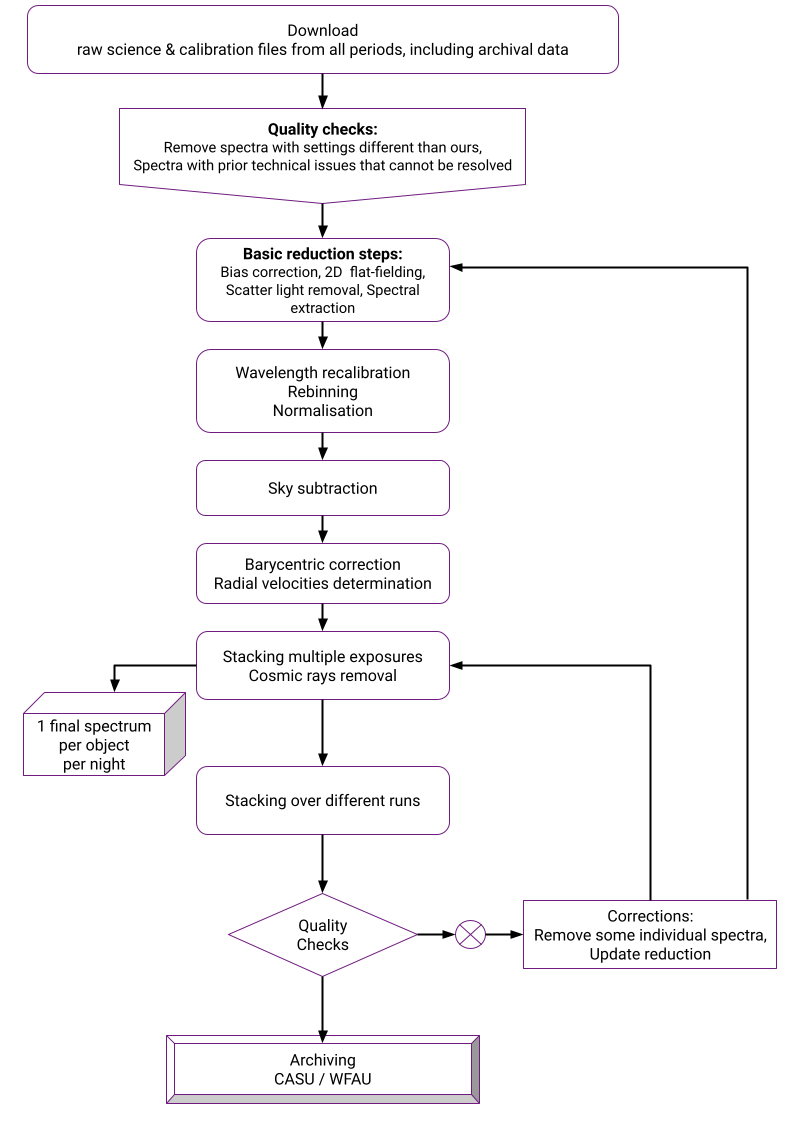}
\caption{Flowchart of the reduction processes for the GIRAFFE spectra.}
\label{FlowchartGiraffe} 
\end{center}
\end{figure}

The remainder of this section describes the GIRAFFE processing pipeline. 
This was developed by the late Jim Lewis of the Cambridge Astronomical Survey Unit (CASU). 
This processing system is the basis for that to be implemented for the future WEAVE and 
4MOST survey facilities. This text description is lightly edited from Jim's final description of this one of his many and much valued 
contributions to astronomy surveys. It provides an experienced perspective and a valuable introduction 
to the challenges in pipeline processing surveys with a very wide range of stellar astrophysical parameters. The logical flow is summarised in Fig~\ref{FlowchartGiraffe}.

\begin{figure}[!h]
\begin{center}
\includegraphics[width=\hsize]{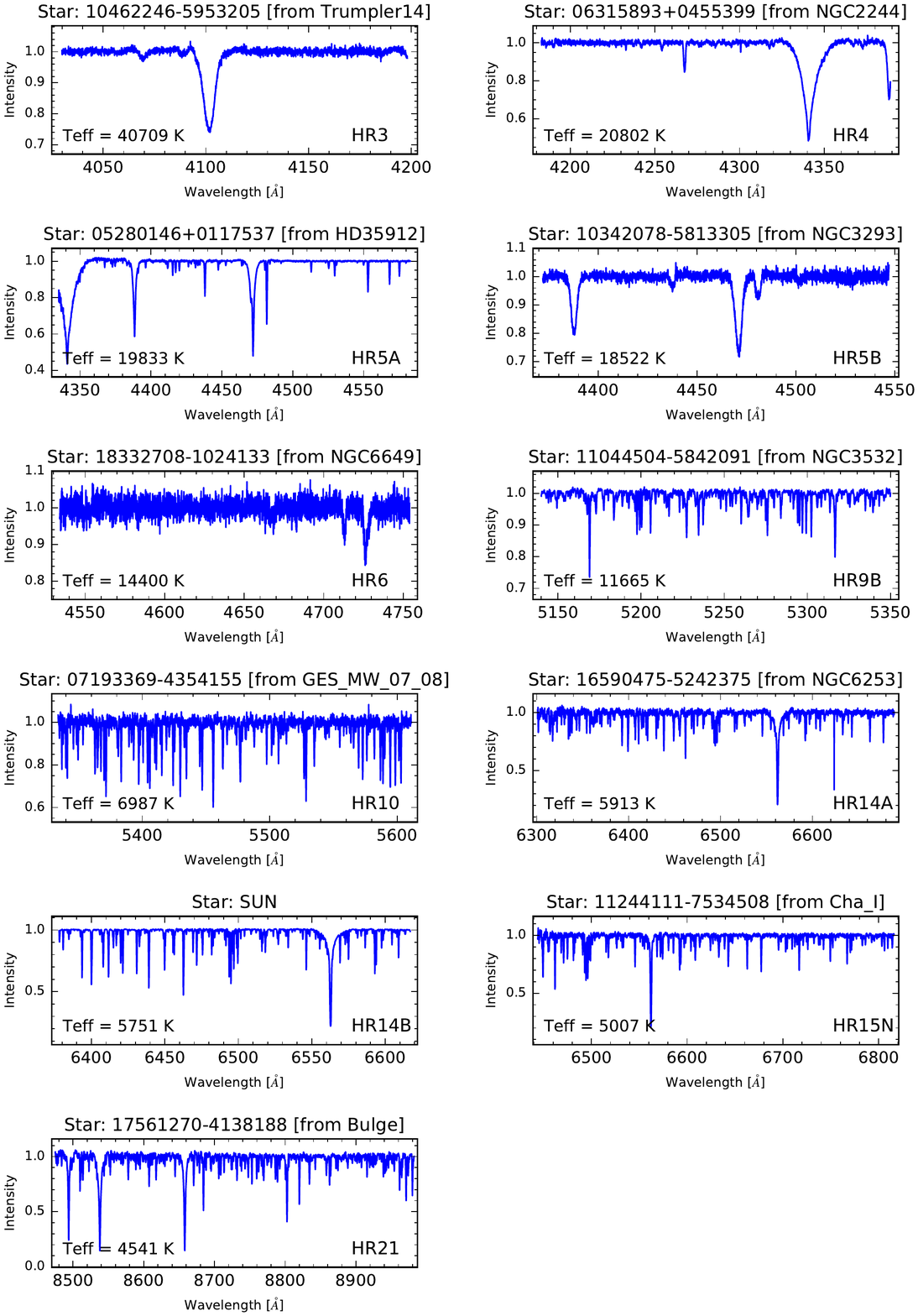}
\caption{ Example spectra from several GIRAFFE setups.}
\label{setups_gir}
\end{center}
\end{figure}

Fig~\ref{setups_gir} shows some example spectra from several of the GIRAFFE setups.

\noindent Reduction of GIRAFFE spectra involves the following steps:
\begin{itemize}
\item Default basic processing, including bias
subtraction; cross-talk and scattered light removal; bad pixel masking;
flat-fielding; wavelength calibration; extraction.
\item Wavelength recalibration of each extracted spectrum using sky
  emission lines, for red wavelength settings, or almost-simultaneous short simcal observations for the bluer settings.
\item Combining sky fibres for the determination of the master sky
  spectrum for each integration and subtracting (master or local, as
  scientifically appropriate) from individual extracted
  objects.
\item Combining the individual integrations into  co-added
  spectra, after measuring the RV, checking for
  binarity, and velocity shifting to the  heliocentric frame. Co-added
  spectra are used then to reject cosmic rays from
  individual spectra.
\end{itemize}

\subsection{Bias correction}

The bias for a given row in the detector is determined from the mean of the
same row in the detector underscan region. At this point the input data are 
integers and hence a clipped mean is used rather than the median in order 
to avoid quantisation of the bias estimate at the 1 ADU level.

\subsection{2-D flat field}

Flat field correction is done for a number of reasons. First of all it
removes the inter-pixel quantum efficiency variations in the
detector. It is also useful for taking out the large scale background
variations that arise due to the camera optics.  For the latter it is
common to do dome flat observations with the same grating, filter and
fibre module as the science exposure. The flat field spectra can be
extracted in the same way as the science spectra and then divided into
the science spectra. This should result in an object spectrum with a
background shape that is much closer to the true continuum of the
object. However as 1d spectra are the result of a summation across the
point-spread function of the fibre, any information on the quantum efficiency
variation of the individual pixels is lost. What is needed are dome
flat exposures that are taken with the grating and filter set as for
the science observations but without the fibre module. This would help
ensure uniform illumination of the entire detector, including the
inter-fibre regions where the scattered light contribution is
estimated. This is common practice with longslit spectroscopic
observations.

Unfortunately this sort of longslit flat field is not something that
is done with GIRAFFE and it appears not to be possible. There is a
maintenance slit in GIRAFFE that is used to do instrumental
health-check exposures and dome flat sequences are often done as
part of a rolling program of maintenance observations. These, however
are not dispersed, nor are they filtered and, as flat fields are
wavelength dependent, will not be a very accurate model of the
inter-pixel sensitivity variation in a science observation for a given
wavelength setup. Nevertheless, applying these flats to the science
data delivers a very evident reduction in noise. Tests have
also shown that these dome flats are quite stable over the course of
several months. With that in mind it was decided to apply these flats
to the science and calibration observations before extraction.

\subsection{Spectral extraction and 1-D flat field}

The spectra are extracted using the optimal extraction method outlined
in \citet{Marsh89} which is designed to work on 2d spectral profiles that
are highly curved. The extraction profiles are fitted to dome flat
exposures that are done during the daytime using the same grating
angle and filter setup as the science exposures done the
previous night. During the fitting procedure the contribution of
scattered light is estimated from the inter-fibre regions and
subtracted off at each point in the raw fibre data. 

The dome-flat spectra are extracted using the fitted extraction 
profiles. The average flux in all of the extracted dome flat 
spectra in a given exposure is calculated and then each of the 
flat spectra is divided by this average. This ensures that the 
differences in fibre throughput are corrected. 

Science spectra are extracted using profiles that are appropriate
for the GIRAFFE setup used in the observations. The 1d flat field
is then used to do the large scale continuum and fibre throughput
correction mentioned above.

\subsection{Wavelength solution and corrections}

The wavelength solution for a given grating/filter setup is tied to 
Thorium-Argon exposures done during the day. The solution for each
spectrum is determined independently through identification of arc
lines and doing a standard polynomial fit to the pixel positions and
wavelengths. 

The wavelength solutions are quite stable, but some drift in the zero
point does exist.  To measure this we use the aforementioned simcal
observations for most of the setups employed. Simcal observations 
consist of five fibres which are illuminated by the arc lamp and located
roughly uniformly across the detector. The arc lines are identified
exactly the same way as for the arc exposures and a mean shift 
is calculated from the change in pixel positions of the arc lines 
between the two exposures for those fibres. The correction for a
spectrum is done as an interpolation of the simcal shifts for the 
nearest two simcal fibres.

For the HR21 setup where we do not use the simcal arcs, it is possible
to monitor the wavelength drift by using the night sky lines that
exist in the individual science spectra. These can be identified in
the extracted spectra and the inferred wavelength from the standard
solution can be compared with a set of known sky line wavelengths to
measure the drift.

\subsection{Rebinning}

All spectra for a given setup are rebinned using a linear interpolation
onto a common wavelength scale with common endpoints. In doing the rebinning 
we take into account the wavelength solution found from the arc exposures, 
the wavelength solution corrections found from either simcal or skyline 
offset measurements and the  heliocentric correction for each 
object. This results in spectra on a uniform linear wavelength grid with 
0.05\AA \, pixels,
centred on the {heliocentric} system. Thus the spectra have
only been interpolated once, which minimises the correlation between 
adjacent pixels.

\subsection{Normalisation}

Within the analysis groups of the Gaia-ESO Survey there was a desire to have spectra
normalised such that the continuum shape was completely removed. 
This is not something
that can be done in a general way with the Gaia-ESO Survey as there is such a large
variety of stellar types in the samples. The pipeline produces a normalised 
spectrum
using a sliding window filter to model out the continuum. The results are
reasonably good and are certainly sufficient for the purpose of measuring
cross correlated velocities.  However serious
analysis that requires a normalised spectrum should include a normalisation
procedure that is optimised for the type of star being investigated. 
Results from the specialised working group analyses produce results very different from the
pipeline outcomes for stars with non-typical properties. These include common types,
OBA stars and young pre-main sequence stars, where considerable specialist processing is 
critical for reliable analyses.

\subsection{Sky subtraction}\label{sec:sky}
All spectra need to have a background (sky) correction applied. What
can be difficult is determining exactly how much to subtract off. The
OBs for the Gaia-ESO Survey are designed so that there are between 10-20 sky
fibres per field and of course these are observed simultaneously with
the science objects. What
one gets in each fibre can depend on spatial variations in the night sky as
well as the presence of real astronomical background (e.g. emission
nebulosity). The Gaia-ESO Survey spectra are taken in a variety of astronomical
environments and hence working out a general solution to the
background subtraction is not a straight-forward problem. In practise we
 combine all the sky fibres in a given
exposure into a single high signal master sky spectrum. If the observation
is done in a wavelength region where there are many sky lines, then 
for each science spectrum a scale factor is calculated, which is applied
to the master sky before it is subtracted from the science spectrum.
This scaling only applies to the HR21 setup.
In setups with wavelength regions without sky emission lines, this scale 
factor is assumed to be one.  

There is a need to provide for cases where there is a strong
variation in the  astronomical environment of the stars being observed. 
This is often the case when observing in active star forming regions.
To this end more complex sky subtraction algorithms are required.

\subsection{Sky background estimation }

It is not obvious that there is a single
"best" way to correct Gaia-ESO Survey spectra for background emission. The idea in a
nutshell is to form an estimate of the background emission from both
astronomical and atmospheric sources that is affecting the spectrum of
a science object. This estimated background spectrum can then be
subtracted on a pixel by pixel basis from the input science spectrum
to give an output spectrum that consists of emission from the object
alone. Unfortunately getting enough information to form that
estimated background spectrum can in some cases be extremely difficult
and it might well be that obtaining a result in which one can have full
confidence is impossible. 

When assessing the best way to remove background emission from Gaia-ESO Survey
spectra, it is important to understand the nature and origin of the
features in any background spectrum. If we look at a Gaia-ESO Survey spectrum that
has not yet been background corrected, but which is on a linear
wavelength scale and not corrected to  heliocentric, that spectrum will consist of contributions
from the following sources:

\begin{itemize}
\item The object spectrum – all features and continuum are shifted by the
star's line-of-sight velocity combined with its heliocentric
correction.  
\item Unresolved stellar light – this is generally quite
small (unless we are observing in high density regions such as the
galactic centre) and will mainly consist of continuum emission. Any
such light will be redshifted to a variety of velocities, and this may
result in very broad features.  
\item Emission lines from a surrounding
nebular component – these features originate in HII regions and the
like. They will be shifted by the recessional velocity of the emitting
gas and the heliocentric motion. Furthermore these features may be
broadened and skewed by local internal kinematics.  
\item Emission lines,
absorption (telluric) lines and continuum originating in the earth's
atmosphere – these features are in the earth's rest frame.
\item Solar light – residual solar light comes from moon light and the
zodiacal light and broadly speaking will be in the solar system rest
frame.
\end{itemize}

The method used to sample the background emission is to
assign a number of fibres to (what appear to be) background
positions. These are spread around the field in order to give as broad
a sample as possible of the background conditions in the area.
Correcting the background can be done in several ways. Some
possibilities are: 
\begin{itemize}
\item 1) Create an average background spectrum from all of the
background spectra in a given observation. Subtract this from each
object. This is a 'vanilla'
correction.  This is what must be done in the absence of any sort of
prominent atmospheric emission features in the wavelength
region. Without such features it is impossible to to say with any
certainty whether the amount of continuum emission being subtracted is
correct.
\item 2) Create an average background spectrum from all of the background
spectra in a given observation.  Use prominent emission features to
work out a scaling factor between this mean spectrum and the object
spectrum you wish to correct. Then scale the mean background spectrum
and subtract it from the object spectrum. Note that this scale factor
is not an attempt to correct for fibre throughput – that has been done
at the flat-fielding stage. It is rather an attempt to model out the
differences in the background emission across the field. The scaling
is done so as to minimise the emission lines in the stellar spectrum.
\item  3) Take one or more background spectra that lay in spatial proximity
to an object and average these. If there are emission features, then
scale the mean spectrum before subtracting. This is the  method
used if the background emission is varying rapidly in the
spatial sense. In extreme cases one can allocate a dedicated background
fibre for each object, but this has not been done for the Gaia-ESO Survey.
\item  4) If emission lines exists and they are all atmospheric, then simply
remove them by interpolating between two pixels on either side of
each of the lines. Then remove a scaled amount of the continuum. Such
regions are then flagged with a zero weight to help those fitting the
spectra later on. If the emission is nebular, then this is probably
not a good thing to do. Both the stars and the background may have the
same emission lines and it's important to try and subtract out the
correct amount of each. Also, the continuum emission and the nebular
emission arise from completely different physical properties and to
base the continuum estimation on the flux of the nebular lines risks
getting the continuum subtraction badly wrong. 
\end{itemize}

  The problem becomes
more interesting when the object spectrum that we want to recover has
emission and absorption features of its own in common with the
background (for example, H$\alpha$ ). That is now not a case of scaling the
background spectrum so that we have no more emission features in the
corrected spectrum (as with method number 2 above). We want to subtract just
enough to remove the background contribution, but no more. If the
background emission source is uniform and at exactly the same recessional velocity
as the object, this can be achieved by averaging all the sky
fibres. But this just simply does not happen in real star-forming systems. By and large there will
be gradients in the velocity and in the flux of emission lines across
the field.  When this happens then averaging all the background fibres
together to form a mean sky is clearly incorrect. This situation requires special consideration by users. An example applied to Gaia-ESO spectra of
the young cluster $\gamma$ Vel is given in \citet{Damiani2014}. \citet{Bonito2020}  explore the issue of sky subtraction where nebular emission contributes to emission lines of interest for star forming regions, for NGC 2264 (H$\alpha$, and forbidden emission lines including [SII], [NII]).

A more complex but real-world scenario is where there are background spectra emission
lines from both gas in the vicinity of the object and from the earth's
atmosphere.  The atmospheric lines in the background spectra will
match in wavelength to those in the object spectra. But in general
there will be a wavelength offset between the gas emission lines for a
background spectrum and those in an uncorrected object spectrum as the
fibres used for these spectra will sample the emitting gas in two
separate places where the local kinematics may well differ.
 
\subsection{Sky subtraction approach adopted for each GIRAFFE setup}

\subsubsection  {HR5A, HR6 and HR9B}
Background spectra for HR5A, HR6, and HR9B do not show any emission lines and they
all show some of the characteristic absorption features of the solar
spectrum.  For GIRAFFE setups such as these with no emission
lines, we have to employ option 1 outlined in the previous section. In
general the continuum emission in the background spectra is small and
a 10-20\% error in the background correction does not make a great
deal of difference in the final object spectrum (unless, of course,
that object is very faint).

\subsubsection {HR3 and HR4}
Background spectra for HR3 and HR4 do have some
background  nebulosity and hence there is a small emission line at
$\lambda$4101 which is H$\delta$. As this is a nebular  line it would be wrong to
scale the mean background  to try and remove it. In the absence of
this line a vanilla subtraction is all that can be done. Where the line
exists then one can shift  the mean sky
spectrum  so that the H$\delta$ lines coincide and then do a vanilla
subtraction.

\subsubsection {HR10}
In the HR10 (H548.8) setup there is a very strong emission in the form of
the [OI] $\lambda$5577 atmospheric line, and also three weaker atmospheric OH lines. Although the intensity
of 5577 in general varies across the field of view, the shape of the
line and its central wavelength is reasonably stable (notwithstanding
any changes in the line-spread function across the detector due to the
camera optics). For that reason it is possible to use the intensity of
this line to work out an individual scaling factor that can be applied
to a mean background spectrum before subtracting it from each object
spectrum. Because of the stability of this line's position and shape
it corrects out reasonably well. Because the other lines are so weak
by comparison, we do not include them in the scaling
calculation.

Tests show that correction using the
'vanilla' method described above clearly overcorrects the sky subtraction. 
Correction by scaling to the line provides much better correction. 
The HR10 setup with good signal-noise allows application of method 4,
which is simply to work out a scale factor which we use to correct the
continuum and then just cut out the line residual. This provides a small cosmetic
improvement over intensity-scaling.

\subsubsection {HR21}

In the HR21 (H875.7) setup there is a large number of atmospheric
emission lines. These not only vary in
intensity across the field, the intensity of the lines vary with
respect to each other. The lines at $\lambda$8493 and $\lambda$8505 show 
definite changes in their flux ratios between the concurrently-observed spectra. This
 is a sky background spatial
variation. This effect means that it might not always be possible to
define a single scaling factor which will remove all of the lines
equally well. Also as many of these lines are blends the intensity
variation will alter their blended shapes across the field. This means
that even if a scale factor that is consistent with all the line
intensities can be defined, there may still be significant residual
emission after background correction.
Where appropriate residual line artefacts can be clipped from the 
spectra - the key result is that the weight map contains the 
relevant astrophysical information for later analysis.

\subsubsection {HR14 \& HR15N}
 
The HR14 and HR15N setups have both a large number of atmospheric
lines and (when observing young clusters) nebular lines. As
mentioned before, this has the potential of causing a lot of trouble
because the sets of lines are in two different velocity frames of
reference. Any attempt to shift the mean background spectrum so that
the positions of the nebular lines match will mean shifting the
atmospheric lines out of coincidence. 

Another problem with nebular lines in the background spectrum is that
they will be affected by the internal kinematics of the emitting gas
and this will vary across the field.  As such if we seek to use a master sky
which is a mean of all the sky spectra in an exposure, we will have
mean line profiles that look nothing like the lines we want to
correct.

There is a further issue here on the question of whether we should be
scaling nebular lines to a mean background  spectrum at all. The
objects themselves may well have one or more of these lines and
scaling  a  mean  background   so  that nebular  lines  are  removed
is  in essence  forcing an astrophysical scenario where the objects
cannot have emission lines. This is further complicated by the velocity
offsets that may occur between the background and the object, for
example a star with an expanding circumstellar shell.  An example analysis of 
Gaia-ESO HR15N spectra for the young cluster $\gamma$ Vel is given in \citet{Damiani2014}.

 \subsection{Outcome methods adopted \label{sky}}

 In what follows we summarise how the background subtraction algorithms
 were optimised for the final data releases.

\begin{itemize}
\item HR5A, HR6 and HR9B – We use the vanilla algorithm. 
\item HR3 – The H$\delta$ line that sometimes appears is nebular in origin. If it
does appear then the mean background spectrum can be shifted so
that the lines coincide. Then a vanilla correction can be done. The
continuum in most cases is mainly solar in origin and this has nothing
to do with the physical process that gives rise to a nebular
line. Scaling the spectrum to remove the line would probably cause an
error in the amount of continuum to be removed. In the absence of the
emission line, just subtract a mean background spectrum.
\item  HR10 – The $\lambda$5577 line is atmospheric in origin and hence it is safe
to use it to work out a scale factor. The residual line itself is
removed using the clipping method. The pixels where the line is removed
are given a zero weight. The scale factor calculated from the line
ratio can then be used to subtract out the correct amount of
continuum.
\item  HR21 – All of the lines are atmospheric in origin, hence we 
treat this in the same manner as for HR10.
\item HR14 and HR15 – These can have both atmospheric and nebular
emission. In the case where no nebular emission is present, then we
treat the spectra in the manner outlined for HR10. In the presence
of emission lines we adopt a two-pass procedure. The first is to
create a mean background  spectrum  with nebular lines that are
clipped out. This is scaled by the atmospheric line ratios, which
will correct the continuum. The atmospheric lines are then clipped
out. Then a second background  spectrum with no atmospheric lines and
with a mean  continuum of zero is shifted in wavelength space so
that the nebular lines coincide. These are subtracted  using a scale
factor of 1. This method sometimes leaves ugly remnants in the
spectra, because the line profile shapes are 
different. But this is a scientifically more valid way of dealing with
the problem. 

\item If a science analysis is
interested in the emission features of stars in the Gaia-ESO Survey it will  be
necessary to go back to the spectra  that are  not
background corrected.  Using  multi-component  fits  and  velocity
information  it  may  be possible  to recover the true emission  line
that belongs  to the stellar  object.  This  however  is  not
something that can be done in bulk in a data reduction pipeline. 
\end{itemize}

\subsection{Stack multiple exposures}

The observations are generally broken up into at least two exposures. The
rebinned and sky corrected spectra are coadded  using the information in the variance spectra to form a final spectrum
for each object for a given night. It is at this point that any artefacts
such as cosmic rays are removed.

\subsection{Cross-correlation velocities }\label{subsec:crosscor}

The pipeline also does a simple cross correlation of each spectrum
relative to a selection of model atmospheres. This gives a 
first pass estimate of the radial velocity, surface gravity, metallicity
and effective surface temperature of each star. A much better estimate of these
parameters is generated initially in the Radial Velocities WG (WG8 \& WG9) and finally through the full spectroscopic analyses
in the specific Working Groups.  This sanity checks for variable radial velocity before co-adding.

\subsection{Stacking over different runs}

In order to achieve the desired signal to noise many stars are being 
observed over the course of several runs. The final spectra that are
released to the analysis groups are those that have been coadded over
all the observing runs that have been done, with in all cases check for radial velocity variability before addition. These final stacks are also
cross-correlated as specified in section \ref{subsec:crosscor} to give
initial estimates of velocity, surface gravity, effective temperature and
metallicity.

\section{Radial velocities [WG8] and first-pass classification [WG9]}\label{wg8} 

\subsection{GIRAFFE spectra}

The work of WG8 and WG9 has been led, and largely implemented, by Sergey Koposov. The two WGs were in effect merged in to a single set of operations, as described below.

\citet{Koposov2015}  introduces the Radial Velocity methodology, building on the earlier development of precise radial velocities from GIRAFFE spectra by \citet{Koposov2011}  in studies of the internal kinematics of dSph galaxies. \citet{Jackson15}  investigates the delivered velocity precision as a function of stellar type, signal to noise ratio and observing setup. The maximum achieved precision is of the order of 0.25 km/s, matching the initial goal.

After pipeline processing to remove
instrumental signatures, extracted individual GIRAFFE spectra
with their variance spectra and quality control flags are available. Telluric lines, cosmic rays in cases of single exposures
and known defects are masked. Astrophysical emission lines are detected and fitted, but not considered during the radial velocity processing.
The method utilises the direct pixel-fitting process described by \citet{Koposov2011}.  This is not a cross-correlation method, which is known to be non-optimal, but proceeds by fitting observed data by model spectra derived from a spectral library. The essence is to ensure one is working with a minimally rebinned spectrum with reliable variance associated with each pixel. As described in \citet{Koposov2015}, which this description follows,  the Gaia-ESO spectra are rebinned by the pipeline, which generates correlated noise in the spectra. It is necessary to utilise appropriately the full covariance spectrum calculated during reduction, or an appropriate validated scaling of the errors. Each spectrum is then fitted by a suitable subset of a large spectral library. This provides a first-pass estimate of both the stellar parameters ($T_{eff}$, log $g$, [Fe/H], [$\alpha$/H]) and the radial velocity.

For performance reasons, two iterations are repeated. First, to optimise the radial velocity, keeping the stellar astrophysical parameters fixed, followed by an iteration to optimise the stellar parameters while keeping the radial velocity fixed. We also iterate starting from random stellar parameters to reduce the likelihood of being trapped in a local minimum in the fitting surface. Where the signal-noise ratio is sufficiently high, the stellar Vsin$i$ is also fitted.

Given these initial estimates, the next step is to adopt a Gaia-ESO optimised grid of model spectra. The Gaia-ESO model grid was calculated for the survey using the same methodology as that for the AMBRE project, which is described by \citet{Laverny2012}. This is a high-resolution ($ {\mathcal R} \ge 300,000$) grid using Turbospectrum \citep{Turbo1998}, MARCS model atmospheres \citep{MARCS2008} , and the dedicated Gaia-ESO line list \citep{Heiter2021}. The grid covers $3000K \le T_{eff} \le 8000K$, $0 \le {\rm log} g \le 5$, $-5 \le$ [Fe/H] $\le +0.5$, and $0.0 \le [\alpha/Fe] \le 0.8$.

Given the first-pass parameter estimates, for a given set of stellar parameters ${\omega = \{T_{\rm eff},\log{g},{\rm [Fe/H]},[\alpha/{\rm Fe}]\}}$ we first produce a flux-normalised synthetic spectrum $S(\lambda,\omega)$ at wavelengths $\lambda$ by interpolating spectra from a surrounding grid. We redshift our interpolated spectrum by velocity $V$ such that the normalised synthetic flux at an observed point $\lambda$ is given by $S\left(\lambda \left[1 + \frac{V}{c}\right],\omega\right)$, where $c$ is the speed of light. The observed continuum is modelled as a low-order polynomial, with order set to a higher value for spectra with signal-noise $\ge 20$,   with coefficients $b_j$ which enter multiplicatively:
\begin{equation}
    M(\lambda,\omega,v,\{b\}) = \sum_{j=0}^{N-1}b_{channel,j}\lambda^{j} \times S\left(\lambda\left[1 + \frac{V}{c}\right],\omega\right)
\end{equation}

The continuum in each observed channel is modelled separately. We convolve the model spectrum with a Gaussian line spread function (LSF) (with free parameter $\mathcal{R}$) to match the resolving power in each channel, and resample the model spectrum to the observed pixels $\{\lambda\}$. Although the spectral resolution $\mathcal{R}$ in each channel is reasonably well-known, refocusing of the GIRAFFE spectrograph during the survey improved the quoted spectral resolution. For this reason we chose to include the spectral resolution $\mathcal{R}$ as a nuisance parameter with reasonable priors and marginalise them away.  After convolution with the LSF, binning to the observed pixels $\{\lambda\}$ and assuming Gaussian error $\sigma_i$, the probability distribution $p\left(F_i|\lambda_i,\sigma_i,\omega,V,\{b\},\{\mathcal{R}\}\right)$ for the observed spectral flux $F_i$ is:
\begin{equation}
p\left(F_{i}|\lambda_i,\sigma_i,\omega,V,\{b\},\{\mathcal{R}\}\right) = \frac{1}{\sqrt{2\pi\sigma_{i}^{2}}}\exp{\left(-\frac{\left[F_i - M_i\right]^2}{2\sigma_{i}^{2}}\right)}.
\end{equation}

Under the implied assumption that the data are independently drawn, the likelihood of observing the data $D$, given our model, is found by the product of individual probabilities:

\begin{equation}
    \mathcal{L} = \prod_{i=1}^{N}\,p\left(F_i|\lambda_i,\sigma_i,\omega,V,\{b\},\{\mathcal{R}\}\right)
\end{equation}

\noindent{}and the probability $\mathcal{P}$ of observing the data is proportional up to a constant such that:

\begin{eqnarray}
\mathcal{P}     & \propto   & \mathcal{L}\left(D|\theta\right) \times \mathcal{P}r\left(\theta\right) \nonumber \\
\ln\mathcal{P}  &   =       & \ln\mathcal{L}\left(D|\theta\right) + \ln\mathcal{P}r\left(\theta\right) 
\end{eqnarray}

\noindent{}where $\mathcal{P}r(\theta)$ is the prior probability on the model parameters $\theta$. 

This analysis describes the joint probability distribution of an identified template and an associated target star radial velocity. In practice we evaluate the probability for a grid of radial velocities - essentially a uniform prior on radial velocity - and for our grid of templates, again assuming a uniform prior. The resulting 2-D probability distribution can be marginalised over  appropriate parameters  to determine the maximum  likelihood a posteriori estimate of the velocity and the velocity error, and similarly to determine the parameters of the best-fitting template.

\subsection{GIRAFFE radial velocity accuracy}

\begin{figure}
\begin{center}
\includegraphics[width=7.5cm]{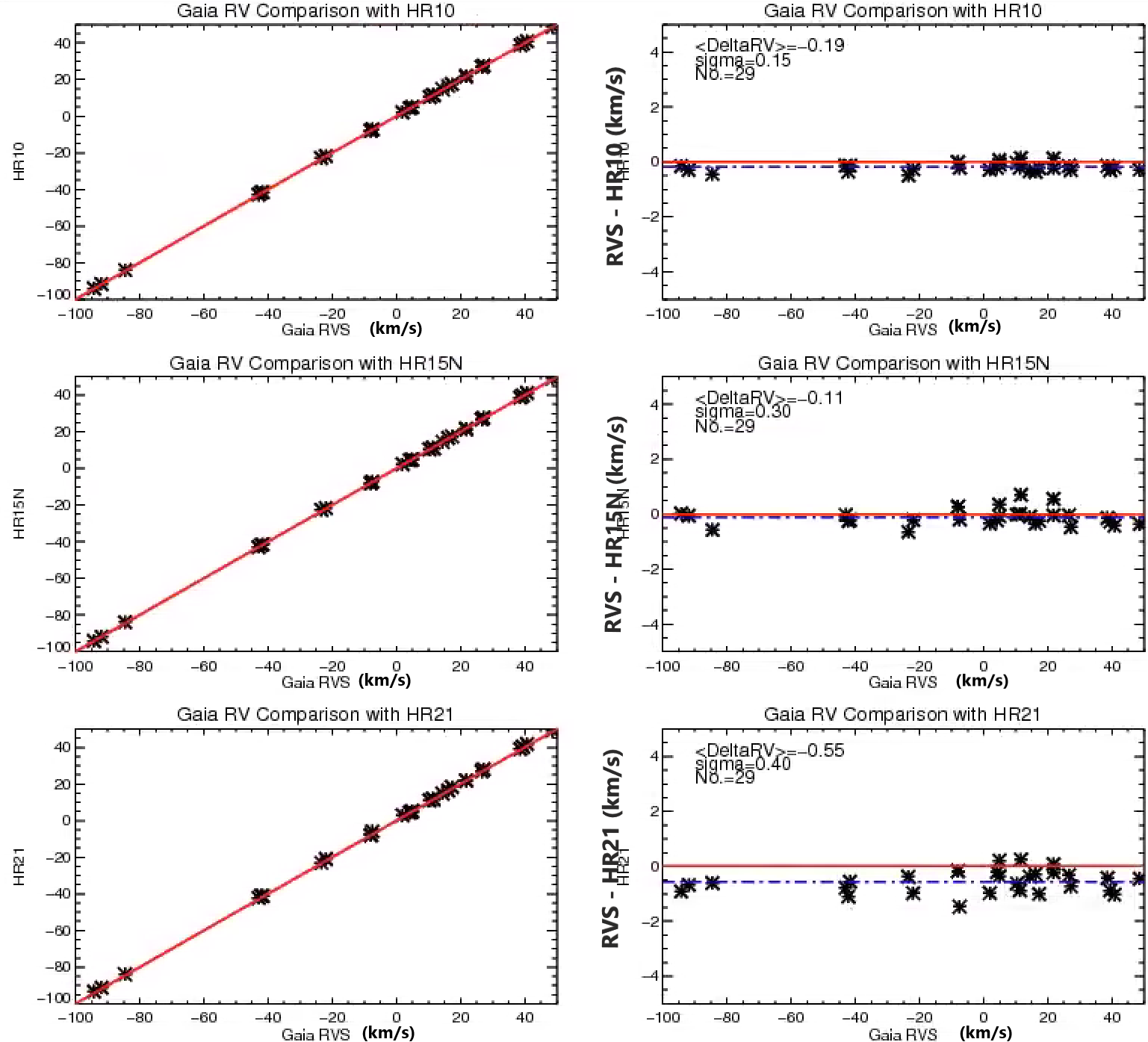}
\caption{Comparison of radial velocities determined for the Gaia Radial Velocity Standards by the Gaia ESO Survey pipeline for various GIRAFFE instrumental setups with the reference values for the same stars from \citet{SoubiranRV}.
 The red lines are the one-one relation, blue lines fitted offsets.}
\label{RV-1}
\end{center}
\end{figure}

\begin{figure}
\begin{center}
\includegraphics[width=8.0cm]{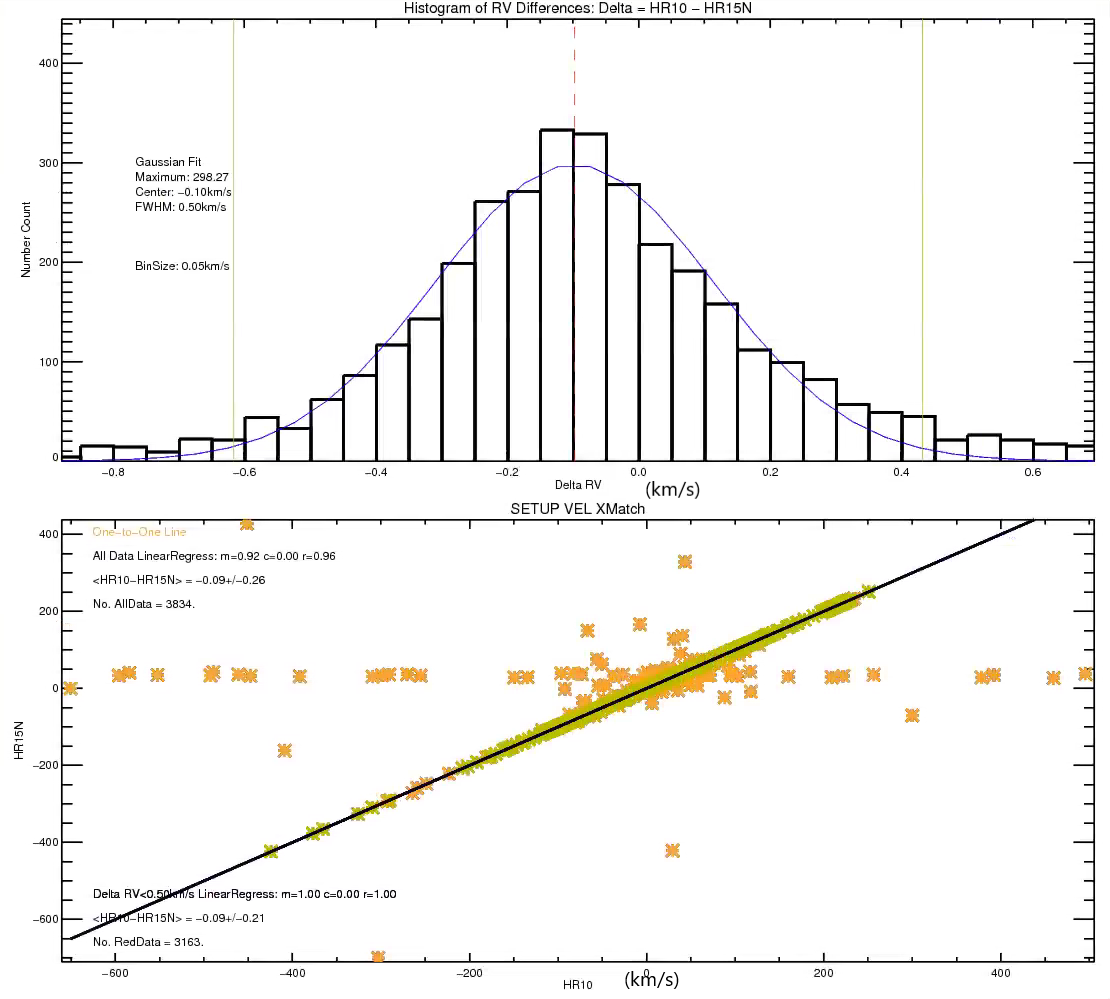}
\caption{A comparison of the Gaia ESO Survey radial velocity pipeline results for the stars in common between the GIRAFFE HR15N and HR10 instrumental setups. The agreement is very good, with the offset between the two scales determined to be -0.09$\pm$0.26 km/s.}
\label{RV-2}
\end{center}
\end{figure}

 It is well-known that although the formal radial velocity precision derived from cross-correlation or pixel-fitting methods can be almost arbitrarily small for sufficiently high S/N spectra, the actual precision achievable with most spectrographs is generally limited by systematic effects. This includes instrument flexure, uncertainties in the wavelength calibrations, Line Spread Function (LSF) variation/asymmetry and template mismatches. This systematic component has to be included in the total error budget. We have found this systematic error between the HR21 and HR10 setups to be around 300\,m s$^{-1}$ from large numbers of Gaia-ESO Milky Way spectra. It is important to note that this systematic component is not expected to be present when comparing RVs obtained from spectra using the same setup in sequential exposures, but it becomes important when comparing RVs from different nights, or between setups. We include this systematic error floor in suitable comparisons hereafter.

The range of instrumental setups in which a star is observed, and hence the number of available spectra with associated radial velocities, varies per star. Calibrators, for example, have typically been observed with a broader range of instrumental configurations and will thus have a relatively greater number of radial velocity determinations than a typical field star. Additionally, particular analysis nodes and Working Groups deliver revised estimates of the radial velocities for their targets of interest that they determine during their specialised analysis for the parametrisation of these spectra. Thus, as with most of the quantities derived from the spectral analysis, multiple radial velocity results are available per star, and these need to be homogenised to produce a single recommended radial velocity per star. 

As part of the homogenisation by WG15 of the available radial velocities for each star, different instrumental setups were compared, and offsets were applied to bring the radial velocities on to the same scale. The radial velocities from the HR10 setup were used to establish the zeropoint of the radial velocity scale due to their good agreement with the literature values of the Gaia Radial Velocity Standards (\citet{SoubiranRV}; see Fig~\ref{RV-1}). Fig~\ref{RV-2} shows an example of the cross-match between different instrumental setups and the offset found and applied during homogenisation. A more complete description of the radial velocity accuracy and zeropoint as a function of stellar and observational parameters is presented in \citet{Jackson15}, while an updated version is presented in the accompanying paper \citep{Randich21}.

\subsection{UVES radial velocities}

Determination of radial velocities from the high-resolution UVES spectra is described fully in the UVES processing description paper \citep{Sacco-UVES}, and updated in the companion article \citep{Randich21}. 

We provide a brief summary from that paper here for convenience. \\
Radial velocities are derived by cross-correlating each spectrum with a grid of synthetic template spectra.  The grid is composed of 36 spectra convolved at the FLAMES-UVES spectral resolution. It covers seven effective temperatures (T$_{eff}$ = 3100, 4000, 5000, 6000, 7000, 8000 K), three surface gravities (log $g$ = 2.5, 4.0, 5.0), and two values of metallicity  ([Fe/H] $= 0.0, -1.0$).

Each spectrum is cross-correlated with all the spectra of the grid, using the IRAF task FXCOR masking the Balmer lines (H$\alpha$ and H$\beta$) and regions of the spectra with strong telluric lines. To derive the radial velocity, the cross-correlation function (CCF) with the highest peak is selected and the peak is fitted with a Gaussian function to derive its centroid. This procedure fails for very early-type stars with an effective temperature above the highest temperature of our grid, which are characterised by the presence of no, or very few, absorption lines other than the Balmer lines.  Radial velocities for these stars are treated separately by the dedicated Working Group WG13.

To estimate the precision of the velocities, we used the differences between velocities measured from the lower (RVL) and upper (RVU) spectra, which are measured independently by the pipeline. Assuming identical uncertainties on the velocities from the two wavelength ranges, and since there is no systematic offset between lower and upper spectra (median(RVU - RVL) = 0.007 km/s), the empirical error on the velocities derived by our pipeline is defined to be
${\rm \sigma_{UL} =  (RV_U  - RV_L)/\sqrt{2}}.$ The statistical error on a radial velocity is equal to the 68th percentile rank of the distribution of these empirical errors, after outliers have been removed, and corresponds to $\sigma$ = 0.18 km/s.

Since the upper and the lower spectrum are calibrated using the same arc lamp, our approach for the error estimate does not take into account the error due to the variations of the zero point of the wavelength calibration. In order to estimate this source of uncertainty, we used spectra of targets observed multiple times in different epochs. Similarly to the above case, the empirical error is estimated as  $\sigma = \vert \Delta RV \vert/ \sqrt{2}$ where $\vert \Delta RV\vert$ is the difference between two observations of the same target performed on different nights. The distribution of this empirical error is much wider than the distribution of the errors $\sigma_{\rm UL}$; the 68th percentiles are $\sigma_U$ = 0.38 km/s and $\sigma_L$ = 0.40 km/s for the lower and upper ranges, respectively. This proves that the variations of the zero point of the wavelength calibration are the main source of uncertainty. Therefore, we adopt $\sigma \sim 0.4$ km/s as the typical error for the radial velocities derived from the FLAMES-UVES spectra of the Gaia-ESO Survey.

\section{Spectrum analyses (WG10,WG11,WG12,WG13,WG14)}\label{spectrum_analyses}

The spectrum analyses have been performed in five Working Groups; 
Working Group 10 analysed GIRAFFE  spectra of normal FGK stars; Working Group 11 analysed UVES spectra of normal FGK stars; Working Group 12 analysed cool pre-main sequence stars; Working Group 13 analysed hot OBA stars; Working Group 14 has two roles, one to process unusual and complex objects, such as white dwarfs and spectroscopic binaries, and also to develop project wide data quality flags.

The goal of the working groups was to process  extracted spectra to refine 
astrophysical parameters, to deliver elemental abundances to a level
appropriate for the relevant stellar type and available signal-noise ratio, to derive
stellar properties (e.g. activity, accretion, rotation whenever
relevant) and to provide detailed analysis-level quality-control. Each Working Group had a similar structure, involving a set of ``nodes", each of which processed (a subset of) the spectra using its preferred analysis methods and software.

The structure of the Working Groups for
spectrum analysis provided close coordination between the teams,
ensuring the optimum range of analyses would be applied to the various
stellar and data types as appropriate.  The methodologies are all
established, all publicly well-documented, forming the basis of most
modern spectrum analyses in the literature.  Below we provide a
general description of the input data, as well as of the strategy and
methods followed.

\noindent{\bf Input }
The main input to the spectrum analyses consists of reduced
spectra.  These have been put on a wavelength scale, have been
velocity shifted to a barycentric reference frame, and have been
pipeline normalised. Quality information is also provided, including variance
spectra, Signal-Noise ratio, non-usable pixels, and any other relevant output from the spectrum processing.  Additional inputs are the
radial and rotational velocities derived by Working Group 8, photometric data, and
first guess atmospheric parameters derived by Working Group 9.  For cluster stars,
cluster distances and reddening values are also  available as input
to the spectrum analysis.

Line lists, atomic and molecular data
($gf-$values, broadening constants, etc.) appropriate for the different
categories of targets and spectral intervals are compiled by dedicated efforts and  made available survey-wide to the
analysis nodes. These actions are taken to ensure homogeneity in the
derived quantities. Similarly, all the nodes participating in the
analyses have adopted a fixed set of model atmospheres, which is
sufficiently broad to be optimised for each class of survey
astrophysical targets.

\noindent{\bf Analysis: strategy and methods}
The five Working Groups delivering the spectrum analysis all follow a similar 
approach, summarised in the following.
\begin{itemize}
\item The data analysis was both distributed and duplicated among the nodes
contributing to each working group. Specifically, more than one group
analyses and produces results for (nearly) all relevant survey targets. This
duplication of different methods allows, given  performance comparison of 
the results,
production of a set of recommended parameters. It also, through
rigorous quality control, provided  a quantitative estimate
of both random and method-dependent uncertainties. In the event of
discordant results for a specific star, individual checks could be
conducted.  Quality monitoring and outlier detection were performed
throughout the survey. Monitoring of homogeneity was conducted by
the analysis teams, in coordination with Working Group~15 (see below).  The
decision on the final recommended values was the
responsibility of Working Group~15.
\item A first pass analysis was performed,
followed by a more refined analysis. The first pass analysis 
quality checked the preliminary classification parameters and 
provided astrophysical parameters, which, together with the
information on photometry from the target selection procedure, 
were input to subsequent analyses.
\item Depending on the star's spectral-type and characteristics, 
appropriate optimal tools, software, and model atmospheres were used;
however, some methodologies in common to all Working Groups could be identified.\\
The methods to derive astrophysical parameters and abundances could be
roughly divided in two broad categories. The first one
includes the main types of parameterisation methodology, such as exhaustive 
search algorithms, global optimisation methods, projection algorithms, 
pattern-recognition methods, and Bayesian parameterisation
approaches. The second one consists of more classical approaches, 
based on measurements of equivalent widths of absorption lines and
inversion codes (spectral synthesis), or use of curves of growth for particular 
lines/elements (e.g., Li).
Equivalent widths were measured with (semi-)automatic codes by fitting Gaussian
profiles to the lines. The available codes included DAOSPEC,
ARES, and SPECTRE. \\
Specific methods were used in 
special subsets of the sample (e.g., H$\alpha$ wings, line-depth-ratios).
In most cases the codes were automatic, and proven to be able to handle
large scale data volumes.\\
\end{itemize}
More specific details on the analysis of the different types of stars/spectra 
are given in the following sections. The WG10 homogenisation  was carried out with reference to the results of the WG11 homogenisation, since many stars were in common, thus WG11 is described first.

\subsection{Working Group 11: UVES spectra of ``normal'' FGK stars}

The operation of Working Group 11 (WG11) is described more fully in the article "The Gaia-ESO Survey: The analysis of high-resolution UVES spectra of FGK-type stars", \citet{Smiljanic14}. Further details on updates implemented for the final Gaia-ESO data release are given in \cite{Worley19_wg10}. A total of 13 different nodes participated in the analysis described in \citet{Smiljanic14}.

The methodologies are described in the Appendix A of \citet{Smiljanic14} and summarised in Table \ref{tab:wg11.nodes}. An exception is the Arcetri node which did not participate in earlier WG11 analysis cycles and is described in \citet{Lanzafame15}. Here, we give just a brief summary of the codes used by each node for completeness:

\paragraph{Arcetri:} Equivalent widths of the Li and the nearby Fe line were measured with Gaussian fitting. Abundances were determined with a set of curves of growth \citep{Franciosini_prep} determined from a grid of synthetic spectra computed with the methods and tools described in \citet{Laverny2012} and in \citet{Laverny2013} and adopting the AMBRE linelist of \citet{Guiglion2016}. 

\paragraph{CAUP:} Equivalent widths were measured with the Automatic Routine for line Equivalent widths in stellar Spectra code \citep[ARES,][]{ARES1,ARES2}. Atmospheric parameters and chemical abundances were determined with MOOG \citep{MOOG}.

\paragraph{EPINARBO} The EPINARBO node was composed of sub-nodes located at several institutes that contributed during the different phases of the project, including: Laura Magrini (INAF-Arcetri); Angela Bragaglia and Paolo Donati (INAF-Bolgna); Antonella Vallenari, Tristan Cantat-Gaudin and Rosanna Sordo (INAF-Padova); and University of Indiana (Eileen Friel and Heather Jacobson). Equivalent widths were measured with DOOp \citep{Cantat14}, a wrapper that allows improved and automated use of DAOSPEC \citep{DAOSPEC}. Atmospheric parameters and abundances were determined with the Fast Automatic MOOG Analysis code \citep[FAMA,][]{FAMA} which automates the use of MOOG \citep{MOOG}. 

\paragraph{IAC-AIP:} The code FERRE \citep[see][and references therein]{2014A&A...568A...7A} was used. The strategy was to find, for each observed spectrum, the atmospheric parameters of the best fitting model among a grid of pre-computed synthetic spectra.

\paragraph{LUMBA:} The LUMBA UVES analysis pipeline is fully described in \citet{LUMBA}. The pipeline made use of the Spectroscopy Made Easy code \citep[SME,][]{SME,SME.evolution} for computing on-the-fly synthetic spectra that were used to determine atmospheric parameters and chemical abundances. Departures from Boltzmann and Saha statistics were considered for iron lines which are prominently used in the derivation of stellar parameters. 

\paragraph{Nice:} The MATrix Inversion for Spectral SynthEsis \citep[MATISSE,][]{MATISSE} algorithm was used. The code used a method that projects the observed spectrum onto functions determined from the linear combination of a pre-computed grid of synthetic spectra. The UVES analysis of the Gaia-ESO setups proceeded as described in \cite{worley2016}.

\paragraph{OACT:} The OACT node used the code ROTFIT \citep{2003A&A...405..149F,2006A&A...454..301F}. The method consisted in a $\chi^2$ minimisation of the residuals between the observed spectrum and a set of reference spectra. In this case, a library of observed spectra, from the ELODIE archive \citep{ELODIE}, was used as reference.

\paragraph{UCM:} The code STEPAR \citep{STEPAR} was used to determine atmospheric parameters based on equivalent widths automating the use of MOOG \citep{MOOG}. Equivalent widths were measured with the Tool for Automatic Measurement of Equivalent width \citep[TAME,][]{TAME}.

\paragraph{Vilnius:} Equivalent widths were measured with DAOSPEC \citep{DAOSPEC}. The node developed its own wrapper to automate the use of the MOOG code \citep{MOOG} for the determination of atmospheric parameters and chemical abundances. Although most abundances were computed from the equivalent widths, a few selected species were analyzed with spectrum synthesis. The Vilnius node was the only provider of carbon and nitrogen abundances from the synthesis of C$_{2}$ and CN molecular features, and oxygen abundances from the synthesis of the forbidden oxygen line at 6300 \AA. \\

\begin{table*}
	\caption{Overview of the WG11 Nodes that participated in the analysis of the last Gaia-ESO data release.}
	\label{tab:wg11.nodes}
	\centering
	\small
	\begin{tabular}{lp{3.9cm}p{3.9cm}p{3.9cm}}
		\hline\hline
		Node &  Method & Analysis lead & Data products \\
		\hline
		Arcetri  & Equivalent widths & Elena Franciosini & Lithium abundances \\
		CAUP     & Equivalent widths & Vardan Adibekyan, Elisa Delgado-Mena, Sergio Sousa & Stellar parameters, abundances \\
        EPINARBO & Equivalent widths & Laura Magrini & Stellar parameters and abundances\\
        IAC-AIP  & Library of synthetic spectra & Michael Weber & Stellar parameters \\
        LUMBA    & On-the-fly spectrum synthesis & Alvin Gavel & Stellar parameters and abundances\\
        Nice     & Library of synthetic spectra & Clare Worley & Stellar parameters \\
        OACT     & Library of observed spectra & Antonio Frasca and Katia Biazzo & Stellar parameters and activity \\
        UCM      & Equivalent widths & Hugo Tabernero & Stellar parameters \\
        Vilnius  & Equivalent widths, spectrum synthesis & Arnas Drazdauskas, Gražina Tautvaišienė & Stellar parameters and abundances \\
		\hline
	\end{tabular}
\end{table*}

In total, WG11 analyzed 8175 sets of UVES spectra from 6987 individual stars. This number includes all the UVES spectra also analyzed by WG12 (pre-main sequence stars). A fraction of the stars observed in young open clusters as candidate pre-main sequence stars are indeed normal FGK-type stars, which are better suited for analysis by the WG11 methodologies. Their inclusion among the WG11 sample ensures they receive similar treatment as the remaining sample. The decision about which final result to adopt (from WG11 or WG12) is a later step performed by WG15.

The final data products resulting from this analysis, with a breakdown by contributing node, include: lithium abundances (from the Arcetri node only); chromospheric activity indicators (from the OACT node only); abundances of carbon, nitrogen, and oxygen (from the Vilnius node only); atmospheric parameters (from the nodes CAUP, EPINARBO, IAC-AIP, LUMBA, Nice, OACT, and UCM); and other chemical abundances (from the nodes CAUP, EPINARBO, LUMBA, and Vilnius).

With respect to the description presented in \citet{Smiljanic14}, the most important change in WG11 procedures concerns the homogenisation process, i.e., the process of combining the multiple measurements of a certain quantity into a final recommended value, with well characterised uncertainties. For this final release, we made use of a hierarchical Bayesian inference method. Details will be given in \cite{Worley19_wg10}, here we only summarise its most important aspects.

The concept and first implementation of the Bayesian approach to combine the multiple measurements was developed internally to Gaia-ESO by Andrew Casey (2014-2017, private communication). His implementation was used for producing the WG11 results available in the Gaia-ESO internal data release 5\footnote{\url{https://github.com/andycasey/ges-idr5}}. For the final analysis (which corresponds to the Gaia-ESO internal data release 6), we built upon his ideas and initial work to develop a slightly different implementation of the Bayesian approach.

The concept is the following. Let us assume that a certain star ``$n$" is characterised by a parameter with value {\bf true.param$_{n}$}. What a node ``$i$" returns after the analysis is a noisy measurement of that parameter, {\bf param$_{i,n}$}. We assume that the effects of the node measuring that parameter can be separated into two components: a stochastic error of Gaussian nature ({\bf random.err$_i$}, which is a property of the node) and a systematic offset ({\bf bias$_{i,n}$}, which is given by a function that is a property of the node and whose value was computed for that star). In that case, we can write something like:

\begin{equation}
    \mathrm{param}_{i,n}\,\sim\,dnorm(\mathrm{true.param}{_n},\mathrm{random.err_i}) + \mathrm{bias}_{i,n}
\end{equation}


\noindent where dnorm($\mu$,$\sigma$) stands for the normal distribution of mean = $\mu$ and standard deviation = $\sigma$. The symbol ``$\sim$'' indicates that the left part of the equation ({\bf param$_{i,n}$}) is a result of a random draw from the right part of the equation. Numerically, we actually add the bias term to {\bf true.param$_n$} inside the Normal distribution. If a node is affected by a certain bias, it is not really measuring from a distribution centred around the true value, but from a distribution centered around the biased value. We further assume that we can parametrise the bias as a quadratic function of the measured parameter, to capture its variation across the parameter space. 

Using a multi-dimensional Normal distribution, we generalise this idea for the case where several nodes are providing measurements of the same quantity. Instead of a standard deviation, the multi-dimensional Normal is characterised by a covariance matrix with terms that take into account the stochastic error of each node and also the correlations among the nodes. 

This is all written as a Bayesian hierarchical model where the parameters are inferred using MCMC simulations. The first step is to estimate the terms of the covariance matrix and the coefficients of the quadratic bias function of each node. That is done using a set of calibrators for which reference values, and associated uncertainties, of the parameter in question are known. In practice, we write the reference value, and its uncertainty, as a Gaussian prior to the true value of the parameter. The second step consists in applying the biases and the covariance matrix to the remaining sample to invert, from the measurements, the most likely true value of the parameter (and its uncertainty).

For obtaining values of $T_{\rm eff}$, we used as reference calibrators the \textit{Gaia} benchmark stars \citep{Heiter15,Jofre18}. For the case of $\log~g$, in addition to the \textit{Gaia} benchmark stars, we also included the sample of giants with asteroseismic gravities obtained from K2 \citep{Worley2020} and CoRoT data (Masseron et al., in preparation). For the metallicities, stars from the sample of open and globular clusters listed in Tables 7 and 8 of \citet{Pancino17} were used in addition to the \textit{Gaia} benchmark stars. Cluster members were either adopted from \citet{DR4GCs} or defined using radial velocities. 

By applying the same biases and the covariance matrix to the calibrators, we estimated our capacity of recovering the reference values. The comparison shows our estimate of the systematic errors in $T_{\rm eff}$, $\log~g$, and [Fe/H] to be of the order of 85 K, 0.14 dex, and 0.09 dex, respectively. However, we remark that this budget necessarily includes the errors in the reference scales themselves. Comparisons between the $T_{\rm eff}$ and $\log~g$ values obtained from the Bayesian analysis for cluster members against theoretical isochrones demonstrate that robust parameter values were obtained. We defer a more detailed description of these (and other) tests to \cite{Worley19_wg10}.

For the chemical abundances, the first step for determining biases and the covariance matrix is not possible, because of the lack of a reference sample. In this case, we implemented a simplified inference model that estimates at the same time the best abundance and the covariance matrix. In this sense, the method essentially returns a weighted mean, but where the weights (and possible correlations) are estimated simultaneously through the node to node comparisons. For the abundances, we work on a line-by-line basis, i.e., taking into account separately each spectral line measured by each node. We adopt the solar abundances of \citet{Grevesse07} as a strong solar prior (to be reproduced within 1$\sigma$ = 0.01 dex). When available, chemical abundances for the  \textit{Gaia} benchmark stars \citep{Jofre15} are used as weaker priors.

\subsection{Working Group 10: GIRAFFE spectra of ``Normal'' FGK stars}

The operation of Working Group 10 is described more fully in  \cite{Worley19_wg10},
 \cite{Recio-Blanco14} and \cite{Worley2020}. The analysis teams (nodes) that participated in the final Gaia-ESO data release are summarised in Table~\ref{tab:wg10.nodes}. 

\begin{table*}
	\caption{Overview of the WG10 Nodes that participated in the analysis of the final Gaia-ESO data release.}
	\label{tab:wg10.nodes}
	\centering
	\small
	\begin{tabular}{lp{3.9cm}p{3.9cm}p{3.9cm}l}
		\hline\hline
		Node &  Method & Analysis responsible & Data products & Setups Analysed \\
		\hline
		Arcetri  & Equivalent widths & Elena Franciosini & Lithium abundances & HR15N \\
		CAUP     & Equivalent widths & Andressa Ferreira &  Abundances & HR10, HR21, HR15N \\
        EPINARBO & Equivalent widths & Laura Magrini & Stellar parameters and abundances & HR15N, HR9B \\
        IAC  & Library of synthetic spectra & Carlos Allende-Prieto & Stellar parameters & HR10, HR21 \\
        LUMBA    & On-the-fly spectrum synthesis & Diane Feuillet and Karin Lind & Stellar parameters and abundances & HR10, HR21, HR15N \\
        MaxPlanck     & Neural networks & Maria Bergemann and Mikhail Kovalev & Stellar parameters and activity & HR10, HR21\\
        OACT     & Library of observed spectra & Antonio Frasca and Katia Biazzo & Stellar parameters and activity & HR15N, HR9B \\
        Vilnius  & Equivalent widths and spectrum synthesis &  \u{S}ar\={u}nas Mikolaitis & Stellar parameters and abundances & HR10, HR21 \\
		\hline
	\end{tabular}
\end{table*}

There is a significant overlap between the WG10 and WG11 nodes hence the description of the methodologies of Arcetri, CAUP, EPINARBO, IAC (see IAC-AIP), Lumba, OACT and Vilnius can be found in the WG11 section above. The additional WG10 node is MaxPlanck:

\paragraph{MaxPlanck:} This method determined stellar parameters and magnesium abundance through the use of neural networks described in \cite{Kovalev2019}. A training set of synthetic spectra was generated using the MARCS stellar atmosphere models and the Gaia-ESO line list \citep{Heiter2021}.

In total, WG10 analyzed 158809 GIRAFFE spectra from 92348 individual stars. The summary of the number of spectra and number of stars per GIRAFFE setup is given in Table~\ref{tab:wg10.starsspectra}. 

\begin{table}
	\caption{Summary of the number of spectra and stars per GIRAFFE setup.}
	\label{tab:wg10.starsspectra}
	\centering
	\small
	\begin{tabular}{lll}
		\hline\hline
		Setup &  No. Stars & No. Spectra \\
		\hline
		HR15N & 25785 & 26550 \\
        HR9B & 3473 & 4161 \\
        HR10 & 59722 & 60579   \\
        HR21 & 66542 & 67519 \\
        Total & 92348 & 158809  \\
		\hline
	\end{tabular}
\end{table}

For the parameter analysis, the nodes were able to analyse each setup individually but also determine parameters by analysing the HR10 and HR21 spectra together. Thus the homogenisation was carried out as four setups: HR15N, HR9B, HR10|HR21, HR21-Bulge. For the HR10|HR21 setup all results from HR10+HR21 and HR10-only were combined. For HR21-Bulge only the results for the bulge fields and the calibration stars (FGK benchmarks, CoRoT, K2, Globular Clusters, Calibrating Open Clusters) were homogenised.

The homogenisation per setup was determined using the bayesian inference method developed for the WG11 analysis (as described above) and adapted for WG10. See \cite{Worley19_wg10} for a description of this method. The reference set in each case was based on the cross-match of each WG10 setup to the WG11 sample, the FGK Benchmarks, and the other WG10 setups, primarily HR15N, in a bootstrapping approach. Thus the WG10 setups were homogenised directly onto WG11 and to the other WG10 setups. This gave the per star homogenisation per setup.

Quality diagnostics showed that the agreement between the WG10 setups and WG11 was excellent and so no further calibrations were needed for the parameters. To generate the final per star dataset for WG10, when a star was present in multiple setups, a priority procedure was implemented to take the parameters from the most appropriate setup based on the observing programme and science goals. See \cite{Worley19_wg10} for the quality diagnostics and priority procedure. The final WG10 homogenisation of the node analyses resulted in 78005 stars with stellar parameters.

The chemical abundance homogenisation was carried out combining all node results for all setups for each element for each star. In this way the separate setups were treated together as a single non-continuous spectrum per star. The reference set was the cross-match to WG11 which as a sample included a range of standards and science targets. This meant the WG10 abundances were put directly onto the WG11 abundance scale.

\subsection{Working Group 12: Cool pre-main sequence stars}

\begin{table}[ht]
\centering
\caption{Output parameters of the Gaia-ESO PMS analysis (WG12). The number of stars for which each parameter was derived is indicated for the GIRAFFE/HR15N and UVES/U580 setup separately. For the Li equivalent width, only the number of stars with $W({\rm Li})>$ 100\,\AA,  is reported. For the veiling, only stars for which $r>0$ are counted. See text for a description of the symbols used.}
\label{tab:wg12pars}
\begin{tabular}{lrr}
\hline
\hline
Parameter & GIRAFFE/HR15N & UVES/U580 \\
\hline
\multicolumn{3}{c}{raw} \\
\hline
$W({\rm H} \alpha)$            & 2826  &  476  \\
$W({\rm Li})$\, ($>$100\,\AA)   & 3570  &   46  \\
${\rm H}\alpha\,10\%$         & 2149  &   47  \\
FWZI                          & 4783  &    0  \\
\hline
\multicolumn{3}{c}{fundamental} \\
\hline
$T_{\rm eff}$                 & 14211 & merged with WG11 \\
$\log g$                      &  6499 & merged with WG11 \\
$\gamma$                      & 14166 & \dots   \\
$[{\rm Fe/H}]$                        & 11018 & merged with WG11 \\
$\xi$                         &  \dots & merged with WG11 \\
$v\sin i$                     & 12753  & merged with WG11 \\
$r>0$                         &   352  & \dots\\
\hline
\multicolumn{3}{c}{derived} \\
\hline
$\Delta W({\rm H}\alpha)_{\rm chr}$   &  2826  & 476 \\
$\Delta W({\rm H}\beta)_{\rm chr}$    &  \dots & 256 \\
$F({\rm H}\alpha)_{\rm chr}$          &  2726  & 436 \\
$F({\rm H}\beta)_{\rm chr}$           &  \dots & 224 \\
\hline
\end{tabular}
\end{table}

Working Group 12 made use of specialised methods for analyzing spectra of young, low-mass stars, particularly not-embedded pre-main sequence stars. 
The analysis took chromospheric activity and possible mass accretion into account. 
Both UVES/U580 and GIRAFFE/HR15N spectra were analyzed.

Six nodes contributed to the analysis: INAF–Osservatorio Astrofisico di Arcetri, Centro de Astrofisica de Universidade do Porto (CAUP), Università di Catania and INAF–Osservatorio Astrofisico di Catania (OACT), INAF–Osservatorio Astronomico di Palermo (OAPA), Universidad Complutense de Madrid (UCM), and Eidgenössische Technische Hochschule Zürich (ETH).

The operation of Working Group 12 is described more fully in the article "Gaia-ESO Survey: Analysis of pre-main sequence stellar spectra" \cite{Lanzafame15}.
Here, a brief summary and some updates are reported.

A summary of the parameters obtained by WG12 is reported in Table\,\ref{tab:wg12pars}.
A distinction is made amongst  {\it raw}, {\it fundamental}, and {\it derived} parameters.
The H$\alpha$ emission equivalent width ($W({\rm H} \alpha)$), the Li absorption equivalent width ($W({\rm Li})$), the H$\alpha$ width at 10\% of the line peak \citep{2004A&A...424..603N}, and the H$\alpha$ full width at zero intensity (FWZI)  are considered {\it raw} parameters as these are directly measured on the input spectra and do not require any prior information.
They are used to identify pre-main-sequence (PMS) stars and optimise the evaluation of the fundamental parameters in \texttt{ROTFIT} \citep{2006A&A...454..301F}, one of the two methods used.
In addition to $T_{\rm eff},  \log g$, and [Fe/H], the fundamental parameters derived include also micro-turbulence velocity ($\xi$), projected rotational velocity ($v\sin i$), veiling \citep[$r$, see, e.g., ][]{1988BAAS...20R1092H}, and a gravity-sensitive spectral index \citep[$\gamma$, see][]{2014A&A...566A..50D}.
Finally, the derived parameters are those whose derivation requires prior knowledge of the fundamental parameters: chromospheric activity indices ($\Delta W({\rm H}\alpha)_{\rm chr}$ and $\Delta W({\rm H}\beta)_{\rm chr}$) taken as equivalent width difference with respect to an inactive template with the same fundamental parameters, and chromospheric line fluxes ($F({\rm H}\alpha)_{\rm chr}$ and $F({\rm H}\beta)_{\rm chr}$) evaluated from the chromospheric activity index and the expected flux at the continuum.

Fundamental parameters were inferred by the OACT node using \texttt{ROTFIT} \citep{2006A&A...454..301F} and by the OAPA node using a spectral indices method \citep{2014A&A...566A..50D}.
The spectral indices results are considered valid below a given threshold of blending due to rotational broadening.
In the case at hand, the limits are $v \sin i < 90$\,km/s for $T_{\rm eff}$, $v \sin i < 30$\,km/s for $\log g$, and $v \sin i < 70$\,km/s for [Fe/H]. 
Comparison with benchmark stars have shown the \texttt{ROTFIT} application to HR15N spectra produces strong biases for $T_{\rm eff} > 6200$\,K, and therefore all \texttt{ROTFIT} results are disregarded in this range.
When both methods produce valid results, it was imposed that  substantial agreement exists before  averaging them. 
Comparison with benchmarks and results from photometry have led to the following homogenisation criteria:
\begin{itemize}
\item The spectral index method was taken as reference $T_{\rm eff}$. The two methods were averaged when $\Delta T_{\rm eff} < \pm 220$\,K for $T_{\rm eff}<4000$\,K, $\Delta T_{\rm eff} < \pm 360$\,K for $4000<T_{\rm eff}<5000$\,K, and $\Delta T_{\rm eff} < \pm 300$\,K for $T_{\rm eff}>5000$\,K. Otherwise only the spectral index value is recommended.
\item $\log g$ are averaged only if the two results differ by $<0.3$\,dex. Otherwise, only the \texttt{ROTFIT} results is given if $\log g_{\rm OACT} >4.2$ and $\log g_{\rm OAPA} >5.0$. In all other cases, no $\log g$ is given. 
\item When both methods produce [Fe/H] the values are averaged. 
\end{itemize}

Besides the ($T_{\rm eff}$, $\log g$, [Fe/H]) triad, WG12(OACT) also recalculated radial velocities for clusters badly affected by nebular lines (NGC2264, NGC6530, Trumpler14, NGC2451) after applying an alternative different sky background subtraction.

The projected rotational velocity $v \sin i$ from HR15N was also recalculated by OACT using a cross-correlation technique and masking emission features in young stars.
Changes in the spectrograph resolution were also taken into account. 
Upper limits are given when the error is larger than the $v \sin i$ value itself or when $v \sin i \leqslant 7$ km s$^{-1}$, which is the limit imposed by the GIRAFFE resolution.

\subsection{Working Group 13: OBA stars} 

The operation of Working Group 13 is described more fully in the article
"The Gaia-ESO Survey: The analysis of hot-star spectra"  \cite{Blomme21}.

In Working Group 13 (WG13), eight Nodes worked on the analysis of the O-, B- and A-type spectra. These were selected from observations made with the HR3/4/5A/6/14A GIRAFFE setups and their corresponding UVES data. Some archive data were also processed.
The temperature range covered by WG13 is large, and most of the Nodes cannot fully cover that range.
Two types of analysis were used by the Nodes. The first one was based on a carefully selected set of diagnostic photospheric spectral lines, and radiative transfer calculations were used to fit the profiles in detail. The second one used the full observed spectrum, comparing it to theoretically generated ones. In the comparison, weights can be used to stress the importance of certain spectral regions. After the Nodes have derived the stellar parameters, their results are homogenised and the abundances are determined. More details of the WG13 work are given in \citet{Blomme21}.

\begin{table*}
\caption{Overview of the WG13 Nodes. 
Listed are the effective temperature range covered by the Node, the 
spectral analysis
technique used, the stellar parameters that are determined, the elements for which
abundances are determined, and the number of spectra processed.}
\centering
\begin{tabular}{lllllr}
\hline\hline
Node & \multicolumn{1}{c}{$T_{\rm eff}$ range} & Technique & Parameters determined & Abundances & \# spectra \\
\hline
ROBGrid    & $\hphantom{1}3000\, - 50\,000$\,K    & $\chi^2$ minimisation with grid  & $T_{\rm eff}$, $\log g$, [M/H], $v_{\rm rad}$, $v \sin i$\tablefootmark{a} & -- & 8667  \\
           &                     &  of theoretical spectra \\
ROB        &  $\hphantom{1}6000\, - 12\,000$\,K  & Fe - Fe$^{+}$ ionisation balance  & $T_{\rm eff}$, $\log g$, [Fe/H], $\xi$,  $v \sin i$\tablefootmark{a}   & C, O, Mg, Al, Sc, Fe & 517 \\
           &                     &  of diagnostic photospheric lines \\
MGNDU      & $\hphantom{1}5000\, - 15\,000$\,K    & PCA and SIR                      & $T_{\rm eff}$, $\log g$, [M/H] , $v_{\rm rad}$, $v \sin i$                           & -- & 186 \\
Li{\`e}ge  & $10\,000 - 32\,000$\,K & $\chi^2$ minimisation with grid  & $T_{\rm eff}$, $\log g$, $v_{\rm rad}$, $v \sin i$\tablefootmark{a}           & He, C, N, Ne, Mg, Si & 696 \\
ON         & $14\,000 - 33\,000$\,K & Non-LTE synthesis and            & $T_{\rm eff}$, $\log g$, $v \sin i$\tablefootmark{a}                            & C, O, Si & 184 \\
           &                     & Si ionisation balance\\
IAC        & $22\,000 - 55\,000$\,K & $\chi^2$ minimisation with grid  & $T_{\rm eff}$, $\log g$, $v \sin i$, $v_{\rm macro}$  & He & 268 \\
           &                     &  of FASTWIND models  \\
Mntp       & $30\,000 - 45\,000$\,K & $\chi^2$ minimisation with grid & $T_{\rm eff}$, $\log g$, $v \sin i$, $v_{\rm macro}$                  & -- & 55 \\
           &                     &  of CMFGEN models  \\
Li{\`e}geO & $20\,000 - 45\,000$\,K  & CMFGEN & $T_{\rm eff}$, $\log g$, $v \sin i$, $v_{\rm macro}$                  & He, C, N & 293 \\
\hline
\end{tabular}
\tablefoot{
\tablefoottext{a}{What is listed here as $v \sin i$ is actually the total line-broadening
parameter, which can include other effects, such as macroturbulence.}
}
\end{table*}

\paragraph{ROBGrid Node.}
The ROBGrid Node used theoretical spectra from the literature and compared them to the observations, to derive the stellar parameters ($T_{\rm eff}$, $\log g$, and metallicity -- if not too different from solar metallicity). Each observed spectrum was compared to all theoretical spectra to find the best fit. For each comparison, the radial and projected rotational velocities are determined using a cross-correlation technique \citep[][their Eq. 7]{wg13:David+14} with a rotationally broadened theoretical spectrum. To judge the goodness-of-fit the $\chi^2$ for that comparison is calculated. The final stellar parameters (as well as the radial and projected rotational velocity) are determined by the theoretical spectrum having the minimum $\chi^2$.
In a larger loop around this fitting procedure, we also refined the normalisation of the observed spectrum. ROBGrid does not determine errors of the stellar parameters, but assigned errors that were derived in the homogenisation phase. The Node also does not provide abundances of individual elements, because the abundances that went into the literature models cannot be changed.

\paragraph{ROB Node.}
The ROB Node used a three-step approach to determine the stellar parameters and abundances. First the stellar parameters were estimated from a limited number of diagnostic H Balmer, Fe, and Mg absorption lines. Next, the detailed profiles of a more extensive set of diagnostic lines were fit by an iteration over $T_{\rm eff}$, surface gravity ($\log g$), line-of-sight microturbulence velocity ($\xi$), and metallicity ($[$M/H$]$). Finally, abundances were  measured from selected sets of sufficiently (medium-strong to) strong lines. To explore the parameter space, we calculated a large homogeneous grid of synthetic spectra with the LTE radiative transfer code {\sc Scanspec}\footnote{\tt http://alobel.freeshell.org/scan.html} \citep{wg13:Lobel11a}. The error bars range from $\sim\pm$150 K for $T_{\rm eff} < 8500$~K to $\sim\pm$250 K for $T_{\rm eff} > 11\,000$ K. The metallicities and element abundances have error bars ranging from $\pm$0.05 dex to $\pm$0.1 dex,

\paragraph{MGNDU Node.}
The MGNDU node analysed data from the UVES 520 setup. The spectra were renormalised as in \cite{wg13:PCA-1}, and the radial velocities redetermined because a high radial velocity accuracy is important for the MGNDU procedure \citep{wg13:PCA-2,wg13:PCA-1}. The procedure used is a combination of Principal Component Analysis (PCA) and a Sliced Inverse Regression (SIR). It starts by compiling a learning database using synthetic spectra. The model atmospheres for this database were calculated using the latest version of the ATLAS9 code \citep{wg13:ATLAS9-1,wg13:CastelliKurucz03,wg13:ATLAS9-3}. Synthetic spectra were calculated from these models, using the SYNSPEC48 LTE code \cite{wg13:SYNSPEC} and the line list of \cite{wg13:PCA-1}. Once the learning database has been constructed, it can be applied to any observed spectrum to derive the stellar parameters \citep{wg13:SIR}. The average uncertainties for the inverted parameters are around 150 K for $T_{\rm eff}$, 0.35 for $\log g$, 0.15 for [M/H], and 2 km\,s$^{-1}$ for $v \sin i$. No elemental abundances, other than Fe, are determined by MGNDU. 

\paragraph{Li{\`e}ge Node}
The Li{\`e}ge Node covers the temperature range 10\,000 to 32\,000 K, corresponding to the range of B-type stars. A least-square minimisation was used to fit the observed normalised spectra with a grid of solar-metallicity, synthetic spectra computed with the SYNSPEC program on the basis of non-LTE TLUSTY \citep{wg13:Lanz+Hubeny07} and LTE ATLAS \citep{wg13:kurucz93} model atmospheres. We first determined the radial velocity and projected rotational velocity (macroturbulence is not included). The full wavelength domain was used to determine the effective temperature and surface gravity. The typical 1-$\sigma$ uncertainties are $\sim$750 K for $T_{\rm eff}$, $\sim$0.15 for  $\log g$, $\sim$15 km\,s$^{-1}$ for $v \sin i$ and $\sim$2 km\,s$^{-1}$ for the radial velocity. For the abundance determination, we considered He, C, N, Ne, Mg, and Si. We found the best $\chi^2$ fit of a grid of rotationally-broadened synthetic spectra to the observed line profiles of \ion{He}{i} $\lambda$4471, \ion{C}{ii} $\lambda$4267, \ion{N}{ii} $\lambda$4630, \ion{Ne}{i} $\lambda$6402, \ion{Mg}{ii} $\lambda$4481, \ion{Si}{ii} $\lambda$6371, and \ion{Si}{iii} $\lambda$4568-4575, from which we derived the abundances. The line profile calculations used the non-LTE code DETAIL/SURFACE originally developed by \citet{wg13:butler84}. More details about the version of the code currently used and the model atoms implemented are given in \citet{wg13:morel06} and \citet{wg13:morel08}. 

\paragraph{ON Node}
The ON Node started by determining $v \sin i$ from the \ion{He}{i} lines.
Stellar parameters and abundances were then determined using theoretical spectra calculated with the fully non-LTE spectral synthesis code \texttt{SYNSPEC} and a new grid of line-blanketed non-LTE model atmospheres calculated with \texttt{TLUSTY} \citep{wg13:hubeny1995,wg13:hubenylanz2017}.  The grid covers $T_{\rm eff}$ between $14\,000$ and  $33\,000$\,K, and surface gravity between 3.0 and 4.5. The Hydrogen lines of the GIRAFFE spectra are compared to these theoretical spectra to find the combinations of $T_{\rm eff}$ and $\log g$ that can reproduce the observed H profiles. The effective temperature was then determined from the ionization balance of \ion{Si}{ii}-\ion{Si}{iii}-\ion{Si}{iv}. Next, a range of microturbulence velocity $\xi$ is explored and the corresponding abundances of C, O, and Si are determined. By requiring the abundances to be independent of line strength, their final values are derived. The uncertainties on the individual parameters are $1000$\,K for $T_{\rm eff}$, $0.15$\ for $\log g$, $15$\,\% for $v \sin i$, and $2$\, km\,s$^{-1}$ for $\xi$.

\paragraph{IAC Node}
The IAC node analyses the O and early B-type stars using large grids of synthetic spectra computed with the non-LTE FASTWIND stellar atmosphere code \citep{wg13:santolaya97, wg13:puls05}.
First, a Fourier transform plus goodness-of-fit method determines the projected rotational velocity and the macroturbulent broadening. A $\chi^2$ approach is then used to compare the observed diagnostic lines (H$\alpha$, H$\gamma$, H$\delta$, \ion{He}{i}~$\lambda$4387, \ion{He}{i}~$\lambda$4471, \ion{He}{i}~$\lambda$4713, \ion{He}{i+ii}~$\lambda$6678, \ion{He}{ii}~$\lambda$4541 and \ion{He}{ii}~$\lambda$4686) to the theoretical spectra. From this we determine effective temperature ($T_{\rm eff}$), surface gravity (log $g$), Helium abundance ($N_{\rm He}/N_{\rm H}$), microturbulence ($\xi$), wind-strength
parameter\footnote{$Q$ is defined as $Q = \dot{M}/(v_{\infty}R)^{1.5}$ \citep{puls96}, where $\dot{M}$ is the mass-loss rate, $v_{\infty}$ the terminal
	velocity of the wind, and $R$ the stellar radius.}
($Q$) and the exponent of the wind 
velocity-law\footnote{The velocity in the stellar wind is given by $v(r) = v_{\infty}(1 - R_{\ast} /r)^{\beta}$, where $R_{\ast}$ is the stellar radius of the star.} ($\beta$). Typical uncertainties are of the order of $\pm$1000 K in $T_{\rm eff}$,  $\pm$0.10 in log $g$, $\pm$0.15 in log $Q$ and $\pm$0.03 dex in Helium abundance.

\paragraph{Mntp Node}
The Mntp node relied on a pre-computed grid of synthetic spectra, calculated with the non-LTE code CMFGEN \citep{wg13:hm98}. The projected rotational velocity of each star was derived from the Fourier transform of \ion{Si}{iii}~$\lambda$4552 and/or \ion{He}{i}~$\lambda$4713. By convolving the theoretical spectra with the instrumental profile, the rotational profile and a radial-tangential macroturbulence profile, the value for the macroturbulence was determined. A $\chi^2$ approach was used to find the best-fit spectrum, concentrating on those spectral lines that are sensitive to effective temperature and surface gravity. Typical uncertainties are 2500 K on $T_{\rm eff}$ and 0.15 on $\log g$.

\paragraph{Li{\`e}geO Node}
The Li{\`e}geO Node used the CMFGEN non-LTE atmosphere code \citep{wg13:hm98} to analyze the O- and early B-type stars. A grid of CMFGEN models was constructed covering the range $T_{\rm eff}=27,000$ to $36,000$ K, and surface gravity 3.0 to 4.3. These models include the effect of the stellar wind. The spectra are convolved with a rotation profile and a radial-tangential macroturbulence profile. For the O-type stars, the ionization balance between He {\sc i} and He {\sc ii} is used to determine the effective temperature. For B-type stars, we used Si {\sc iii} and Si {\sc iv}, with He {\sc i} $\lambda$ 4471 and Mg {\sc ii} $\lambda$ 4481 as a secondary diagnostic. Carefully selected lines of Carbon and Nitrogen were used to determine the abundances.

\paragraph{Homogenisation}
Once the different nodes have delivered the stellar parameters, the results were homogenised. Node results for the benchmark stars are used to introduce corrections, but this could only be applied for the ROBGrid results as the other nodes did not analyse a sufficient number of benchmark stars. The recommended values for the stellar parameters are then determined as a weighted sum of the different node results (for details, see \citep{Blomme21}). Contrary to the practice in other Working Groups, these recommended stellar parameters values are not used in determining the abundances; instead the Node stellar parameters are used. Abundances are determined for only a small number of stars, with little overlap between the nodes. Where there was overlap, the recommended abundances are determined by a straight average of the node abundances.

\subsection{Working Group 14: Non standard objects, quality flags}

The aim of Working Group 14 (WG14) was to identify and characterise outlier objects. The majority of outlier objects were initially steps up the data processing and verification learning curve, and after improvement and iteration became obsolete. This was an invaluable contribution to survey quality control. For further information see \cite{SVE19}.
 WG14 scanned all Gaia-ESO Survey spectra, using a mix of goodness-of-fit spectrum-matching outputs, and diagnostic analyses such as the shape of cross-correlation functions, looking for outlying data. the most significant are outlined below. In order to detect and characterise an anomaly most efficiently, a dedicated dictionary has been defined. This dictionary is briefly described in Sect.~\ref{Sect:dico}. WG14 could then systematically investigate specific types of outlying features potentially affecting all Gaia-ESO data (i.e. whatever the object spectral type is), namely emission line objects and binary/multiple objects, as described below 
(Sect.~\ref{Sect:emission} and Sect.~\ref{Sect:binarity}, respectively). 
Science verification illustrations of the value of the work of WG14 are available in:
"The Gaia-ESO Survey: Catalogue of H$\alpha$ emission stars" \citep{Traven-2015}, 
"The Gaia-ESO Survey: double-, triple-, and quadruple-line spectroscopic binary candidates" \citep{Merle17}, and "The Gaia-ESO Survey: detection and characterisation of single-line spectroscopic binaries" \citep{MerleSB1}.

\subsubsection{Gaia-ESO Survey dictionary}
\label{Sect:dico}
\paragraph{\bf The complete dictionary}
During analyses many nodes and Working Groups were flagging various types of outliers but with idiosyncratic naming conventions. It seemed desirable to use a common dictionary. This dictionary should convey information that is at the same time useful, exact and exhaustive for Gaia-ESO Survey-consortium data-processing
users, but also for later use external to the survey consortium, without demanding too-much specialist project-specific knowledge.
It was primarily the task of WG14 to flag outliers in such a way. 

This dictionary has been defined with three main sections.
First, a technical section (TECH) listing mainly data and data-product issues: data reduction, S/N, analysis issues and result quality issues, separately for parameter and abundance determination issues.
These flags have been very useful to trace back data reduction problems or main analysis problems encountered by the nodes, and so in many cases the problematic flagged data has been able to be replaced by improved-quality data.
Second, a peculiarity (PECULI) section, listing peculiarities possibly affecting the spectra, was introduced, listing for example binarity or emission line characterisation, as well as anomalous line and molecular absorption.
Third, a comment (Remark) section was included, listing specific classes of stars, so that if any WG and nodes, in the course of their analysis, recognised a peculiar type of object, they could flag it as such. This section was however not used as much as the two other ones.
A 3-level confidence flag (A,B,C), complementing each flag, has been introduced. For each data release, a first-pass analysis was performed beforehand, flagging most prominent issues, so that this information could be conveyed to the various nodes during their analysis phase.

\paragraph{\bf The simplified dictionary}
In the final data release of recommended results, a significant number of flags are potentially available. A compromise is needed between providing too-much information to be helpful for an end-user, and failing to flag important information. In the event WG15 compromised with a simplified set of flags. These focus on the processing and quality-control issues noted above, and some astrophysically-interesting products, discussed below. The larger set of information is available in the full node-level information also made available.

\subsubsection{Emission line detection and classification}
\label{Sect:emission}
Automated fits to the profile of detected emission lines are feasible. The most relevant for Gaia-ESO is H$\alpha$. The fitting used two independent Gaussian profiles and a third component accounting for nebular emission allowed distinction of several morphological types of H$\alpha$ line profiles with the introduction of a simplified classification scheme. The spectra were sorted into eight distinct morphological categories: single component emission, emission blend, sharp emission peaks, double emission, P-Cygni, inverted P-Cygni, self-absorption, and emission in absorption. A quantitative discussion of the degree of variability of H$\alpha$ emission profiles, which is expected for young, active objects, was also possible thanks to multiple observations. 
As proof of principle of the utlity of this approach,
\cite{Traven-2015} discussed the properties of H$\alpha$ emission stars across the sample of 22035 spectra from a Gaia-ESO Survey internal data release after 22 months of observation. These are observed with the GIRAFFE instrument and largely to stars in young open clusters. A catalogue of stars with properties of their H$\alpha$ emission line profiles, morphological classification, analysis of variability with time and the supplementary information from the SIMBAD, VizieR, and ADS databases was published. The relevant flags and parameters are published.

\subsubsection{Binarity and radial-velocity variability detection}
\label{Sect:binarity}
\paragraph{SB1}
Although Gaia-ESO was not designed as a radial-velocity survey, most stars are observed more than once, with the subsequent radial velocity determinations being checked, and as appropriate flagged, for variability at the pipeline processing stage (Sect~\ref{wg8}. In WG14 a more careful analysis was carried out. The aim, apart from intrinsic interest,  is to identify multi-component spectra and to alert other WG/nodes analyses that atmospheric parameters should not be used for these system components; SB-specific pipelines should be used instead.
This proceeded in two stages. The first involved analysis of the shape of the pipeline-generated cross-correlation functions (CCF). The CCF will be distorted from symmetry if stacking spectra uncorrected for unrecognised but true radial-velocity variations. This prompted an investigation of the individual spectra radial velocity determinations, looking for evidence for spectroscopic binaries with one visible component (SB1).
The Gaia-ESO Survey internal data releases were investigated to identify and characterise SB1. A statistical $\chi^2$-test was performed on the CCF's for stars characterised by at least two observations and a signal-to-noise ratio larger than three. The resulting sample of candidate RV variables was cleaned from contamination by pulsation/convection-induced variables using Gaia DR2 parallaxes and photometry. Monte-Carlo simulations using the NASA/HEASARC SB9 catalogue of spectroscopic orbits allowed estimation of the detection efficiency. Thus we could correct the SB1 rate to evaluate the Gaia-ESO SB1 binary fraction and its dependence on effective temperature and metallicity. 
This analysis remains to be completed for the full Gaia-ESO dataset, but for a proof of principle analysis using part of the data  \citep{MerleSB1} found 641 (resp., 803) FGK SB1 candidates at the 5$\sigma$ (resp., 3$\sigma$) level.  The orbital-period distribution was estimated from the RV standard-deviation distribution. After correcting for the detection efficiency and selection biases, the SB1 frequency could be estimated, and its dependence on metallicity and spectral type could be constrained .

\paragraph{SB2}
For double-lined and higher-multiplicity spectroscopic binaries one similarly
can detect and flag binarity/multiplicity (SBX, X$\ge 2$) for stars from the cross-correlation function (CCFs) of the Gaia-ESO spectra with spectral templates. Due to the large number of spectra, WG14 automated the task analysing the successive derivatives of the CCF. The use of successive derivatives allowed the de-blending of multicomponent CCFs. The code written to perform this task provides: i) the number of peaks detected in the CCF, ii) radial velocities corresponding to the peak maxima.

As a proof of principle this method was applied \citep{Merle17} on the fourth Gaia-ESO internal data release and allowed the detection of 354 SBn candidates (342 SB2, 11 SB3, and even one SB4), of which only nine candidates were known in the literature. This implies that about 98\% of these SBn candidates are new, illustrating the known incompleteness of such studies at faint magnitudes.   Among the SB2 candidates, detailed analysis (including follow-up observation on the SALT telescope) of the unique SB4 (four peaks in the CCF) reveals that CNAME 08414659-5303449 (HD 74438) in the open cluster IC 2391 is a physically bound stellar quadruple system. This very rare discovery indicates the success of teh approach.

Building on this success,
    \citet{Vanderswaelmen-prep}  (see also 
\citealt{Vanderswaelmen-2018})
 improved the detection by designing specific cross-correlation masks (called NACRE masks) which produce more narrow CCFs, and therefore allow detection of SB2 with a smaller radial velocity difference. Our investigation showed that the HR21 wavelength range, around the near-infrared Ca II triplet, tends to host numerous strong and saturated lines (compared to the HR10 wavelength range) that broaden the CCFs. The new NACRE masks exclude these strong features to keep only weak, mildly-blended atomic lines and produce more narrow CCFs.  We therefore re-computed the CCFs for the approximately 150000 individual HR10 and HR21 spectra and analysed them with DOE. The resulting flags are in the released dataset.

\subsubsection{t-SNE flagging}
The t-SNE node employed a semi-automatic approach to classification of spectra. This means that we manually assigned labels to groups of spectra that share a similar morphology, where the groups of spectra are revealed by the dimensionality reduction (clustering) technique t-SNE (for more details see \citealt{Traven-2017}). Our intention was not to provide highly reliable classification labels for every spectrum, rather we identified most of the common outstanding peculiarities in the whole dataset, with spectra labelled up to the limits of our semi-automatic procedure.
The benefit of this method is the ability to discover different kinds of  unexpected peculiarities/features/issues as well as provide a clear overview of the structure of the whole spectral dataset (distinction between dwarfs/giants, hot/cool stars, etc.).

We produced classification results for the four Gaia-ESO setups (HR14A, HR15N, HR21, UVES-U580) that displayed at least a few well defined and interesting groups of spectra after analysis. We emphasise  that this is not an exhaustive classification, since a person is needed to identify and name the different peculiar groups, however apart from that the t-SNE framework developed allows for a very efficient inspection and also updates to the classification as the dataset grows with time.

One of the strongest and most abundant features that we recognised in the t-SNE spectral classification are incorrectly subtracted nebular emission
lines (HR14A, HR15N), along with other reduction issues. Some interesting objects with intrinsic emission in the H$\alpha$ and CaII lines are likewise revealed by t-SNE. We also provided an interactive interface to the t-SNE maps dubbed the ``GES Explorer'', which is available internally to the Gaia-ESO collaboration, and can serve as a powerful exploratory tool for the Gaia-ESO dataset. In that way this t-SNE effort provides flags of value, particularly for analyses of young stars.

\section{Survey parameter homogenisation WG15} \label{WG15}

Fuller information on the work of WG15 can be found in \citet{Hourihane21}. The core team members were Patrick Francois, Anais Gonneau, Anna Hourihane, Laura Magrini and Clare Worley.

The aim of this WG was to ensure that the data products generated by the spectrum analyses are
coherent, the resulting stellar atmospheric parameters and abundances
homogenous, the parameters are calibrated onto an identified (set
of) external calibrator objects, and the process is fully documented,
supporting the data releases. This homogenisation process is a key
aspect of analysis quality control, and proceeds through a double
iteration. The first part of this homogenisation process is done inside each Working Group, where in
particular the outputs of the different node-based analysis tools are compared
and combined. Bringing together a final ``best'' parameter set
requires, in addition to the internal analyses, careful analysis of
calibration targets, and targets observed in more than one setting and
instrument.  
 In general systematic calibration scale and zero offsets between analysis groups, and between target classes and fundamental benchmark stars were found to be small by the end of the survey, for the final processing iteration. during the survey teams identified many causes of systematics and mitigated them. Some remain, as for example in radial velocity zero point offset between setups as discussed in Sect~\ref{wg8} and illustrated in Fig~\ref{RV-1} and Fig~\ref{RV-2}. All systematic remaining offsets are corrected. The sanity check on the outcome of this on the HRD scatter pre- and post-processing is shown in Fig~\ref{hrd-homog} above.

The homogenisation process is based on the homogenised results issued by the working groups WG10, WG11, WG12 and WG13 and the flags from WG14. 
During the first internal release, the results released by the WGs were obtained per spectrograph setup and not per individual object (cname). 
We verified that the benchmark star results (stellar parameters)  coming from the different setups gave  consistent results. 

For later data releases the homogenisation process was performed in two steps. 
First we homogenised the stellar parameters,  namely  $T_{ \rm eff}$, $\log g$, ${\rm [Fe/H]}$  and $\xi$, and delivered to the WGs a file 
containing all the targets with an homogenised set of stellar parameters. From that parameter set, updated abundances were computed by the nodes
 of the different WGs and the resulting abundances were set on the same scale by the WGs. 
 The resulting updated abundance files then came back  to WG15 to perform the  element by element abundance homogenisation.  
 Two more steps are needed to produce the final WG15 file,  radial velocity homogenisation and the homogenisation of the flags. 
 
\subsection{ Stellar parameter homogenisation}

The aim of the homogenisation  process is to put the astrophysical parameter results of the analyses of the different Working Groups WG10, WG11, WG12 and WG13  on a common scale. 
To allow a calibration  of the work of the different WGs, Gaia-ESO has been designed to have several sets of stars in common amongst the analysis WGs.  
A set of FGK benchmark stars  is analysed by all WGs except WG13 (hot stars) for which a special set of benchmark stars has been used.
 The benchmark stars have their atmosphere fundamental parameters  determined 
by independent means.  They are used as reference for the parameter scale.  
In addition to using the stars in common between WGs to perform the homogenisation process,
we also used the objects (open and globular clusters) that have been observed and processed  by several WGs. 
In particular, a set of clusters referenced as calibration clusters has been observed in a range of instrumental setups to aid in inter-setup and inter-WG calibration. 
For example,  the  stars in NGC~6705 are observed in all setups, and are in common to all WGS, thus making this 
cluster  a fundamental inter-calibrator for the whole survey.

\subsubsection{Absolute parameter scale}

The first step in the homogenisation process is to map the results of each WG on to
 the absolute external scale represented by the benchmark stars and to ensure internal consistency of the WG results with this scale. 
 For the WG10 results, checks are performed to ensure that good agreement is found between the joint analysis  of the HR10-HR21 setups  and the literature
 results for the benchmark stars.  Care has also been taken to verify that the other WG10 setups (HR21 alone, HR15N and HR09B) are mapped onto the HR10-HR21 scale. 
For WG11, we checked the agreement of the results for both  setups U520 and U580. We also searched for potential corrections among the three setups U520 U580 and HR15N
 that have been used by WG12 for the analysis of the benchmark stars.  Several warm benchmark stars added during the project have been used by WG13  to set their scale on the 
 literature data. Stars in common amongst the WG13 setups were also considered to check the internal consistency  of the WG13 results.  
 A priority order has been recommended by WG13, based on their evaluation of the reliability of the outcomes, and adopted by WG15 for their results. 
 The adopted sequence is" HR3|HR5A|HR6|HR9B|HR14", "HR5A|HR6|HR9B|HR14A","HR3|HR15A|HR6","HR14","HR9B" and "U520".

\subsubsection{Relative parameter scale}

Once the WGs results have been mapped onto the benchmark scale, 
the results of each WG for the whole sample of stars need be checked with respect to each other. 
As the full set of stars observed by Gaia-ESO  covers a larger range of stellar parameters than the benchmark stars, it is 
mandatory to perform a series of checks. To evaluate any possible offsets between WGs we  use the stars in common between different WGs
that give us a direct estimate of any differences between WGs' results. 

A second test is to plot the Hertzprung Russell diagram of  Milky Way stars from the different WGs  in the same metallicity range and compare 
these distribution with the theoretical : see Fig~\ref{hrd-homog} above for an example.  

A third test uses the member stars in open and globular clusters, which are both considered 
to be composed of chemically homogenous populations. 
Clusters are particularly important in the process of homogenisation as they allow us to put stars that are not common between 
WGs on a common scale as hot, massive cluster stars and pre-main sequence stars. 

Globular clusters are important as they cover a wide range in metallicity for which both GIRAFFE and UVES observations were completed. 
They were investigated for $T_{ \rm eff}$, $\log g$, ${\rm [Fe/H]}$  offsets between U580 and HR10+HR21  samples. 

Open clusters analysed in Gaia-ESO can be divided into two categories, intermediate-age/old clusters (with ages $\le $ 100 Myr up to several Gyr) and young clusters 
with younger ages that may have massive stars (mass $\le$ 8 $M_{\odot}$). To allow a comparison of the results of the different WGs and a final homogenisation of the whole Gaia-ESO Survey results, 
 several  open clusters are observed in more than one setup and  are analysed by several WGs.  These so-called intercalibration clusters give a solid basis to perform the comparison 
 of the results between different WGs and different setups. The best example is the open cluster NGC~6705 analysed by the four WGs as noted above. 
 For both open and globular cluster, an important check is to estimate qualitatively the agreement with a theoretical isochrone (PARSEC).

 The homogenisation and the offsets applied to the different WGs  results  follow a sequence described in detail in \citet{Hourihane21}. 
 If a star has results in more than one WG, then the selection of the recommended set of parameters is based on the setup used and/or the 
 competence of each on the analyses on that type of star. An example is that  the results  for stars with $T_{eff} \ge 7000 K$ all come from WG13. 
Another example  is to give  priority to the results coming from UVES over GIRAFFE spectra.  For multiple exposures of the
same benchmark stars, we have selected the one with the highest signal-to-noise ratio (S/N).

\subsection{ Abundance homogenisation}

The task of the abundance homogenisation is based on the
determination of the offsets in the abundances of the individual elements
between the different WGs. Stars in common between WGs or cluster member
stars observed by several WGs are used to evaluate these offsets.

The abundances were determined by the analysis nodes and WGs based on the recommended parameters that were the product of the WG15 homogenisation. 
 As the large wavelength range and the high resolution of the UVES
spectrograph permit a more precise determination of the stellar parameters
and abundances than GIRAFFE spectra, the homogenisation process for the WG10 abundances uses the WG11 results as the baseline. WG13 results are treated separately as the hot stars are suspected to have abundance anomalies and do not follow the general 
trends found in FGK stars.  The lithium abundance  is also treated separately. 
For the other elements, the first step is to check the solar and the benchmark abundances and look for possible offsets with respect to the literature values. 
The abundances measured in the stars in common between WGs are then compared to check for possible offsets. 
The abundance ratios vs $[Fe/H]$ determined in the open clusters (both calibration and science open clusters) are then computed  for the different WGs to 
check for anomalies.  The same work is also performed on globular clusters. Median elemental differences between WG10 and WG11 are determined 
and used to find offsets and/or trends as a function of metallicity.
The  Milky Way stars abundances are also compared to literature data \citep{Bensby14,Battistini15,Battistini16, Reddy06, Pereira17,Takeda16}
 and checked for offsets and trends as a function of metallicity.

\subsection{ Radial velocity homogenisation}

The Gaia Radial Velocity standard stars have been investigated as part of the homogenisation of the Gaia-ESO radial velocities and the Gaia-ESO HR10 RVs have been found to have zero offset from the reference values for these stars \citep{SoubiranRV}. Therefore, HR10 RVs were taken as the zero point for the Gaia-ESO RVs. The HR10 RVs come from the pipeline of Sect~\ref{wg8}. Only RVs from the stacked, singlespec spectra from which the stellar parameters and abundances were determined are considered here. For investigation of the RV variations between individual or nightly stacked spectra, see \citet{Jackson15} and Sect~\ref{Sect:binarity}.

Offsets were calculated between each of the setups and the zeropoint of the Gaia-ESO RV scale, HR10. The offsets were then applied to put the other setups onto the HR10 scale. For WG13, RVs have been   based on a combination of WG13 setups.

Where the HR10 setup was not available, the radial velocity was sourced from the same setup as that from which the parameters were selected. Potential offsets were applied when the RVs were taken from a different setup to HR10. These offsets are calculated on the basis of stars in common between the setups. The order of priority in which RVs from the different setups were selected was:  HR10, HR15N, U580, HR21, HR9B, HR3, HR5A, HR6, HR14A, HR14B, U520, HR4, HR5B.

\subsection{Flag homogenisation}

The homogenisation is based on the dictionary of flags produced by  {\sc WG14}.
We compared the flags produced by the different {\sc WGs} and searched for possible conflicts.
In general,  all the flags from the WG Recommended files are included with any duplicates removed.
Details can be found in \citet{Hourihane21}.

\section{Survey progress monitoring and data publication (WG16, WG17, WG18) \label{monitor}}

Survey progress monitoring is a major task, sufficiently critical to
efficient survey progress that the relevant WG (WG16) was led by the
Co-PIs directly, ably supported by the dedicated Project Office team. Survey progress is a complex mix of management,
communications, and book-keeping. Management involves monitoring the
progress of all WGs involved in data preparation, processing, and
analysis. The key aspect for this WG was the book-keeping. Reliable and
quantitative information must track, for every target star, the number
of observations attempted, achieved, and still awaited. The processing status
of each observation must be tracked, and updated as each WG deposits
data in the operational database. SNR data for each spectrum, SNR data
for each object, including repeats and different settings, and
necessary additional information prior to object readiness for science
analysis, must all be maintained for all $10^5$ targets. For the clusters
the same information must be maintained both for the individual targets
and for the clusters as a whole. In particular, the fractional completion
of data taking and processing for each cluster must be tracked. 

Internal data management of data products was designed around the ESO
FITS raw data structure. The extracted, wavelength calibrated and sky
subtracted spectra are in a 2d ``image'' with the corresponding
fibre information in binary table extensions. Processing and quality control
information is propagated through the FITS header. The outputs
from all later stages of processing are incorporated in further
binary table extensions. This model, where all relevant information
about an observation is kept in the same container file, was developed by the Cambridge Astronomical Survey Unit (CASU)
team for the VISTA pipelines, and cuts bookkeeping tasks down to a minimum.

\subsection{Operational database (WG17)} 

The Gaia-ESO project utilised both an operational database, based in Cambridge at CASU,  and a
dedicated archive, based in Edinburgh, to hold all relevant information, to complement the
primary public stable archive, which is that of ESO.  This approach builds on
the operational VISTA (and other projects) systems hosted at CASU, Cambridge,
and at the Wide Field Astronomy Unit (WFAU), Edinburgh. It is a proven and successful model, and proved
essential to support the survey data flow. In particular, this approach
kept separate the day-to-day processing, the spectrum analyses, and
all activities in which data are being determined or updated, from all
those science activities which should be based on readily accessible
static information.

The operational database held all data while it remained incomplete, or
subject to change. All observation preparation WGs submitted all relevant
data associated with target selection, up to and including OB
preparation, to the operational database [WG1-WG6]. Raw data from ESO
were added. Pipelines [WG7-WG9] operated on the raw data, generated
pipeline reduced and extracted spectra, with associated variances, quality control (QC)
info, RVs and classification outputs, and wrote these back to the
operational database. An aspect of the QC is the quantitative SNR data, also
supplied to ESO through the regular survey progress reports. Spectrum
analysis groups [WG10-WG15] read, but do not modify, these spectra,
carry out their analyses, and deliver back FITS-table results to the
operational database. These tables are attached to each relevant
spectrum, allowing the progress monitor to be updated.

Subsequent processing stages collated and incorporated available 
existing information such as photometric indices, proper motions, and
also appended derived radial velocities and stellar atmosphere parameters
in additional FITS table extensions. 
Further processing stages involving common tasks such as continuum 
estimation and normalisation to unity are readily  
incorporated by including extra FITS extensions.
The philosophy is to keep the same file architecture throughout and
minimise bookkeeping and versioning issues by always attaching information 
directly to the files.  This also has the advantage of removing direct 
dependency on availability of access to external databases.
To minimise resampling of spectra we also maintained extracted
spectra in natural units (e.g. pixel-based fluxes) and use FITS table 
extensions and/or FITS header information to specify conversions to physical 
units, should this ever be appropriate and required, and/or to 
zero-velocity systems.

We use the Gaia DPAC model - all data are stored in the central
repository, taken out for use/analysis, and have parameters
returned. No process adjusts the spectra, except by resetting the
master spectrum if needed when the whole process restarts from
scratch. No data processing WG talks directly to the archive. When the
spectrum teams agree their job is converged for some source, all
relevant data become fixed. 

At this stage all data are copied to the permanent Gaia-ESO archive,
hosted at WFAU Edinburgh, and become available internally to the survey consortium to begin science quality control. 

\subsection{Internal survey archive (WG18)}\label{archives} 
After pipeline processing, including atmospheric parameter and
abundance determination etc., the spectroscopic survey data were
made available to the consortium for quality control, science
verification and preliminary analysis via a bespoke archive system.
This system acts not only as the `internal' archive system for the
consortium, but also as a publicly accessible portal that provides
enhanced database--driven products to facilitate world community
exploitation of prepared static releases of survey data.  Metadata
associated with the products available in the archive complies with
the corresponding VO Data Models, in particular with the VO Spectrum
Data Model. This follows the tried--and--tested Vista Data Flow System
(VDFS) model developed by Cambridge and Edinburgh, for UKIRT--WFCAM,
VISTA-VIRCam and VST survey data, and supports both internal team
science verification, and provides a global archive system
complementary to that provided by ESO.

The Gaia-ESO Survey archive design follows proven models, and 
includes the following features:
\begin{itemize}
  \item back--end relational database management system store
  \item a near--normalised relational design to track all data, 
metadata and provenance through the pipeline and subsequent analysis stages
  \item simple interface applications for the novice user
  \item tabular data (i.e. catalogues) available through  
ConeSearch protocols
  \item SQL interface and relational model exposed to users through 
interactive web forms for more complicated usage scenarios
  \item integration of multi--wavelength catalogue data and image thumbnails
  \item publication of spectroscopic data to the VO through the 
Simple Spectral Access Protocol
  \item publication of all tabular data (e.g.~input catalogues, 
derived physical quantities for targets etc.) to the VO through the 
Table Access Protocol
(thereby making them accessible to sophisticated client--side exploration 
and analysis utilities such as TOPCAT)
   \item finally, cross-linking to the Gaia EDR3 star identifications is
 provided in the ESO SAF, while updated cross-matching to the Gaia-ESO CNAME identifiers will be provided as a Gaia data product with future Gaia data releases.
\end{itemize}

The internal CASU Gaia-ESO archive is hosted at Cambridge/CASU. \footnote{ http://casu.ast.cam.ac.uk/gaiaeso/}

The public Gaia-ESO Survey archive is Edinburgh/WFAU. \footnote{ http://ges.roe.ac.uk/}

The primary public data archive is the ESO Science Archive Facility. \footnote{ http://archive.eso.org/cms.html}

\section{Gaia-ESO Survey public data products}\label{ESODR}

This section introduces the final data release content. A fuller description of this is provided in the companion paper \citep{Randich21}, while the detailed final set of homogenisation processes is described in the WG15 paper \citet{Hourihane21}.

All spectra and derived parameters and abundances are available through the ESO Science Archive Facility \footnote{http://archive.eso.org/cms/data-portal.html} and the dedicated WFAU portal. \footnote{ http://ges.roe.ac.uk/pubs.html}

Preliminary data releases have been available through these archives during the survey. The set of reduced spectra, apart from data for 592 objects now also released, have been available through the ESO SAF since December 2020.

Summary survey statistics are provided here in Table~\ref{Spectra-origin} for the relative numbers of survey and archive spectra, Table~\ref{Spectra-summary} for the total number of spectra by spectrograph setup, and Table~\ref{Spectra-crosses} for the cross-matches between setups.

\begin{table*}
\begin{center}
\caption{Gaia-ESO Survey Final Data Set: numbers of spectra  }
\begin{tabular}{|l|lcc|}
\hline
 & Gaia-ESO & ESO Archive  & Total Spectra  \\
\hline
UVES & 14484 & 1954   & 16438 \\
\hline
GIRAFFE & 178698 &  7097 & 185795 \\
\hline
TOTAL & 193182 & 9051  & 202233 \\
\hline
\end{tabular}
\label{Spectra-origin}
\end{center}
\end{table*}

\begin{table*}
\begin{center}
\caption{Gaia-ESO Survey Final Data Set: numbers of unique sources by setup }
\begin{tabular}{l|ccccccccccccc}
\hline
Total & U580 & U520 & HR10 & HR21 & HR15N & HR3 & HR4 & HR5A & HR5B & HR6 & HR9B & HR14A & HR14B \\
115614 & 6641 & 500 & 59722 & 66542 & 40973 & 2228 & 1253 & 2072 & 107 & 2121 & 3873 & 2252 & 107 \\
\hline
\end{tabular}
\label{Spectra-summary}
\end{center}
\end{table*}

\begin{table*}
\begin{center}
\caption{Gaia-ESO Survey Final Data Set: number of cross-matches by setup}
\begin{tabular}{l|cccccccccccc}
\hline
 & U520 & HR10 & HR21 & HR15N & HR3 & HR4 & HR5A & HR5B & HR6 & HR9B & HR14A & HR14B \\
U580 & 110 & 171 & 249 & 869 & 31 & 9 & 26 & 1 & 26 & 162 & 70 & 1 \\
U520 & - & 55 & 61 & 77 & 22 & 0 & 22 & 1 & 22 & 104 & 22 & 1 \\
HR10 & - & - & 58354 & 3834 & 181 & 167 & 181 & 1 & 181 & 253 & 211 & 1 \\
HR21 & - & - & - & 3805 & 181 & 167 & 181 & 1 & 181 & 253 & 211 & 1 \\
HR15N & - & - & - & - & 259 & 245 & 259 & 1 & 259 & 794 & 286 & 1 \\
HR3 & - & - & - & - & - & 1250 & 2069 & 107 & 2114 & 468 & 2045 & 107 \\
HR4 &  - & - & - & - & - & - & 1198 & 106 & 1249 & 289 & 1185 & 106 \\
HR5A &  - & - & - & - & - & - & - & 57 & 2066 & 469 & 2046 & 57 \\
HR5B &  - & - & - & - & - & - & - & - & 107 & 55 & 57 & 107 \\
HR6 &  - & - & - & - & - & - & - & - & - & 471 & 2047 & 107 \\
HR9B &  - & - & - & - & - & - & - & - & - & - & 481 & 55 \\
HR14A &  - & - & - & - & - & - & - & - & - & - & - & 57 \\
\hline
\end{tabular}
\label{Spectra-crosses}
\end{center}
\end{table*}

\begin{table*}
\begin{center}
\caption{Gaia-ESO Survey Final Data Set: elemental abundances determined }
\begin{tabular}{c|c|c|c|c|c|c|}
\hline
He1 & C1 & C2 & C3 &  C-C2 & N2 & N3 \\
N-CN & O1 & O2 & Ne1 & Na1 & Mg1 & Mg2 \\
Al1 & Al2 & Si1 & Si2 & Si3 & Si4 & S1 \\
Ca1 & Ca2 & Sc1 & Sc2 & Ti1 & Ti2 & V1 \\
Cr1 & Cr2 & Mn1 & Co1 & Ni1 & Cu1 & Zn1 \\
Sr1 & Y2 & Zr1 & Zr2 & Mo1 & Ba2 & La2 \\
Ce2 & Pr2 & Nd2 & Sm2 & Eu2 &  & \\
\hline
\end{tabular}
\label{elements}
\end{center}
\end{table*}

The survey yielded 185795 GIRAFFE spectra for 108473 unique stars and
16438 UVES spectra for 7141 stars (see Table~\ref{Spectra-origin}) , most of
which were observed at two different epochs.   The products delivered at the end of the survey include all the
extracted spectra, with relevant ancillary information, and value
added deliverables.

Each source has a unique CNAME identifier, in standard coordinate format, and is cross-matched to the Gaia EDR3 identifiers. The Gaia cross-match will be maintained as part of future Gaia data releases.

The final data release includes, for
all targets with completed observations:

\noindent {\bf Advanced Data Products - reduced data }
These are the outputs of Working Groups 0-9.
\begin{itemize}
\item
One-dimensional, wavelength calibrated, sky-subtracted, normalised,
UVES and/or GIRAFFE spectra for each survey target.  Where no RV variability is
detected, co-added sum spectra are provided, in addition to
single-epoch spectra. UVES spectra are provided as sets of
single echelle orders and a merged spectrum
\item The associated variance spectrum
\item Associated quality control information 
\item Supplementary value-added data include:
 \item
The photometry (and additional membership information for
clusters) used to select the targets 
\item The class of target - cluster, standard, etc
\item Selected matched
multi-wavelength photometric data where available
 \item Object classification 
 \item Radial velocity and its error distribution function
 \item Analysis for RV variability
 \item Projected rotational velocity and error estimate (where relevant)
\end{itemize}

  {\bf Advanced Data Products - astrophysical parameters}
Advanced data products from expanded and refined spectral analysis,
calibrated using the current Gaia-ESO calibrations. These are the
outputs of iterative homogenisation and quality control involving
Working Groups 10-15, and survey consortium science verification analysis. 

\begin{itemize}
\item Whenever possible stellar astrophysical parameters: 
effective temperature, surface gravity
\item Whenever possible stellar metallicity [Fe/H]
\item Whenever possible [$\alpha$/Fe] ratios
\item Measurements of stellar activity or mass accretion/ejection rates, for cluster members (where relevant)
\item Quantitative mass loss estimates, for early-type stars
\item Elemental abundances, with the specific elements depending on
target type. 
\item Quantitative uncertainties on the delivered quantities, derived
  from the multiple reduction systems implemented
\end{itemize}

The final data-release includes homogenised values for all deliverables listed above for all stars,
  calibrated onto the final Gaia-ESO calibration system. Halpha, tSNE and BIN flags are allocated to 20,019, 17,408 and 1865 sources respectively.
Elemental abundance determinations are provided as listed in Table~\ref{elements}. Note that in the final 
abundance release, we do not provide Fe1 and Fe2 abundances
and we leave only [Fe/H] to indicate the iron abundance. The [Fe/H] parameter is homogenised together with
the other stellar parameters ($T_{\rm eff}$, $\log g$, [Fe/H], $\xi$,  $v \sin i$), while Fe1 and Fe2 are re-computed by
Nodes keeping the input stellar parameters fixed. Since the stellar parameters, including
[Fe/H], are homogenised using external calibrators, in some cases, the recomputed
abundances do not completely fulfill the ionization balance, and there might be differences in
Fe1 and Fe2 (in some specific areas of the parameter space, in particular at high metallicity).
To avoid misleading results, we provide only [Fe/H].

Discussion of the astrophysical quality and value of the parameters is described in \citet{Randich21}, while an analysis of the 
precision of the parameters, and external comparisons is in \citet{Hourihane21}. An example of the power of combining Gaia-ESO and Gaia information in astrophysical analyses is 
provided by \citet{Jackson2021}, who provide membership analysis for 63 open clusters and 7 globular clusters.

\section {Conclusions} \label{Conclusions}

The Gaia-ESO Public Spectroscopic Survey has been a large ambitious VLT spectroscopic survey of representative samples of the main Galactic stellar populations. One primary aim was to provide high-quality astrophysical parameters, radial velocities to complement Gaia kinematics, and elemental abundances for stars of all accessible ages and abundances, with parameters firmly anchored to open and globular clusters. This links field star properties to tested isochrones, establishing a basis to map temporal as well as spatial evolution in the Galaxy. A second aim was to provide consistent-quality spectroscopic studies of a large sample of open clusters, mapping accessible age-metallicity-location space. For each cluster a wide mass-range is accessed, linking the complementary analysis approaches required for hot and cool stars, (pre-)main sequence and evolved, ensuring consistency. A third aim was to provide a large sample, of order $10^5$, field stars, sufficient to map populations across chemical abundance and kinematic properties. A fourth aim was to establish a robust set of calibration stars, suitable for use both in Gaia calibration, and for cross-calibrations between the several recent, current or planned stellar spectroscopic surveys and asteroseismic projects. Another ambition was to bring together the many European stellar spectroscopy groups into successful partnership, providing a robust community foundation for those forthcoming survey projects. Finally, a requirement is that all the data, calibrated spectra and derived parameters, be available in a free public archive for future uses.

Gaia-ESO succeeded in all these ambitions. The ambitious target of $10^5$ field stars and $\sim 60$ open clusters were surveyed, with 114325 stars and 62 open clusters newly surveyed. A very wide range of parameter space was mapped. Substantial scientific advances have been made, with over 100 team science verification refereed papers already published. The spectra are already widely accessed from the ESO SAF, with the astrophysical parameters now also available. The community did come together, learning how to homogenise data from very many different analysis pipelines in a way consistent with the primary Gaia Benchmark Star calibrations. A community of some 400 scientists from more than 110 institutions came together to deliver the survey. Significant scientific results have already been delivered, some of which are presented in the companion paper \citet{Randich21}.

\begin{acknowledgement}
Based on data products from observations made with ESO Telescopes at the La Silla Paranal Observatory under programme ID 188.B-3002. These data products have been processed by the Cambridge Astronomy Survey Unit (CASU) at the Institute of Astronomy, University of Cambridge, and by the FLAMES/UVES reduction team at INAF/Osservatorio Astrofisico di Arcetri. Public access to the data products is via the ESO SAF, and the Gaia-ESO Survey Data Archive, prepared and hosted by the Wide Field Astronomy Unit, Institute for Astronomy, University of Edinburgh, which is funded by the UK Science and Technology Facilities Council.
This work was partly supported by the European Union FP7 programme through ERC grant number 320360 and by the Leverhulme Trust through grant RPG-2012-541. We acknowledge the support from INAF and Ministero dell' Istruzione, dell' Universit\`a' e della Ricerca (MIUR) in the form of the grant "Premiale VLT 2012". The project presented here benefited in development from discussions held during the Gaia-ESO workshops and conferences supported by the ESF (European Science Foundation) through the GREAT Research Network Programme. 
This research has made use of the SIMBAD database,
operated at CDS, Strasbourg, France.
R.Smiljanic acknowledges support from the National Science Centre, Poland (2014/15/B/ST9/03981).
This work was partly supported by the INAF grant for mainstream projects: "Enhancing the legacy of the Gaia-ESO Survey for open cluster science". F.J.E. acknowledges financial support from the Spanish MINECO/FEDER through the grant AYA2017-84089 and MDM-2017-0737 at Centro de Astrobiología (CSIC-INTA), Unidad de Excelencia María de Maeztu, and from the European Union’s Horizon 2020 research and innovation programme under Grant Agreement no. 824064 through the ESCAPE - The European Science Cluster of Astronomy and Particle Physics ESFRI Research Infrastructures project. T.B. was funded by the “The New Milky Way” project grant from the Knut and Alice Wallenberg Foundation. S.R.B. acknowledges support by the Spanish Government under grants AYA2015-68012-C2-2-P and PGC2018-093741-B-C21/C22 (MICIU/AEI/FEDER, UE). W. J. S. acknowledges CAPES for a PhD studentship. J.M.A. acknowledges support from the Spanish Government Ministerio de Ciencia e Innovación through grants AYA2013-40611-P, AYA2016-75931-C2-2-P, and PGC2018-095049-B-C22. T.M. and others from STAR institute, Liege, Belgium  are grateful to Belgian F.R.S.-FNRS for support, and are also indebted for an ESA/PRODEX Belspo contract related to the {\em Gaia} Data Processing and Analysis Consortium and for support through an ARC grant for Concerted Research Actions financed by the Federation Wallonie-Brussels.. This research has been partially supported by the following grants: MIUR Premiale "{\em Gaia}-ESO survey" (PI S. Randich), MIUR Premiale "MiTiC: Mining the Cosmos" (PI B. Garilli), the ASI-INAF contract 2014-049-R.O: "Realizzazione attivit\`a tecniche/scientifiche presso ASDC" (PI Angelo Antonelli), Fondazione Cassa di Risparmio di Firenze, progetto: "Know the star, know the planet" (PI E. Pancino), and Progetto Main Stream INAF: "Chemo-dynamics of globular clusters: the {\em Gaia} revolution" (PI E. Pancino). V.A.acknowledges the support from Funda\c{c}\~ao para a Ci\^encia e Tecnologia (FCT) through Investigador FCT contract nr. IF/00650/2015/CP1273/CT0001. AJK acknowledges support by the Swedish National Space Agency (SNSA). AB acknowledges support by ANID, -- Millennium Science Initiative Program -- NCN19\_171, and FONDECYT regular 1190748. E. M. acknowledges financial support from the Spanish State Research Agency (AEI) through project MDM-2017-0737 Unidad de Excelencia "Maria de Maeztu" - Centro de Astrobiología (CSIC-INTA). T.Z. acknowledges financial support of the Slovenian Research Agency (research core funding No. P1-0188) and the European Space Agency (Prodex Experiment Arrangement No. C4000127986). P.J. acknowledges support FONDECYT Regular 1200703. The work of I.N. is partially supported by the Spanish Government Ministerio de Ciencia, Innovación y Universidades under grant PGC2018-093741-B-C21 (MICIU/AEI/FEDER, UE). Funding for this work has been provided by the ARC Future Fellowship FT160100402. CAP acknowledges financial support from the Spanish Government through research grants MINECO AYA 2014-56359-P, MINECO AYA2017-86389-P, and MICINN PID2020-117493GB-I00. S.F. was supported by the grants 2011-5042 and 2016- 03412 from the Swedish Research Council and the project grant "The New Milky Way" from the Knut and Alice Wallenberg Foundation.  CASU is supported through STFC grants: ST/H004157/1, ST/J00541X/1, ST/M007626/1, ST/N005805/1, ST/T003081/1. Work reported here benefited from support through the GREAT-ITN FP7 project Grant agreement ID: 264895. DKF acknowledges funds from the Alexander von Humboldt Foundation in the framework of the Sofja Kovalevskaja Award endowed by the Federal Ministry of Education and Research and the grant 2016-03412 from the Swedish Research Council. A.H. acknowledges support from the Spanish Government Ministerio de Ciencia e Innovación and ERD Funds through grants PGC-2018-091 3741-B-C22 and CEX2019-000920-S. X.F. acknowledge the support of China Postdoctoral Science Foundation 2020M670023. M.~L.~L.~Dantas acknowledges the Polish NCN grant number 2019/34/E/ST9/00133. Part of this work was funded by the Deutsche Forschungsgemeinschaft (DFG, German Research Foundation) -- Project-ID 138713538 -- SFB 881 (``The Milky Way System'', subproject A09). MZ acknowledge support from the National Agency for Research and Development (ANID) grants: FONDECYT Regular 1191505, Millennium Institute of Astrophysics ICN12-009, BASAL Center for Astrophysics and Associated Technologies AFB-170002. R.B. acknowledges support from the project PRIN-INAF 2019 "Spectroscopically Tracing the Disk Dispersal Evolution". HMT acknowledges financial support from the Agencia Estatal de Investigación of the Ministerio de Ciencia, Innovación y Universidades through projects PID2019-109522GB-C51,54/AEI/10.13039/501100011033, and the Centre of Excellence ``María de Maeztu'' award to Centro de Astrobiología (MDM-2017-0737). JIGH acknowledges financial support from the Spanish Ministry of Science and Innovation (MICINN) project AYA2017-86389-P, and also from the Spanish MICINN under 2013 Ramon y Cajal program RYC-2013-14875. V.P.D. is supported by STFC Consolidated grant ST/R000786/1. N.L. acknowledges financial support from “Programme National de Physique Stellaire” (PNPS) and the “Programme National Cosmology et Galaxies (PNCG)” of CNRS/INSU, France. A.~R.~C. is supported in part by the Australian Research Council through a Discovery Early Career Researcher Award (DE190100656). Parts of this research were supported by the Australian Research Council Centre of Excellence for All Sky Astrophysics in 3 Dimensions (ASTRO 3D), through project number CE170100013. PSB is Supported by the Swedish Research Council through individual project grants with contract Nos. 2016-03765 and 2020-03404. AM acknowledges funding from the European Research Council (ERC) under the European Union’s Horizon 2020 research and innovation programme (grant agreement No. 772293 - project ASTEROCHRONOMETRY). JP was supported by the project RVO: 67985815. E.D.M. acknowledges the support from FCT through the research grants UIDB/04434/2020 \& UIDP/04434/2020 and through Investigador FCT contract IF/00849/2015/CP1273/CT0003. This work was (partially) supported by the Spanish Ministry of Science, Innovation and University (MICIU/FEDER, UE) through grant RTI2018-095076-B-C21, and the Institute of Cosmos Sciences University of Barcelona (ICCUB, Unidad de Excelencia ’Mar\'{\i}a de Maeztu’) through grant CEX2019-000918-M. SLM acknowledges the support of the UNSW Scientia Fellowship program and the Australian Research Council through Discovery Project grant DP180101791. GT acknowledges financial support of the Slovenian Research Agency (research core funding No. P1-0188) and the European Space Agency (Prodex Experiment Arrangement No. C4000127986). S.G.S acknowledges the support from FCT through Investigador FCT contract nr. CEECIND/00826/2018 and POPH/FSE (EC). H.G.L. acknowledges financial support by the Deutsche Forschungsgemeinschaft (DFG, German Research Foundation) -- Project-ID 138713538 -- SFB 881 (``The Milky Way System'', subproject A04). This work was (partially) supported by the Spanish Ministry of Science, Innovation and University (MICIU/FEDER, UE) through grant RTI2018-095076-B-C21, and the Institute of Cosmos Sciences University of Barcelona (ICCUB, Unidad de Excelencia ’Mar\'{\i}a de Maeztu’) through grant CEX2019-000918-M. T.K. is supported by STFC Consolidated grant ST/R000786/1. MV acknowledges the support of the Deutsche Forschungsgemeinschaft (DFG, project number: 428473034). T.M. is supported by a grant from the Fondation ULB. We acknowledge financial support from the Universidad Complutense de Madrid (UCM) and by the Spanish Ministerio de Ciencia, Innovación y Universidades, Ministerio de Economía y Competitividad, from project AYA2016-79425-C3-1-P and PID2019-109522GB-C5[4]/AEI/10.13039/501100011033. U.H. acknowledges support from the Swedish National Space Agency (SNSA/Rymdstyrelsen). D.G. gratefully acknowledges support from the Chilean Centro de Excelencia en Astrof\'isica y Tecnolog\'ias Afines (CATA) BASAL grant AFB-170002. D.G. also acknowledges financial support from the Direcci\'on de Investigaci\'on y Desarrollo de la Universidad de La Serena through the Programa de Incentivo a la Investigaci\'on de Acad\'emicos (PIA-DIDULS). A. Lobel acknowledges support in part by the Belgian Federal Science Policy Office under contract No. BR/143/A2/BRASS. We acknowledge financial support from the Universidad Complutense de Madrid (UCM) and by the Spanish Ministerio de Ciencia, Innovación y Universidades, Ministerio de Economía y Competitividad, from project AYA2016-79425-C3-1-P and PID2019-109522GB-C5[4]/AEI/10.13039/501100011033. AM acknowledges the support from the Portuguese Funda\c c\~ao para a Ci\^encia e a Tecnologia (FCT) through the Portuguese Strategic Programme UID/FIS/00099/2019 for CENTRA. TM acknowledges financial support from the Spanish Ministry of Science and Innovation (MICINN) through the Spanish State Research Agency, under the Severo Ochoa Program 2020-2023 (CEX2019-000920-S). EJA acknowledges funding from the State Agency for Research of the Spanish MCIU through the Center of Excellence Severo Ochoa award to the Instituto de Astrofisica de Andalucia (SEV-2017-0709).

\end{acknowledgement}


\bibliographystyle{aa} 
\bibliography{biblio}

\end{document}